\documentclass[11pt,a4paper]{article} 
\usepackage{jheppub}
\usepackage{amsthm}
\usepackage{xcolor}
\usepackage{hyperref} 
\usepackage{url}
\usepackage{microtype}
\usepackage{subcaption}
\usepackage{empheq}
\usepackage{booktabs}
\usepackage[makeroom]{cancel}
\usepackage{dsfont} 
\usepackage{ytableau} 
\usepackage{setspace}
\usepackage[skins]{tcolorbox}
\usepackage{mathrsfs}

\newcommand{\beq}{\begin{equation}}
\newcommand{\eeq}{\end{equation}}

\newcommand{\E}{{\epsilon}}

\newcommand{\fg}{\mathfrak{g}}

\newcommand{\fq}{\mathfrak{q}}

\newcommand{\cC}{\mathcal{C}}
\newcommand{\cN}{\mathcal{N}}
\newcommand{\cW}{\mathcal{W}}

\newcommand{\cY}{\mathcal{Y}}

\newcommand{\sh}{\mbox{sh}}

\newtheoremstyle{fullit}
  {\topsep}      
  {\topsep}      
  {\normalfont}  
  {0pt}          
  {\itshape}     
  {.\ }          
  {0pt}          
  {\thmname{#1} \thmnumber{#2}}             

\theoremstyle{definition}

\theoremstyle{fullit}

\DeclareCaptionSubType*[alph]{figure}
\title{Non-Perturbative Schwinger-Dyson Equations for 3d ${\cal N} = 4$ Gauge Theories}
\author{Nathan Haouzi}
\affiliation{Simons Center for Geometry and Physics, Stony Brook University,\vspace{-0.2cm}
		
	Stony Brook, NY 11794, USA}
	\emailAdd{nhaouzi@scgp.stonybrook.edu}

\abstract{We analyze symmetries corresponding to separated topological sectors of 3d ${\cal} N=4$ gauge theories with Higgs vacua, compactified on a circle. The symmetries are encoded in Schwinger-Dyson identities satisfied by correlation functions of a certain gauge-invariant operator, the ``vortex character." Such a character observable is realized as the vortex partition function of the 3d gauge theory, in the presence of a 1/2-BPS line defect. The character enjoys a double refinement, interpreted as a deformation of the usual characters of finite-dimensional representations of quantum affine algebras.
We derive and interpret the Schwinger-Dyson identities for the 3d theory from various physical perspectives: in the 3d gauge theory itself, in a 1d gauged quantum mechanics, in 2d $q$-Toda theory, and in 6d little string theory. We establish the dictionary between all approaches. Lastly, we comment on the transformation properties of the vortex character under the action of 3-dimensional Seiberg duality.}

\begin{document}
		\maketitle
		\clearpage


\newpage

\section{Introduction}

Since its inception, supersymmetry has been a formidable tool to understand the dynamics of gauge theories in various dimensions. More recently, the success of localization methods \cite{Pestun:2016zxk} has brought about a flourish of new results, often shedding light on hidden structures and symmetries in the strongly coupled regime. Among those, novel symmetries of gauge theories in 4 dimensions were uncovered by exhibiting non-perturbative Schwinger-Dyson type equations satisfied by certain correlators of the quantum field theory \cite{Nekrasov:2015wsu,Nekrasov:2016qym}.

Concretely, a correlator is defined via a path integral in quantum field theory. Schwinger-Dyson equations can be understood as constraints that must be satisfied by such a correlator. This comes about from demanding that the path integral be invariant under an infinitesimal shift of the  contour (under the condition that the integral measure is left invariant by such a shift). The particular question asked in \cite{Nekrasov:2015wsu} was to determine what type of constraints must be satisfied by correlators in Yang-Mills theory, when a contour gets shifted from a given topological sector to another distinct topological sector of the theory, related to the former by a large gauge transformation. 
Recall that the connected components of the space of gauge fields are labeled by an integer called the instanton charge: $\frac{-1}{8\pi^2}\int\text{Tr}F\wedge F$, where $F$ is the field strength and the domain of integration is the spacetime. Then, the problem can be recast in a simple way:  what symmetries of the gauge theory are made manifest when the instanton charge varies?

The question was answered in the context of supersymmetric $\cN=2$ Yang-Mills, on a regularized spacetime called the $\Omega$-background \cite{Nekrasov:2002qd,Losev:2003py,Nekrasov:2010ka} on $\mathbb{C}^2$. On this background, the instanton number can be changed by adding and removing point-like instantons in a controlled way, and the shift of contour in the definition of the path integral turns into the discrete operation of adding and removing boxes in a Young tableau \cite{Nekrasov:2003rj}.

To mediate the change in instanton number of the theory, it is convenient to construct a local 1/2-BPS codimension-4 ``$Y$-operator," as a function of an auxiliary complex parameter $M\in\mathbb{C}$. Then, Schwinger-Dyson equations are understood as regularity conditions for the vev  $\left\langle Y(M)\right\rangle$ in the parameter $M$. Put differently, the correlator typically has poles in the fugacity $M$, but the Schwinger-Dyson equations tell us that there exists a precise sum of $Y$-operator vevs which is pole-free in $M$, nicknamed the $qq$-character observable. Here, each ``$q$" stands for one of the two parameters of the $\Omega$-background on $\mathbb{C}^2$, and the term ``character" is used because the observable is a (deformed) character of a finite dimensional representation of a Yangian algebra. 

The above construction can be generalized in many ways, for example by considering additional defects in the background \cite{Nekrasov:2017rqy,Jeong:2018qpc}, by studying different gauge groups \cite{Haouzi:2020yxy,Haouzi:2020zls}, or by going away from 4 dimensions: the case of a 5-dimensional gauge theory compactified on a circle has been an particularly fruitful area of research \cite{Tong:2014cha,Kim:2016qqs,Kimura:2015rgi,Kimura:2017hez,Mironov:2016yue,Assel:2018rcw,Haouzi:2019jzk,Bourgine:2017jsi,Bourgine:2019phm,Chang:2016iji}, where the $qq$-character observable arises not as an object defined in the representation theory of Yangians, but instead in the representation theory of quantum affine algebras. Likewise, in the case of a 6-dimensional gauge theory compactified on a 2-torus  \cite{Kimura:2016dys,Agarwal:2018tso}, the $qq$-character observable becomes an object in the representation theory of quantum elliptic algebras.   Remarkably, equivariant localization on the $\Omega$-background can be performed to yield exact expressions for the $qq$-character observables in all of the above cases.\\

Meanwhile, supersymmetric gauge theories in codimension-2 lower dimensions share many common features with their higher-dimensional counterparts, but have not yet been studied in any systematic way. Most notably, there exist once again distinct topological sectors of the theory, this time labeled by an integer called the vortex charge: $\frac{1}{2\pi}\int\text{Tr}F$, where $F$ is the field strength and the integration is over the two real dimensions transverse to the vortex. By the logic we reviewed above, one should then expect non-perturbative Schwinger-Dyson equations to exist also in dimensions 2, 3 (compactified on a circle) and 4 (compactified on a 2-torus).
This time around, invariance under a slight shift of contour in the path integral should translate to a change in vortex number. To mediate such a shift, one could hope to construct as before a 1/2-BPS $Y$-operator, this time around of codimension-2, as a function of (at least) one auxiliary parameter $M$. Then, Schwinger-Dyson equations would again be understood as regularity conditions that the vev  $\left\langle Y(M)\right\rangle$ needs to satisfy in the parameter $M$.

Indeed, the existence of such non-perturbative equations has been anticipated for 2-dimensional gauged linear sigma models with $\cN=(4,4)$ supersymmetry: to exhibit the equations and their associated symmetries, a new vortex $qq$-character observable was conjectured to exist \cite{Nekrasov:2017rqy,Haouzi:2019jzk}, with the same Yangian symmetry as its 4-dimensional counterpart, but involving different twists.

The aim of this paper is to give a first-principles construction of low-dimensional non-perturbative Schwinger-Dyson equations, and interpret them from various physical perspectives. We find it convenient to work in a K-theoretic framework, i.e. we study 3-dimensional $\cN=4$ gauge theories $G^{3d}$ on the manifold $\mathbb{C}\times S^1$. Results for 2-dimensional gauged linear sigma model with $\cN=(4,4)$ supersymmetry can be obtained by reducing the 3d theory on the circle $S^1$. The theories 
$G^{3d}$ we focus on will be of quiver-type, labeled by an $ADE$ Lie algebra, with unitary gauge groups  and fundamental flavors. We require the amount of flavors to be large enough in order for $G^{3d}$ to be Higgsable, and introduce non-abelian versions of Nielsen–Olesen vortex solutions at the Higgs vacua. Our investigations leads us to define of a vortex character observable with quantum affine symmetry\footnote{The literature regarding the representation theory of quantum affine algebras is rich. As a short guide,  quantum affine algebras were introduced by Jimbo \cite{jimbo} and Drinfeld \cite{Drinfeld:1987sy}. The systematic study of their representations was initiated by Chari and Pressley \cite{Chari:1994pf, Chari:1994pd}. Characters of finite-dimensional representations of quantum affine algebras, dubbed ``$q$-characters," were first constructed by Frenkel and Reshetikhin in the 90's \cite{Frenkel:qch}. They were later rediscovered in a physical context when discussing the quantum geometry of 5d supersymmetric quiver gauge theories \cite{Nekrasov:2013xda,Bullimore:2014awa}. A deformed character depending on two parameters was  introduced in \cite{Shiraishi:1995rp,Awata:1995zk,Frenkel:1998} (for related  work on $t$-analogues of $q$-characters, see also \cite{Nakajima:tanalog}). This ``$qq$-character" was again rediscovered in the study of 5d supersymmetric gauge theories \cite{Nekrasov:2015wsu}. In this paper, we will find that it furthermore arises in the study of 3d supersymmetric quiver gauge theories.}.\\

In this 3-dimensional setting, the codimension-2 operator mediating the change in vortex number is a  1/2-BPS loop defect wrapping the circle $S^1$.  We are able to give four definitions of the vortex $qq$-character observable:\\

-- The vortex character is the Witten index of a 1-dimensional gauged supersymmetric quantum mechanics living on the vortices of $G^{3d}$, which includes additional chiral matter due to the defect.\\

-- The vortex character is a sum of half-indices for the 3d $\cN=4$ gauge theory $G^{3d}$, in the presence of a codimension-2 defect. More precisely, each term in the sum is a 3d/1d half-index, where 1-dimensional degrees of freedom due to the defect are coupled to the bulk 3d theory.\\

-- The vortex character is a deformed $\cW_{q,t}$-algebra correlator on an infinite cylinder, with stress tensor and higher spin current insertions, including a distinguished set of ``fundamental" vertex operators.\\

-- The vortex character is the partition function of the 6-dimensional $(2,0)$ little string theory compactified on the cylinder, in the presence of codimension-4 defects and a point-defect. The defects are all realized as D-branes of type IIB wrapping various 2-cycles of a resolved $ADE$ singularity.\\

We will analyze each perspective in detail, and prove that all four definitions are in fact equivalent. Let us briefly comment on them. The most obvious perspective is perhaps the 1-dimensional one. There, we describe microscopically the gauged quantum mechanics on the vortices in some Higgs vacuum of $G^{3d}$. In 3 dimensions, the vortices and the loop defect are described a 1-dimensional $\cN=4$ theory, whose quantum mechanics captures the corresponding dynamics. We count vortices in this background by computing the Witten index of the theory, with appropriate chemical potentials turned on. We show that this index is a deformed character of the finite-dimensional representation of a quantum affine algebra.

This Witten index can be reinterpreted directly from the perspective of  $G^{3d}$ itself, as a sum of half-indices, or holomorphic blocks \cite{Beem:2012mb}, and where the  loop is treated as a codimension-2 line defect. In this picture, the half-index of the 3d theory is computed via Coulomb branch localization, and the line defect is coupled to the bulk theory via gauging of its flavor symmetries. We show that, up to overall normalization, the vortex $qq$-character constructed from the vortex quantum mechanics is the sum of such coupled 3d/1d indices. In this presentation, each term in the sum stands for a weight in the finite-dimensional representation of a quantum affine algebra.

The 3d perspective turns out to have a very natural realization in terms of certain vertex operator algebras called ${\cW}(\fg)$-algebras. These are labeled by a simple Lie algebra $\fg$, which in this work will be simply-laced, and the choice of a nilpotent orbit, which in this work will be the maximal one. They realize the symmetry of Toda theory, here defined on an infinite cylinder. The particular case $\fg=A_1$ is known as Liouville theory, which enjoys Virasoro symmetry. When $\fg\neq A_1$, the Virasoro stress tensor remains, but there are also higher spin currents. 
In the 90's, Frenkel and Reshetikhin introduced a two-parameter deformation of the ${\cW}$-algebras, denoted as ${\cW}_{q,t}(\fg)$  \cite{Frenkel:1998}, and sometimes referred to as deformed ${\cW}$-algebras. Crucially, while an ordinary ${\cW}$-algebra has conformal symmetry, its deformation does not: instead, it is the symmetry of the so-called $q$-Toda theory on the cylinder. Correlators are defined in the free field formalism, as integrals over the positions of some deformed screening currents on  the cylinder. We show that the vortex $qq$-character of $G^{3d}$ is such a correlator: the 3d gauge content is realized as screening current insertions, the 3d flavor symmetry is realized as fundamental vertex operator insertions, and the loop defect  is realized as the insertion of a generating current operator. This latter type of operator includes the deformed stress tensor, but also ``higher spin" currents of the ${\cW}_{q,t}(\fg)$ algebra; they are all constructed in free field formalism as the commutant of the screening currents. There are ${\text{rank}}({\fg})$ independent generators constructed in this way, with spin $s$ in the range $2\leq s \leq {\text{rank}}({\fg})+1$. The vortex Schwinger-Dyson equations of the gauge theory $G^{3d}$ are now interpreted as Ward identities satisfied by the correlator in the ${\cW}_{q,t}(\fg)$-algebra.

Finally, the various operators appearing in the ${\cW}(\fg)$-algebra construction have a natural interpretation in $(2,0)$ little string theory compactified on the cylinder: they are all D-brane defects at points on this cylinder. Namely, D3 branes realize the screening charges and flavor vertex operators, while a set of  D1 branes realizes the stress tensor and higher spin currents of the $\cW$-algebra.

In hindsight, some of the relations are not too surprising: for instance,
the relation between the partition function of 3-dimensional supersymmetric gauge theories, adequately twisted\footnote{The twist in question is commonly called A-twist in our context.}, and the count of their BPS vortices, is well-established by now \cite{Pasquetti:2011fj,Beem:2012mb,Nekrasov:2008kza,Krattenthaler:2011da,Dimofte:2011ju,Taki:2013opa,Cecotti:2013mba,Fujitsuka:2013fga,Benini:2013yva,Hwang:2012jh,Yoshida:2014ssa,Benini:2015noa,Crew:2020jyf}\footnote{Similar results exist for gauge theories supported on 2-manifolds, see for instance \cite{Doroud:2012xw,Benini:2012ui}.}. Further coupling the 3-dimensional theory to a line defect, it is plausible that a 3d/1d half-index should reproduce the Witten index of our vortex quantum mechanics. Furthermore, the relation from gauge theory supersymmetric indices to $\cW$-algebra correlators is an illustration of the so-called BPS/CFT correspondence \cite{Losev:2003py,Alday:2009aq}. Lastly, it is known that the effective theory on D3 branes in $(2,0)$ little string  is precisely the 3d $\cN=4$ theory under study \cite{Aganagic:2017gsx}. The goal of the present paper is to make use of the 1/2-BPS line defect to flesh out these ideas in detail, and exhibit new non-perturbative physics in the process.\\

As an application, we briefly analyze the action of 3d Seiberg duality on the vortex character observable. This duality relates different 3d gauge theories as defined in the UV, but which flow to the same theory in the IR  \cite{Seiberg:1994pq}. Here, we construct Seiberg-dual characters directly from the vortex quantum mechanics, where the duality manifests itself as a wall-crossing phenomenon in the Witten index \cite{Hwang:2017kmk}. This perspective gives us complete control over the action of the duality in 3 dimensions; we find that vortex characters map to vortex characters under Seiberg duality, labeling one and the same representation of the underlying quantum affine algebra, up to twists.\\

The paper is organized as follows: in Section 2, we construct the vortex quantum mechanics of $G^{3d}$ in the presence of a loop defect, and show its Witten index is a vortex character. We comment on how to interpret our result as a set of non-perturbative Schwinger-Dyson identities. In section 3, we re-derive the vortex character directly from the 3d perspective, coupled to a loop defect.  In Section 4, we make contact with Ward identities in the deformed $\cW$-algebra picture.  In Section 5, we define the vortex character straight from little string theory in the presence of codimension-4 and point-like D-brane defects. In section 6, we discuss Seiberg duality and future directions. In section 7, we showcase all the results of the paper for the case of 3d $\cN=4$ SQCD.

\vspace{16mm}

\section{Schwinger-Dyson Equations: the Vortex Quantum Mechanics Perspective}
\label{sec:1dsection}

We start with a lightning review on 3-dimensional gauge theories with 8 supercharges, along with the 1/2-BPS objects that enter our story.

\subsection{3d $\cN=4$ Gauge Theory, Vortices and Loop Defect}
\label{ssec:review}

We consider a 3d $\cN=4$ quiver gauge theory $G^{3d}$ on the manifold $\mathbb{C}\times S^1(\widehat{R})$, where the quiver is labeled by a simply-laced Lie algebra $\fg$ of rank $n$, of shape the Dynkin diagram of $A_n$, $D_n$ or $E_n$. The radius of the circle is denoted by $\widehat{R}$.  For concreteness, the Lagrangian gauge group is a product of $n$ unitary groups, 
\beq\label{gaugegroup3d}
G=\prod_{a=1}^n U(N^{(a)})\; .
\eeq
We introduce flavor symmetry through the gauge group
\beq\label{flavorgroup3d}
G_F=\prod_{a=1}^n U(N^{(a)}_F)\; ,
\eeq
where the associated gauge fields of $U(N^{(a)}_F)$ are frozen. This produces $N^{(a)}_F$ hypermultiplets on node $a$, in the bifundamental representation $(N^{(a)}, \overline{N^{(a)}_F})$ of the group $U(N^{(a)})\times U(N^{(a)}_F)$. 

Finally,  we have hypermultiplets in the bifundamental representation $\oplus_{b>a}\, \Delta^{ab}\,(N^{(a)}, \overline{N^{(b)}})$ of the group $\prod_{a,b} U(N^{(a)})\times U(N^{(b)})$, where $\Delta^{ab}$ is the incidence matrix of $\fg$:  $\Delta^{ab}$ is equal to 1 if there is a link connecting nodes $a$ and $b$ in the Dynkin diagram of $\fg$, and is 0 otherwise. Then, $G^{3d}$ contains a total of $n-1$ such bifundamental hypermultiplets.\\

The $a$-th gauge group in the quiver contains an abelian factor $U(1)\in U(N^{(a)})$, from which one can define a conserved current $j^{(a)}=\frac{1}{2\pi}* \text{Tr}\, F^{(a)}$; the associated global symmetry makes up the so-called topological symmetry of $G^{3d}$. Coupling this current to a $U(1)$ factor from the gauge group results in a Fayet-Iliopoulos (FI) term for the $a$-th node of the quiver.

The theory $G^{3d}$ has a moduli space of vacua with Coulomb and Higgs branches, and a corresponding $SU(2)_C \times SU(2)_H$ R-symmetry, where each $SU(2)$ acts on the two branches separately. In particular, each of the $n$ FI parameters is a triplet under $SU(2)_H$; we decompose each such triplet into a real FI parameter and a complex one. Under the R-symmetry, the 3d $\cN=4$ Poincar\'{e} supercharges transform in the representation $(\bf{2,2})$. They obey the anticommutator relation
\beq
\{Q^\alpha_{a,a'}, Q^\beta_{b,b'}\}=\epsilon_{a b}\, \epsilon_{a' b'}\,\left(\gamma_\mu\, C\right)^{\alpha \beta} P^\mu\; ,
\eeq
where we introduced $SO(2,1)$ $\gamma$-matrices, the charge conjugation matrix $C$, and the 3-momentum $P^\mu$. The upper index $\alpha$ is a spinor index for $SO(2,1)$, while the lower indices $a, a'$ are indices for $SU(2)_C$ and $SU(2)_H$, respectively.  Additionally, the above supercharges obey a reality condition, which we omit writing explicitly here.\\

The aim of this work is to exhibit certain symmetries associated to finite energy configurations of BPS vortices, which sit at Higgs vacua of $G^{3d}$. Therefore, from now on, we require that all theories under study possess a Higgs branch, and moreover that all vacua we study be Higgs vacua. In other words, the flavor symmetry group $G_F$ should have a large enough rank. The vortices then arise as semi-local non-abelian versions of Nielsen-Olson solutions; they are codimension-2 particles, transverse to the $\mathbb{C}$-line and wrapping $S^1(\widehat{R})$.

Then we tune the moduli to sit at such a Higgs vacuum, and the gauge group $G$ breaks to its $U(1)$ centers. We furthermore turn on the $n$ real FI parameters. The complex FI parameters are set to zero throughout this paper. The R-symmetry is broken to $SU(2)_C \times U(1)_H$, and 1/2-BPS vortices solutions appear in the moduli space. They can be described as a 1-dimensional $\cN=4$ supersymmetric quantum mechanics\footnote{Here, by 1d $\cN=4$ supersymmetry we mean  the reduction of 2d $\cN=(2,2)$ supersymmetry to 1 dimension.}, preserving the supercharges $Q^1_{a,1}$ and $Q^2_{b,2}$ of the 3d theory. Those four supercharges anticommute to the generator of translations along the vortices, which we denote as $H$:
\beq
\{Q^1_{a,1}, Q^2_{b,2}\}=\epsilon_{a b}\, H
\eeq
We introduce a line defect to mediate the change in vortex number of the theory. Just like the vortices, in 3 dimensions the defect is a particle transverse to the $\mathbb{C}$-line wrapping $S^1(\widehat{R})$. It is 1/2-BPS and preserves the 4 supercharges above. We now come to its precise characterization.

\subsection{The Vortex Quantum Mechanics}
\label{ssec:QMvortex}

Let us first consider $G^{3d}$ without any defect. In that case, the 1d $\cN=4$ quantum mechanics on its vortices is well-known \cite{Hanany:2003hp,Shifman:2006kd,Eto:2006uw,Eto:2007yv}\footnote{The description of the moduli space of vortices relies on a D-brane construction in the work \cite{Hanany:2003hp}. However, some of the dynamical degrees of freedom considered there turn out to be non-normalizable zero modes when performing a careful field-theoretic analysis  \cite{Shifman:2006kd,Eto:2006uw,Eto:2007yv}; as a result, the K\"{a}hler potentials on the vortex moduli space are in general different in the brane and field theory approaches, with agreement only on certain BPS solutions \cite{Shifman:2011xc,Koroteev:2011rb}. In our context, the index of the vortex quantum mechanics we compute is insensitive to these discrepancies.}. Just like the bulk theory, it is a $\fg$-type quiver theory of rank $n$ which we call $T^{1d}_{pure}$, where the subscript ``pure" emphasizes the absence of defect for now. The Higgs branch of this quiver theory is the moduli space of $(k^{(1)},k^{(2)},\ldots,k^{(n)})$ vortices of $G^{3d}$, where $k^{(a)}$ is a positive integer denoting the rank of the $a$-th gauge group in the quantum mechanics. Concretely, The gauge group of $T^{1d}_{pure}$ is
\beq\label{gaugegroup1d}
\widehat{G}=\prod_{a=1}^n U(k^{(a)})\; .
\eeq
For a 3d gauge group $U(N^{(a)})$ with field strength $F^{(a)}$, each 1d rank above is identified as the nontrivial first Chern class $k^{(a)}=\frac{-1}{2\pi}\int \text{Tr} F^{(a)}$, where the integral is taken over the $\mathbb{C}$-line transverse to the vortex.

There are chiral multiplets in the bifundamental representation $\oplus_{b>a}\, \Delta^{ab}\,(k^{(a)}, \overline{k^{(b)}})$ and in the bifundamental representation $\oplus_{b>a}\, \Delta^{ab}\,(\overline{k^{(a)}}, k^{(b)})$ of $\prod_{a,b} U(k^{(a)})\times U(k^{(b)})$, where $\Delta^{ab}$ is again denoting the incidence matrix of $\fg$:  $\Delta^{ab}$ is equal to 1 if there is a link connecting nodes $a$ and $b$ in the Dynkin diagram of $\fg$, and is 0 otherwise.

Additionally, there is fundamental and antifundamental chiral matter which manifests itself as additional ``teeth" in the 1d quiver. The precise determination of such matter requires specifying the gauge group $G=\prod_{a=1}^n U(N^{(a)})$ and flavor group $G_F=\prod_{a=1}^n U(N_F^{(a)})$ of the 3d bulk theory. We denote this flavor symmetry by $\widehat{G}_F$. When the rank of $G_F$ is large enough, fully Higgsing the 3d quiver theory is always possible, for any $ADE$ Lie algebra. The resulting 1d theory is then a generic handsaw quiver variety \cite{Nakajima:2011yq}, with 1d chiral matter on all $n$ nodes. 
 
Namely, on the $a$-th node, there are $P^{(a)}$ chirals in the representation  $(k^{(a)},\overline{P^{(a)}})$ and $Q^{(a)}$  chirals  in the representation  $(\overline{k^{(a)}},Q^{(a)})$ of $U(k^{(a)})\times U(Q^{(a)})$.
As we reviewed in the previous section, the R-symmetry group of $T^{1d}_{pure}$ is $SU(2)_C\times U(1)_H$, and the R-charge assignment of the various fields is constrained by the superpotential, readable from the ``closed loops" in the quiver diagram.

The 3d FI parameter $\zeta_{3d}^{(a)}$ of the gauge group $U(N^{(a)})$ sets the BPS tension of the vortex on node $a$. It is related to the 1d gauge coupling $\xi_{1d}^{(a)}$ of the gauge group $U(k^{(a)})$, according to $\xi_{1d}^{(a)}\propto 1/(\zeta_{3d}^{(a)})^2$.
Meanwhile, the 3d gauge coupling $\xi_{3d}^{(a)}$ of the gauge group $U(N^{(a)})$ is related to the 1d FI parameter $\zeta_{1d}^{(a)}$ of the gauge group $U(k^{(a)})$, according to $\zeta_{1d}^{(a)}\propto 1/(\xi_{3d}^{(a)})^2$.\\

\begin{figure}[h!]
	\emph{}
	\centering
	\includegraphics[trim={0 0 0 0cm},clip,width=0.99\textwidth]{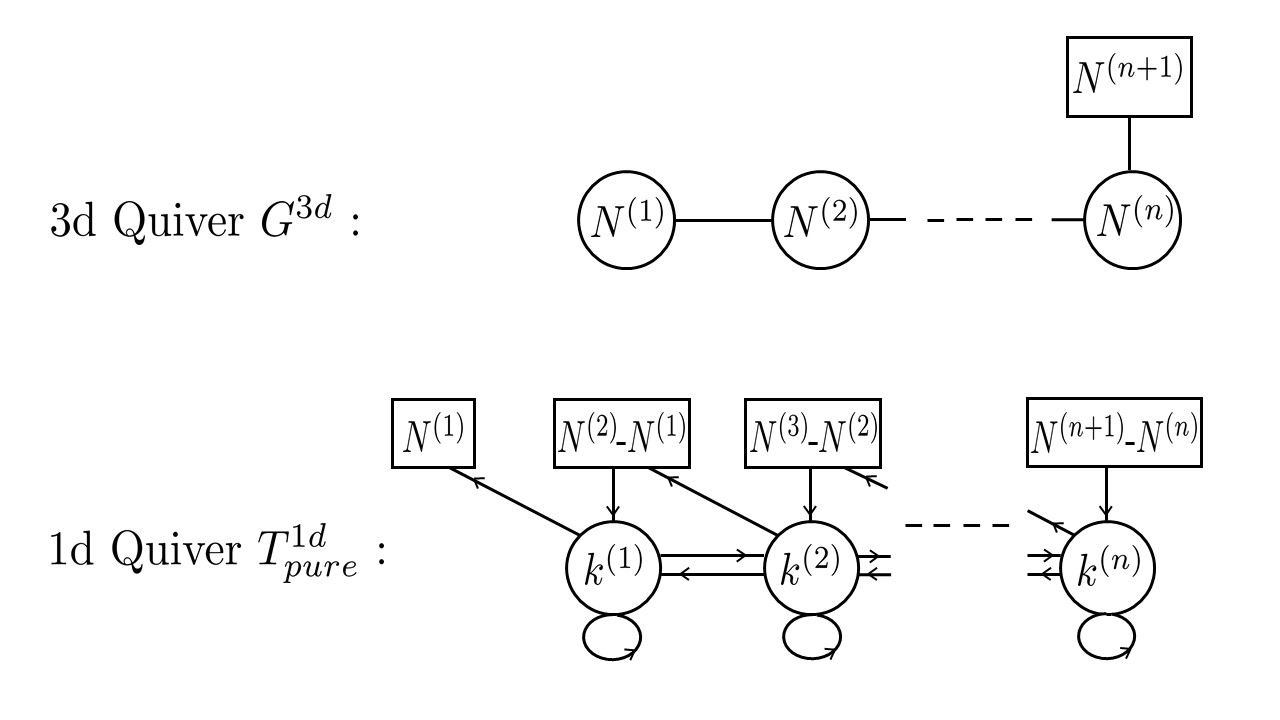}
	\vspace{-12pt}
	\caption{Example of the $G^{3d}$ theory $T_\rho[SU(N^{(n+1)})]$, and its vortex quantum mechanics $T^{1d}_{pure}$. Note we use a 3d $\cN=4$ notation for the quiver on top, and a 1d $\cN=4$ notation for the quiver on the bottom.} 
	\label{fig:flag}
\end{figure}

For a $\fg=A_n$ example, see Figure \ref{fig:flag}. The particular quiver theory $G^{3d}$ on top is sometimes called  $T_\rho[SU(N^{(n+1)})]$, labeled by a partition $\rho=[N^{(1)}, N^{(2)}-N^{(1)}, \ldots ,N^{(n+1)}-N^{(n)}]$ \cite{Gaiotto:2008ak}. The $n$ circles label the gauge group $G=\prod_{a=1}^n U(N^{(a)})$, the box on the right labels a  flavor symmetry group $G_F=U(N_F^{(n)})\equiv U(N^{(n+1)})$. An arrow between two circles labels a $\cN=4$ bifundamental hypermultiplet, while the arrow between the $n$-th circle and the box labels $N^{(n+1)}$ hypermultiplets in the fundamental representation of  $U(N^{(n)})$. The corresponding 1d $\cN=4$ $(k^{(1)},k^{(2)},\ldots,k^{(n)})$ vortex world-line theory $T^{1d}_{pure}$ is shown on the bottom. The circles label the gauge group $G=\prod_{a=1}^n U(k^{(a)})$, the looping arrows label adjoint chiral multiplets, and the straight arrows label fundamental/antifundamental chiral multiplets, which makes up the flavor symmetry $\widehat{G}_F=\prod_{a=1}^{n+1} U(N^{(a)}-N^{(a-1)})$. Specifically, in our previous notation, the number of fundamental chirals at node $a$ is $P^{(a)} = N^{(a)} - N^{(a-1)}$, while the number of antifundamental chirals at node $a$ is $Q^{(a)} = N^{(a+1)} - N^{(a)}$. There are two types of cubic contributions to the superpotential: the first type of terms is due to the bifundamental/adjoint chiral multiplets; meanwhile, the second type of terms is due to the bifundamental/fundamental/antifundamental chirals, meaning the flavor teeth. These superpotential terms can simply be read off the various triplets of arrows making closed loops in the quiver diagram. Mathematically, the theory $T^{1d}_{pure}$ in this example is known as a handsaw quiver, isomorphic to a parabolic Laumon space. This is the moduli space of (based) quasi-maps from $\mathbb{P}^1$ into the flag variety \cite{2010arXiv1009.0676F,Nakajima:2011yq,2013arXiv1301.7052V,Aganagic:2014oia}.\\

\begin{figure}[h!]
	\emph{}
	\centering
	\includegraphics[trim={0 0 0 0cm},clip,width=0.99\textwidth]{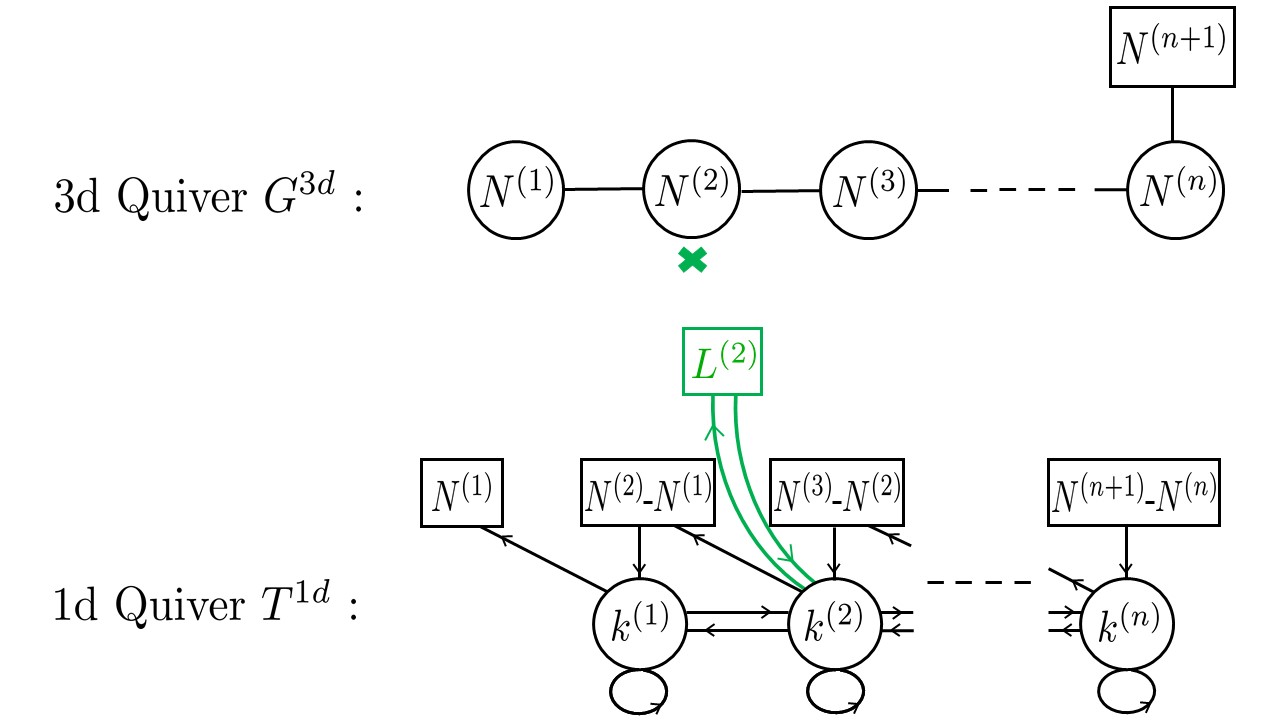}
	\vspace{-12pt}
	\caption{On the top, a loop defect is placed in the 3d theory $T_\rho[SU(N^{(n+1)})]$, with associated defect group $U(L^{(2)})$. We denoted the loop by a green cross. On the bottom, the vortex quantum mechanics $T^{1d}$ is displayed; black and green links  denote 1d $\cN=4$ chiral multiplets obtained by reduction of 2d $\cN=(2,2)$ supersymmetry.} 
	\label{fig:flagdefect}
\end{figure}

We now come to the new physics, and consider the 1/2-BPS loop defect which wraps $S^1(\widehat{R})$ in $G^{3d}$; we call the resulting quantum mechanics $T^{1d}$. The data characterizing a loop on the $a$-th node of the 3d theory, for $a\in\{1,\ldots, n\}$, is a new nondynamical gauge group $U(L^{(a)})$ for the 1-dimensional fields in the chiral multiplets localized on the $S^1(\widehat{R})$. These chiral multiplets transform in the representation $(k^{(a)},\overline{L^{(a)}})$ and $(\overline{k^{(a)}},L^{(a)})$ of $U(k^{(a)})\times U(L^{(a)})$. Note that the various charges of these additional multiplets are constrained by an extra cubic superpotential term on node $a$, due to the new fundamental/antifundamental/adjoint chiral multiplets present there. We thereby refer to the defect group for the entire quiver as: 
\beq
\widehat{G}_{defect}=\prod_{a=1}^n U(L^{(a)}) \;.
\eeq 
We label the Cartan subalgebra of the corresponding background gauge field by the parameters $\{M^{(a)}_\rho\}_{\rho=1,\ldots,L^{(a)}}$. 

On general ground, the Higgs branch of $T^{1d}$ should be understood as describing a generalized vortex moduli space. We leave the precise mathematical characterization of this modified moduli space to future work\footnote{This program was recently carried out successfully in instanton physics in 5 dimensions \cite{Nekrasov:2015wsu,Kim:2016qqs,Chang:2016iji,Haouzi:2019jzk,Haouzi:2020yxy,Haouzi:2020zls}: there, the problem of interest was to count instantons in the presence of a Wilson loop, which is not solved by localizing the loop on usual ADHM solutions \cite{Atiyah:1978ri}, but by defining instead a more general ``crossed instanton" moduli space from the onset. The moduli space of the instanton mechanics is modified in an analogous way to the moduli space of the vortex mechanics we study here. In the end, this is not too surprising, since vortices are ultimately related to instantons on a codimension-2 locus.}. For an explicit example of a quantum mechanics $T^{1d}$ when $\fg=A_n$, see Figure \ref{fig:flagdefect}.

For now, we justify the existence of these extra chiral multiplets inside $T^{1d}$  a fortiori, by showing that they correctly (and uniquely) account for the symmetries of $G^{3d}$ under a shift of vortex number; in other words, they describe the physics of a non-perturbative Schwinger-Dyson identity. We will give a direct proof of the existence of these multiplets in section \ref{sec:littlestringsecion}, where we analyze an underlying string picture.
Having described the quantum mechanics $T^{1d}$, we turn to the definition of its Witten index. As we will soon see, this index has remarkable properties in our context.

\subsection{The Index of the Quantum Mechanics}
\label{ssec:QMindex}

Recall that the gauge theory $G^{3d}$ is defined on $\mathbb{C}\times S^1(\widehat{R})$. Let us denote by $U(1)_{\omega}$ the symmetry associated with rotating the $\mathbb{C}$-line. Then, the global symmetry of the vortex theory $T^{1d}$ is $(\widehat{G}_F/U(1)^n ) \times \widehat{G}_{defect}\times  SU(2)_C \times U(1)_H \times U(1)_{\omega}$, with   $\widehat{G}_F$ the chiral matter producing the teeth of the handsaw quiver, and all other groups as introduced previously. The diagonal combination of $U(1)_H \times U(1)_{\omega}$ commutes with the supersymmetry and acts as a flavor symmetry; we call it $U(1)_{\epsilon_1}$, with generator $J_-$. We further define $r$ as the generator of $U(1)_H$, and $J_3$ as the Cartan generator of $SU(2)_C$.

Then, the refined Witten index of the gauged quantum mechanics $T^{1d}$ has the form \cite{Hori:2014tda,Cordova:2014oxa,Hwang:2014uwa}: 
\begin{align}
\label{index}
\left[\chi\right]^{(L^{(1)},\ldots, L^{(n)})}_{1d} =\text{Tr}\bigg[(-1)^F\, &e^{-\widehat{R}\{Q,\overline{Q}\}}\, e^{\widehat{R}\,\epsilon_2 (2 J_3 - r)}\,e^{2\widehat{R}\,\epsilon_1 J_-}\\
&\times\prod_{a=1}^n e^{\widehat{R}\sum_{d} m^{(a)}_d\Pi^{(a)}_d}\,e^{\widehat{R}\sum_{\rho} M^{(a)}_{\rho}\Lambda^{(a)}_{\rho}}\,\bigg] \; .\nonumber
\end{align}
This index has a path integral interpretation as a twisted partition function on $S^1(\widehat{R})$. The trace is over the Hilbert space of the theory, and the index counts states in $Q$-cohomology, where we have defined the supercharges as $Q\equiv Q^1_{1,1}$ and $\overline{Q}\equiv Q^2_{2,2}$ in the notation of section \ref{ssec:review}. $F$ is the fermion number.   $\{\Pi^{(a)}_i\}$ and $\{\Lambda^{(a)}_{\rho}\}$ are Cartan generators of the flavor group $\widehat{G}_F$ and the  line defect group $\widehat{G}_{defect}$, respectively. We have also defined conjugate variables for these generators: the fundamental/antifudamental chiral multiplet masses $\{m^{(a)}_d\}$, and the defect masses $\{M^{(a)}_{\rho}\}$.  Finally, we have introduced the variables $\epsilon_1$ and $\epsilon_2$, respectively conjugate to $J_-$ and $2 J_3 - r$.

The fugacity $e^{\widehat{R}\, \E_1}$ is well-known in the context of the 3d gauge theory on $\mathbb{C}\times S^1(\widehat{R})$, where it is called the $\Omega$-background \cite{Nekrasov:2002qd,Nekrasov:2003rj,Losev:2003py,Nekrasov:2010ka}. We will analyze it in detail when discussing the 3d gauge theory perspective. In the rest of this paper, the following redefined fugacities will come in handy:
\beq
\label{epsilons}
\epsilon_+\equiv \frac{\epsilon_1+\epsilon_2}{2}\; , ~~~~~ \epsilon_-\equiv \frac{\epsilon_1-\epsilon_2}{2}\; .
\eeq

The index is the grand canonical ensemble of vortex BPS states. There is a natural grading by the vortex number  $k^{(a)}= \frac{1}{2\pi}\int \text{Tr} F^{(a)}$, which is the topological $U(1)$ charge for the $a$-th gauge group, conjugate to the vortex counting fugacity  $\zeta_{3d}^{(a)}$, which is also a real FI parameter in 3d (recall that the complex FI parameters are set to zero throughout). Then, the Witten index can be organized as a sum over vortex sectors $(k^{(1)}, k^{(2)}, \ldots, k^{(n)})$.\\

By standard arguments \cite{Witten:1982df,AlvarezGaume:1986nm}, the Witten index does not depend on the circle scale $\widehat{R}$. In particular, we can work in the limit  $\widehat{R}\rightarrow 0$, where it reduces to Gaussian integrals around saddle points.  These saddle points are parameterized by  $\phi^{(a)}=\widehat{R}\, \varphi^{(a)}_{1d}+ i\, \widehat{R}\, A^{(a)}_{t, 1d}$, with $A^{(a)}_{t, 1d}$ the gauge field and $\varphi^{(a)}_{1d}$ the scalar in the $a$-th vector multiplet of the quantum mechanics. The (complexified) eigenvalues of $\phi^{(a)}$ are denoted as $\phi^{(a)}_1, \ldots, \phi^{(a)}_{k^{(a)}}$. Performing the Gaussian integrals over massive fluctuations, the index reduces to a zero mode integral of various 1-loop determinants, which we write schematically as:
{\allowdisplaybreaks
	\begin{align}
	\label{vortexintegral}
	&\left[\chi\right]^{(L^{(1)},\ldots, L^{(n)})}_{1d}  =\sum_{k^{(1)},\ldots, k^{(n)}=0}^{\infty}\;\prod_{a=1}^{n}\frac{e^{\widehat{R}\,\zeta_{3d}^{(a)}\, k^{(a)}}}{k^{(a)}!}\\
	&\qquad\qquad\qquad\;\;\times \oint  \left[\frac{d\phi^{{(a)}}_I}{2\pi i}\right]Z^{(a)}_{pure,  vec}\cdot Z^{(a)}_{pure,  adj}\cdot Z^{(a)}_{pure,  teeth}\cdot  \prod_{b>a}^{n} Z^{(a,b)}_{pure,  bif}\cdot \prod_{\rho=1}^{L^{(a)}}Z^{(a)}_{defect}\; ,\nonumber \\
	&Z^{(a)}_{pure, vec} = \frac{\prod_{\substack{I\neq J\\ I,J=1}}^{k^{(a)}}\sh\left(\phi^{(a)}_I-\phi^{(a)}_J\right)}{\prod_{I, J=1}^{k^{(a)}} \sh\left(\phi^{(a)}_{I}-\phi^{(a)}_{J}+\E_2 \right)}\nonumber\\
	&Z^{(a)}_{pure, adj} = \prod_{I, J=1}^{k^{(a)}}\frac{\sh\left(\phi^{(a)}_I-\phi^{(a)}_J+\E_1+\E_2\right)}{\sh\left(\phi^{(a)}_{I}-\phi^{(a)}_{J}+\E_1 \right)}\nonumber\\
	&Z^{(a)}_{pure, teeth} = \prod_{I=1}^{k^{(a)}}\prod_{i=1}^{P^{(a)}}\frac{\sh\left(\phi^{(a)}_I-\mu^{(a)}_i+(\E_1-\E_2)/2+\E_2\right)}{\sh\left(\phi^{(a)}_{I}-\mu^{(a)}_i+(\E_1-\E_2)/2\right)}\prod_{j=1}^{Q^{(a)}}\frac{\sh\left(-\phi^{(a)}_I+\widetilde{\mu}^{(a)}_j+(\E_1+\E_2)/2+\E_2\right)}{\sh\left(-\phi^{(a)}_{I}+ \widetilde{\mu}^{(a)}_j+(\E_1+\E_2)/2 \right)}\nonumber\\
	&Z^{(a, b)}_{pure, bif} =\left[\prod_{I=1}^{k^{(a)}}\prod_{J=1}^{k^{(b)}} \frac{\sh\left(\phi^{(b)}_{I}-\phi^{(a)}_{J} + \E_2 \right)}{\sh\left(\phi^{(b)}_{I}-\phi^{(a)}_{J} \right)} \frac{\sh\left(-\phi^{(b)}_{I}+\phi^{(a)}_{J}- \E_1  \right)}{\sh\left(-\phi^{(b)}_{I}+\phi^{(a)}_{J}- \E_1 - \E_2 \right)}\right]^{\Delta^{ab}}\nonumber\\
	&Z^{(a)}_{defect} =  \prod_{I=1}^{k^{(a)}} \frac{\sh\left(\phi^{{(a)}}_I-M^{(a)}_\rho- (\E_1 - \E_2)/2 \right)\, \sh\left(-\phi^{{(a)}}_I + M^{(a)}_\rho- (\E_1 - \E_2)/2\right)}{\sh\left(\phi^{{(a)}}_I-M^{(a)}_\rho- (\E_1 + \E_2)/2\right)\, \sh\left(-\phi^{{(a)}}_I + M^{(a)}_\rho - (\E_1 + \E_2)/2 \right)}\, .\nonumber
	\end{align}}
Some comments are in order:\\

We use the convenient notation $\sh(x)\equiv 2 \sinh(\widehat{R}\,x/2)$.   $\mathcal{M}_k$ is the set of poles enclosed by the contours, which we will characterize below. The prefactor $\prod_{a=1}^{n} 1/k^{(a)}!$ is the Weyl group order of the 1d gauge group $\widehat{G}=\prod_{a=1}^n U(k^{(a)})$.

The factor $Z^{(a)}_{pure, vec}(\{\phi^{(a)}_I\}, \epsilon_2)$ is the contribution of the 1d $\cN=4$ vector multiplet on node $a$.

The factor $Z^{(a)}_{pure, adj}(\{\phi^{(a)}_I\}, \epsilon_1, \epsilon_2)$ is the contribution of the  1d $\cN=4$ adjoint chiral multiplet on node $a$.

The factor $Z^{(a)}_{pure, teeth}(\{\phi^{(a)}_I\}, \{\mu^{(a)}_i\}, \{\widetilde{\mu}^{(a)}_i\}, \epsilon_1, \epsilon_2)$ is the contribution of the  1d $\cN=4$ flavors on node $a$. Note that the numbers $P^{(a)}$ of fundamental chirals (with corresponding masses $\{\mu^{(a)}_i\}$) and  $Q^{(a)}$ of  antifundamental chirals (with corresponding masses $\{\widetilde{\mu}^{(a)}_i\}$) are fully determined in terms of the ranks of the 3d gauge group $G$,  the 3d flavor group $G_F$, and the choice of the 3d vacuum\footnote{This is the data that determines the Higgsing of the 3d theory, which is to say the fundamental mass each Coulomb modulus is frozen to on the Higgs branch. In our 1d notation, the masses and $\{\mu^{(a)}_i\}$ and $\{\widetilde{\mu}^{(a)}_i\}$ should eventually be expressed in terms of the 3d masses $\{m^{(a)}_d\}$.}. For instance, consider the 3d theory $T_\rho[SU(N^{(n+1)})]$, where the fundamental matter ranks are $N_F^{(a)}=0$ for $a=1,\ldots,n-1$, and $N^{(n)}_F=N^{(n+1)}$ on the last node, with corresponding masses $\{m^{(n)}_d\}_{d=1,\ldots,N^{(n+1)}}$. Then, in the quantum mechanics, $P^{(a)} = N^{(a)} - N^{(a-1)}$, while $Q^{(a)} = N^{(a+1)} - N^{(a)}$, and the matter factor becomes: 
\beq
Z^{(a)}_{pure, teeth} = \prod_{I=1}^{k^{(a)}}\prod_{i=N^{(a-1)}}^{N^{(a)}}\frac{\sh\left(\phi^{(a)}_I-m^{(n)}_i+\E_-+\E_2\right)}{\sh\left(\phi^{(a)}_{I}-m^{(n)}_i+\E_-\right)}\prod_{j=N^{(a)}}^{N^{(a+1)}}\frac{\sh\left(-\phi^{(a)}_I+m^{(n)}_j+\E_+ +\E_2\right)}{\sh\left(-\phi^{(a)}_{I}+ m^{(n)}_j+\E_+ \right)} \; .
\eeq
In particular, the chiral multiplet masses $\{\mu^{(a)}_i\}$ and  $\{\widetilde{\mu}^{(a)}_i\}$ are now written exclusively in terms of the $N^{(n+1)}$ 3d masses  $\{m^{(n)}_i\}$, as they should. The generalization to $D_n$ and $E_n$ algebras is straightforward, even though the Higgsing pattern is more intricate to write down explicitly \cite{Aganagic:2015cta}.

The factor $Z^{(a, b)}_{pure, bif}(\{\phi^{(a)}_I\}, \{\phi^{(b)}_I\},  \epsilon_1, \epsilon_2)$ is the contribution of the 1d $\cN=4$ bifundamental  matter between nodes $a$ and $b$. It is only nontrivial when the incidence matrix $\Delta^{a b}$ is as well. Recall that the matrix $\Delta^{a b}$ equals 1 if there is a link connecting nodes $a$ and $b$, and equals 0 otherwise.\\

The above factors account for all the multiplets present in the vortex quantum mechanics $T^{1d}_{pure}$ of a 3d $\cN=4$ gauge theory in the absence of loop defect. Now, recall that the loop is characterized by the group $\widehat{G}_{defect}=\prod_{a=1}^n U(L^{(a)})$, resulting in additional flavors for the quantum mechanics. The superscript notation we use for the index, $\left[\chi\right]^{(L^{(1)},\ldots, L^{(n)})}_{1d}$, makes the dependence on this defect group explicit.
Then, the factor $Z^{(a)}_{defect}(\{\phi^{(a)}_I\}, \{M^{(a)}_\rho\}, \epsilon_1, \epsilon_2)$ is the contribution of 1d $\cN=4$ chiral multiplets on node $a$.\\

In writing the index integrand, note that all the $\cN=4$ chiral multiplets are decomposed into $\cN=2$ Fermi and chiral multiplets, which make up respectively the numerators and denominators. Because the theory is valued on a circle of radius $\widehat{R}$, it is useful in what follows to introduce K-theoretic fugacities for each of the equivariant parameters:
\begin{align}\label{fugacitiesKtheory}
&\widetilde{\fq}^{(a)}\equiv e^{\widehat{R}\,\zeta_{3d}^{(a)}},\\
& q \equiv e^{\widehat{R}\,\epsilon_1},\qquad t\equiv e^{-\widehat{R}\,\epsilon_2},\qquad
  v\equiv e^{\widehat{R}\,\epsilon_+}=\sqrt{q/t},\qquad u\equiv e^{\widehat{R}\,\epsilon_-}=\sqrt{q\, t},\nonumber\\
&f^{(a)}_d \equiv e^{-\widehat{R}\,\mu^{{(a)}}_d},\qquad \widetilde{f}^{(a)}_d \equiv e^{-\widehat{R}\,\widetilde{\mu}^{{(a)}}_d},\qquad z^{(a)}_\rho=e^{-\widehat{R}\,M^{(a)}_\rho}.\nonumber
\end{align}
All parameters are complexified by the holonomy of the corresponding gauge field around the circle.

Crucially, the Witten index also depends implicitly on additional continuous parameters in a piecewise constant manner: the $n$ FI parameters $\zeta_{1d}^{(a)}$, which are themselves $k^{(a)}$-vectors, one for each abelian factor in $\widehat{G}$. Indeed, when such a parameter changes sign and crosses the value $\zeta_{1d}^{(a)}=0$, a non-compact Coulomb branch opens up, and some vacua may appear or disappear, resulting in wall crossing and a jump in the index. This dependence on the 1d FI parameters is in one-to-one correspondence with the choice of the index integration contours, which we now turn to.

We adopt the so-called Jeffrey-Kirwan (JK) residue prescription \cite{Jeffrey:1993}. It was first popularized in our context in a 2-dimensional setup \cite{Benini:2013xpa}, and used in our quantum mechanical context in \cite{Hwang:2014uwa, Cordova:2014oxa, Hori:2014tda}. Let us briefly review its main features. First note that each $Z^{(a)}$-factor in the integrand has the following general form:
\beq\label{integRand}
\frac{\prod_{i=1}^{n_1}\sh(\vec\rho_i\vec\phi_i+\ldots)}{\prod_{j=1}^{n_2}\sh(\vec\rho_j\vec\phi_j+\ldots)}\; ,
\eeq
where $\vec\rho$ is a $k$-tuple vector, with $k=\sum_{a=1}^n k^{(a)}$. The entries of $\vec\rho$ are the set $\{0,\pm1\}$, and $n_1^{(a)}$ and $n_2^{(a)}$ are positive integers specified by the details of the vortex quantum mechanics. The dots ``$\ldots$" stand for a linear function of the spacetime fugacities $\epsilon_1$, $\epsilon_2$, as well as all the other 1d flavor fugacities. Since $\sinh(0)=\sinh(i\pi)=0$, there can be many poles in \eqref{integRand}. We denote a pole locus as $\vec\phi=\vec\phi_*$.

Now, we assemble the $n$ FI parameters $\zeta^{(a)}_{1d}$ into a vector $\zeta_{1d}$ of size $k=\sum_{a=1}^n k^{(a)}$. As we pointed out, the Witten index depends on the choice of a chamber for $\zeta_{1d}$. Apart from the FI parameter vector $\zeta_{1d}$, the JK prescription instructs us to define yet another auxiliary $k$-vector $\eta$, though the index ultimately does not depend on $\eta$. We are a priori free to work with any $k$-vector $\eta$ we want to carry out the JK residue prescription, but there exists a particularly convenient choice $\eta=\zeta_{1d}$.  Indeed, on general grounds, the index integral $\eqref{vortexintegral}$ has $\phi$-poles at $\pm \infty$ with nonzero residues; one can show that the choice $\eta=\zeta_{1d}$, meaning $\eta$ generic but chosen in the same chamber as $\zeta_{1d}$, guarantees that the contributions of $\phi$-residues at $\pm \infty$ vanish. Unless specified otherwise, in this paper we work in a chamber where all components of $\zeta_{1d}$ are positive. We will work in different chambers when discussing 3d Seiberg duality later on. 
Having defined $\eta$, we are to choose $k$ hyperplanes from the arguments of $\sinh$ functions in the denominator of \eqref{integRand}. Those hyperplanes will take the following form:
\beq
\label{linearsystem}
\vec\rho_j  \cdot  \vec\phi_j + \ldots =0\;,\;\;\;\text{where }j=1,\ldots,k.
\eeq
The contours of the index are then chosen to enclose poles which are solutions of this linear system of equation, but only if the vector  $\eta$ also happens to lie in the cone spanned by the vectors $\vec\rho_j$. A practical way to test this condition is to construct a $k\times k$ matrix ${\bf{Q}}=Q_{ji}=(\rho_j)_i$, where $\vec\rho_j=((\rho_j)_1,\ldots,(\rho_j)_k)$, and test if all the components of $\eta\, {\bf{Q}}^{-1}$ are positive. We collect the poles $\vec\phi_*$ satisfying  the condition in a set $\mathcal{M}_k$.

Summing over all the poles in $\mathcal{M}_k$, the Witten index takes the form
\beq\label{JK}
\left[\chi\right]^{(L^{(1)},\ldots, L^{(n)})}_{1d}=\sum_{k^{(1)},\ldots, k^{(m)}=0}^{\infty}\prod_{a=1}^n\frac{(\widetilde{\fq}^{(a)}){}^{k^{(a)}}}{k^{(a)}!}\sum_{\vec\phi_*}\text{JK-res}_{\vec\phi_*}({\bf{Q}}_*, \eta)\, Z^{(a)}_{integrand}  \; ,
\eeq
where $Z^{(a)}_{integrand}$ is the integrand of \eqref{vortexintegral}, and the JK-residue is defined as
\beq\label{JKrule}
\text{JK-res}_{\vec\phi_*}({\bf{Q}}_*, \eta)\frac{d^k \vec\phi}{\prod_{j=1}^k(\vec\rho_j\cdot\vec\phi_j)}=\begin{cases}
	\frac{1}{\left|\text{det}\left({\bf{Q}}\right)\right|} \;\; \text{if}\;\; \eta\in\text{cone}\left({\bf{Q}}\right)\\
	0 \qquad\qquad\qquad\, \text{otherwise}
\end{cases}
\eeq
The condition $\eta\in\text{cone}\left({\bf{Q}}\right)$ means that the vector $\eta$ should lie in the cone spanned by the rows of the matrix ${\bf{Q}}$.
It can happen that a solution of the system of equations \eqref{linearsystem} yields additional zeroes in the denominator of \eqref{integRand}. This typically results in degenerate poles, which can be dealt with using a constructive definition of the JK residue and the so-called flag method \cite{2004InMat.158..453S,Benini:2013xpa}. This is an involved procedure to implement analytically, and we will refrain from doing so in this paper, treating potential degenerate poles on a case-by-case basis instead.

We now come to our main object of study, the derivation of non-perturbative Schwinger-Dyson equations for the gauge theory $G^{3d}$. As we will show, they arise as a regularity condition of the quantum mechanics index $\left[\chi\right]^{(L^{(1)},\ldots, L^{(n)})}_{1d} $  on the defect masses $\{M^{(a)}_{\rho}\}$.

\subsection{The Index is a Vortex $qq$-character}
\label{ssec:1dqqcharacter}

We evaluate the index, using the JK residue prescription above to define the contours. As a warmup, let us practice with the index of $T^{1d}_{pure}$, which is the vortex quantum mechanics of the ``pure" 3d $\cN=4$ theory, in the absence of line defect. We call the corresponding index:
\begin{align}\label{partpure}
&\left[\chi\right]^{(0,\ldots, 0)}_{1d} =\sum_{k^{(1)},\ldots, k^{(n)}=0}^{\infty}\;\prod_{a=1}^{n}\frac{(\widetilde{\fq}^{(a)}){}^{k^{(a)}}}{k^{(a)}!}\\
&\qquad\qquad\qquad\qquad\;\;\times \oint_{\mathcal{M}^{pure}_k}  \left[\frac{d\phi^{{(a)}}_I}{2\pi i}\right]Z^{(a)}_{pure,  vec}\cdot Z^{(a)}_{pure,  adj}\cdot Z^{(a)}_{pure,  teeth}\cdot  \prod_{b>a}^{n} Z^{(a,b)}_{pure,  bif}\; . \nonumber
\end{align}
Working in the $\zeta_{1d}>0$ chamber, the poles that end up contributing to the $T^{1d}_{pure}$ index make up the set $\mathcal{M}^{pure}_k$. The elements of this set satisfy:
\begin{align}
&\phi^{(a)}_I = \phi^{(a)}_J - \E_1 \; , \label{purepole1}\\
&\phi^{(a)}_I = \phi^{(a)}_J - \E_2 \; , \label{purepole2}\\
&\phi^{(a)}_I = \mu^{(a)}_i - \E_- \; ,\;\; \text{for some $i\in\{1,\ldots, P^{(a)}\}$}\; ,\label{purepole3}\\
&\phi^{(b)}_J=\phi^{(a)}_I \; ,\;\; \text{if there is a link between nodes $a$ and $b>a$}\; , \label{purepole4}\\
&\phi^{(b)}_J=\phi^{(a)}_I + 2\epsilon_+ \; ,\;\; \text{if there is a link between nodes $a$ and $b<a$}\; .\label{purepole5}
\end{align}
The poles \eqref{purepole1} arise from the $\cN=4$ adjoint chiral factor,
\beq
Z^{(a)}_{pure, adj} = \prod_{I, J=1}^{k^{(a)}}\frac{\sh\left(\phi^{(a)}_I-\phi^{(a)}_J+\E_1+\E_2\right)}{\sh\left(\phi^{(a)}_{I}-\phi^{(a)}_{J}+\E_1 \right)} \; .
\eeq
The poles \eqref{purepole2}  arise from the $\cN=4$ vector multiplet,
\beq
Z^{(a)}_{pure, vec} = \frac{\prod_{\substack{I\neq J\\ I,J=1}}^{k^{(a)}}\sh\left(\phi^{(a)}_I-\phi^{(a)}_J\right)}{\prod_{I, J=1}^{k^{(a)}} \sh\left(\phi^{(a)}_{I}-\phi^{(a)}_{J}+\E_2 \right)} \; .
\eeq
The poles \eqref{purepole3} arise from the $\cN=4$ flavor factor, 
\beq
\label{teethagain}
Z^{(a)}_{pure, teeth} = \prod_{I=1}^{k^{(a)}}\prod_{i=1}^{P^{(a)}}\frac{\sh\left(\phi^{(a)}_I-\mu^{(a)}_i+\E_-+\E_2\right)}{\sh\left(\phi^{(a)}_{I}-\mu^{(a)}_i+\E_- \right)}\prod_{j=1}^{Q^{(a)}}\frac{\sh\left(-\phi^{(a)}_I+\widetilde{\mu}^{(a)}_j+\E_+ +\E_2\right)}{\sh\left(-\phi^{(a)}_{I}+ \widetilde{\mu}^{(a)}_j+\E_+ \right)} \; .
\eeq
Specifically, the JK contours enclose poles coming from the fundamental chirals only (the $P^{(a)}$-product), and none of the antifundamental chirals. We wrote the poles in terms of 1d flavor fugacities $\{\mu^{(a)}_i\}$ as a shorthand notation, which are really placeholders for the $\text{rank}(G_F)$ 3d fundamental masses $\{m^{(b)}_d\}$. The poles \eqref{purepole4} and \eqref{purepole5} are due to the bifundamental contributions,
\beq
Z^{(a, b)}_{pure, bif} =\left[\prod_{I=1}^{k^{(a)}}\prod_{J=1}^{k^{(b)}} \frac{\sh\left(\phi^{(b)}_{I}-\phi^{(a)}_{J} + \E_2 \right)}{\sh\left(\phi^{(b)}_{I}-\phi^{(a)}_{J} \right)} \frac{\sh\left(-\phi^{(b)}_{I}+\phi^{(a)}_{J}- \E_1  \right)}{\sh\left(-\phi^{(b)}_{I}+\phi^{(a)}_{J}- \E_1 - \E_2 \right)}\right]^{\Delta^{ab}}\, .
\eeq
Two important remarks are in order. First, even though the contours enclose the JK-poles \eqref{purepole2}, the resulting residues are always trivial, because the numerators in $Z^{(a)}_{pure, teeth}$ create a zero at this locus. Second,  because of the bifundamental factor $Z^{(a, b)}_{pure, bif}$, some of the enclosed poles are non-simple for generic rank $k^{(a)}$. However, a careful application of the flag method to construct the JK-residue shows that the poles we enclose above make up an exhaustive list; this was checked numerically in \cite{Hwang:2017kmk}.
Putting it all together, and writing the 1d fundamental chiral masses $\{\mu^{(a)}_i\}$ in terms of the 3d masses $\{m^{(b)}_i\}$,  the various poles which end up contributing with nonzero residue are of the form
\beq\label{polepure}
\phi^{(a)}_I = m^{(b)}_i - \E_- - (s_i-1) \E_1 + 2\,\#^{(ab)} \E_+ \; , \;\;\;\;  \text{with}\; s_i\in\{1,\ldots,k^{(a)}_i\},\;\; i\in\{1,\ldots,N^{(a)}\},\; ,
\eeq
for some mass index $b\in\{1,\ldots,n\}$, and where $(k^{(a)}_1, \ldots, k^{(a)}_{N^{(a)}})$ is a partition of the vortex charge $k^{(a)}$ into $N^{(a)}$ non-negative integers. The pair of integers $(i, s_i)$ is assigned to one of the integers $I\in\{1,\ldots,k^{(a)}\}$ exactly once, and $\#^{(ab)}$ is a non-negative integer equal to the number of links between nodes $a$ and $b<a$ in the Dynkin diagram of $\fg$ (and $\#^{(ab)}=0$ if $b>a$).\\

As an explicit example, consider the $A_n$ theory $G^{3d}=T_\rho[SU(N^{(n+1)})]$ from figure \ref{fig:flag}. Then, the poles with nonzero residue are all of the form
\beq
\phi^{(a)}_I = m^{(n)}_i - \E_- - (s_i-1) \E_1 \; , \;\;\;\;  \text{with}\; s_i\in\{1,\ldots,k^{(a)}_i\},\;\; i\in\{1,\ldots,N^{(a)}\}\; .
\eeq
and $(k^{(a)}_1, \ldots, k^{(a)}_{N^{(a)}})$ is a partition of $k^{(a)}$ into $N^{(a)}$ non-negative integers, and the pair of integers $(i, s_i)$ is assigned to one of the integers $I\in\{1,\ldots,k^{(a)}\}$ exactly once. Performing the residue integral, one finds the following closed-form expression, which is well-known in physics \cite{Dimofte:2010tz} and in mathematics as the K-theoretic J-function \cite{2001math......8105G}:
\begin{align}\label{examplepure1d}
\left[\chi\right]^{(0,\ldots, 0)}_{1d} =\sum_{k^{(1)},\ldots, k^{(n)}=0}^{\infty}\;\prod_{a=1}^{n}(\widetilde{\fq}^{(a)}){}^{k^{(a)}}&\sum_{\substack{\sum_i k^{(a)}_i = k^{(a)} \\ k^{(a)}_i\geq 0}} \left[\prod_{i,j=1}^{N^{(a)}}\prod_{s=1}^{k^{(a)}_i-k^{(a)}_j}\frac{\sh\left(m_i-m_j+\E_2- (s-1)\, \E_1\right)}{\sh\left(m_i-m_j - (s-1)\, \E_1\right)}\right]\nonumber\\ &\qquad\times\left[\prod_{i=1}^{N^{(a+1)}}\prod_{j=1}^{N^{(a)}}\prod_{p=1}^{k^{(a)}_j-k^{(a+1)}_i}\frac{\sh\left(m_i-m_j+\E_2 + p\, \E_1\right)}{\sh\left(m_i-m_j + p\, \E_1\right)}\right].
\end{align}

We now consider the vortex quantum mechanics of $G^{3d}$ in the presence of the line defect, that is to say the index of $T^{1d}$ \eqref{vortexintegral}. For a given vortex number $k=\sum_{a=1}^n k^{(a)}$, the set of poles to be enclosed is denoted as $\mathcal{M}_k$. This set contains the set  $\mathcal{M}^{pure}_k$ of poles we just reviewed for the theory $T^{1d}_{pure}$ (the index in the absence of defect).  
There are also additional poles depending on the masses $M^{(a)}_\rho$, which make up the elements of the set $\mathcal{M}_k\setminus\mathcal{M}^{pure}_k$. Specifically, the new poles are of the form:
\begin{align}
&\phi^{(a)}_I=M^{(a)}_\rho + \epsilon_+ \; , \label{newpole1}\\
&\phi^{(b)}_J=\phi^{(a)}_I \; ,\;\; \text{if there is a link between nodes $a$ and $b>a$}\; , \label{newpole2}\\
&\phi^{(b)}_J=\phi^{(a)}_I + 2\epsilon_+ \; ,\;\; \text{if there is a link between nodes $a$ and $b<a$}\; .\label{newpole3}
\end{align}
The poles \eqref{newpole1} arise because of the interactions between the vortices and the loop defect,
\beq
\label{defect}
Z^{(a)}_{defect} =  \prod_{I=1}^{k^{(a)}} \frac{\sh\left(\phi^{{(a)}}_I-M^{(a)}_\rho- \E_- \right)\, \sh\left(-\phi^{{(a)}}_I + M^{(a)}_\rho- \E_-\right)}{\sh\left(\phi^{{(a)}}_I-M^{(a)}_\rho- \E_+\right)\, \sh\left(-\phi^{{(a)}}_I + M^{(a)}_\rho - \E_+ \right)} \, .
\eeq
The remaining poles \eqref{newpole2} and \eqref{newpole3} are again due to the bifundamental contributions,
\beq
Z^{(a, b)}_{pure, bif} =\left[\prod_{I=1}^{k^{(a)}}\prod_{J=1}^{k^{(b)}} \frac{\sh\left(\phi^{(b)}_{I}-\phi^{(a)}_{J} + \E_2 \right)}{\sh\left(\phi^{(b)}_{I}-\phi^{(a)}_{J} \right)} \frac{\sh\left(-\phi^{(b)}_{I}+\phi^{(a)}_{J}- \E_1  \right)}{\sh\left(-\phi^{(b)}_{I}+\phi^{(a)}_{J}- \E_1 - \E_2 \right)}\right]^{\Delta^{ab}}\, .
\eeq
For a given vortex number $k$, we now argue that the content of the set $\mathcal{M}_k\setminus\mathcal{M}^{pure}_k$ makes it possible to reinterpret the index as the character of a finite-dimensional representation of a quantum affine algebra. In order to prove this, we define a new quantity, the vacuum expectation value of a loop defect operator on node $a$, with corresponding  mass  $z^{(a)}_\rho\equiv z$:
\begin{align}
\label{Yoperator1d}
&\left\langle \left[Y^{(a)}_{1d}(z)\right]^{\pm 1} \right\rangle \equiv
\sum_{k^{(1)},\ldots, k^{(n)}=0}^{\infty}\;\prod_{b=1}^{n}  \frac{\left(\widetilde{\fq}^{(b)}\right)^{k^{(b)}}}{k^{(b)}!} \; \\  &\;\;\qquad\qquad\times\oint_{\mathcal{M}^{pure}_k}  \left[\frac{d\phi^{{(b)}}_I}{2\pi i}\right]Z^{(b)}_{pure,  vec}\cdot Z^{(b)}_{pure,  adj}\cdot Z^{(b)}_{pure,  teeth}\cdot \prod_{c>b}^{n} Z^{(b, c)}_{pure,  bif}  \cdot \left[Z^{(a)}_{defect}(z)\right]^{\pm 1} . \nonumber
\end{align}
Even though the defect factor $Z^{(a)}_{defect}(z)$ is present inside the integrand, the contour integral is defined to \emph{only} enclose poles in the set $\mathcal{M}^{pure}_k$, the same poles as in the pure index \eqref{partpure}.

Remarkably, the index of $T^{1d}$ can be written as a finite Laurent series in such $Y$-operator vevs. We find it convenient to normalize our expressions by the index of the vortex quantum mechanics $T^{1d}_{pure}$, in the absence of defect:
\beq
\label{normalized}
\left[\widetilde{\chi}\right]^{(L^{(1)},\ldots, L^{(n)})}_{1d} (\{z^{(a)}_\rho\})\equiv \frac{\left[\chi\right]^{(L^{(1)},\ldots, L^{(n)})}_{1d}(\{z^{(a)}_\rho\})}{\left[{\chi}\right]^{(0,\ldots,0)}_{1d}}
\eeq
As a function of the defect masses $\{z^{(a)}_\rho\}$, our first main result is that the normalized index can be written as:
\begin{align}\label{character1d}
\boxed{\left[\widetilde{\chi}\right]^{(L^{(1)},\ldots, L^{(n)})}_{1d}(\{z^{(a)}_\rho\})=\frac{1}{\left[{\chi}\right]^{(0,\ldots,0)}_{1d}}\sum_{\omega\in V(\lambda)}\prod_{b=1}^n \left({\widetilde{\fq}^{(b)}}\right)^{d_b^\omega}\; c_{d_b^\omega}(q, t)\; \left(\mathcal{Q}_{d_b^\omega}^{(b)}(\{z_{\rho}^{(a)}\})\right)\,  \left[{Y}_{1d}(\{z^{(a)}_\rho\})\right]_{\omega} \, .}
\end{align}
We will prove this statement momentarily. For now, let us unpack the notation.

$\{z^{(a)}_\rho\}$ denotes collectively the $\sum_{a=1}^n L^{(a)}$ defect masses $z^{(a)}_\rho\equiv e^{-\widehat{R} M^{(a)}_\rho}$. The sum runs over all the weights $\omega$ of the finite-dimensional irreducible representation $V(\lambda)$ of the quantum affine algebra $U_\hbar(\widehat{\fg})$, of highest (classical) weight $\lambda =\sum_{a=1}^n L^{(a)}\, \lambda_a$. The quantization parameter is $\hbar\equiv v^2=q/t$. Here, $\fg$ is the simply-laced Lie algebra denoting the 3d quiver gauge theory (as well as its vortex quantum mechanics), and $\lambda_a$ the $a$-th fundamental weight of $\fg$. The label $d_b^\omega$ is a positive integer that is determined by solving
\beq\label{sl2string}
\omega=\lambda -\sum_{b=1}^n d_b^\omega\, \alpha_b\; .
\eeq
Namely, a given weight $\omega$ is reached by lowering the highest weight $\lambda$ a finite number of times, using the positive simple roots $\{\alpha_b\}_{b=1,\ldots, n}$. This procedure is referred to as building the weight $\omega$ out of  $sl_2$ strings. The equivariant parameter $\widetilde{\fq}^{(b)}$ is the 3d FI parameter for the $b$-th gauge group. 

The factors $c_{d_b^\omega}(q, t)$ are coefficients depending only on $q$ and $t$.\\

The function $\mathcal{Q}_{d_b^\omega}^{(b)}(\{z_{\rho}^{(a)}\})$ is the residue of $Z^{(b)}_{pure, teeth}$ evaluated at the poles \eqref{newpole1}, \eqref{newpole2}, and \eqref{newpole3}. The function is therefore made up of fundamental and antifundamental chiral multiplet contributions, such as:
\beq
\label{antifund}
\prod_{a=1}^n\prod_{\rho=1}^{L^{(a)}}\prod_{i=1}^{P^{(b)}}\frac{\left(1-  v^{{\#'}^{(b)}_i} f^{(b)}_i/z_{\rho}^{(a)}\right)}{\left(1- v^{{\#'}^{(b)}_i}f^{(b)}_i/z_{\rho}^{(a)}\right)}\prod_{j=1}^{Q^{(b)}}\frac{\left(1- t\, v^{\widetilde{\#'}^{(b)}_j} \widetilde{f}^{(b)}_j/z_{\rho}^{(a)}\right)}{\left(1- v^{\widetilde{\#'}^{(b)}_j}\widetilde{f}^{(b)}_j/z_{\rho}^{(a)}\right)}
\eeq
As usual, $P^{(b)}$ stands for the number of fundamental chirals at node $b$, with masses $\{f^{(b)}_i\}$, while $Q^{(b)}$ stands for the number of antifundamental chirals at node $b$, with masses $\{\widetilde{f}^{(b)}_j\}$. The symbols ${\#'}^{(a)}_{i}$ and $\widetilde{\#'}^{(a)}_{j}$  stand for other non-negative integers, which are fixed by the choice of the 3d Higgs vacuum.

Finally, the operator $\left[{Y}_{1d}(\{z^{(a)}_\rho\})\right]_{\omega}$, for a given weight $\omega$, is the expectation value of a rational function of $Y$-operators  $\left\langle \prod_a \left[Y^{(a)}_{1d}\right]^{\pm 1} \right\rangle $ and derivatives thereof\footnote{An example where such a derivative term can appear is the index of the $D_4$ theory with a $U(1)$-defect insertion on node 2, $\left[\chi^{\fg}\right]^{(0,1,0,0)}_{1d}(z^{(2)}_1)$. The index organizes itself as a Laurent series of 29 $Y$-operator terms, one of which involves derivatives of $Y^{(a)}_{1d}$-operators. Note that the second fundamental representation of $D_4$ is only 28-dimensional. However, finite dimensional irreducible representations of quantum affine algebras are notoriously bigger than their non-affine counterpart. Indeed, the second fundamental representation $V(\lambda_2)$ of $U_\hbar(\widehat{D_4})$ decomposes into irreducible representations of $U_\hbar(D_4)$ as $V(\lambda_2) = \textbf{28}\oplus\textbf{1}$. Put differently, one necessarily has to add the trivial representation \textbf{1},  an extra null weight, to the \textbf{28} in order to obtain an irreducible representation of $U_\hbar(\widehat{D_4})$. This is sometimes called the minimal affinization of $\textbf{28}$ \cite{Chari:1994pd}.}, where each operator $\left[Y^{(a)}_{1d}\right]^{\pm 1}$ is a function of a defect mass $z^{(a)}_\rho$. The arguments of each factor is shifted by powers of $q$ and $t$, determined uniquely from \eqref{sl2string}. 

All in all, the index is a twisted\footnote{The character is \emph{twisted} because of the presence of the 3d FI parameters $\widetilde{\fq}^{(b)}$ and the flavor matter factors $\mathcal{Q}^{(b)}$.} character of a finite dimensional irreducible representation $V(\lambda)$ of $U_\hbar(\widehat{\fg})$, with highest weight $\lambda=\sum_{a=1}^n L^{(a)}\, \lambda_a$. Starting with the highest weight $\lambda$, each term in the character can then be obtained by successive ``vortex-Weyl" reflections, which generalize the usual Weyl group action of the Lie algebra $\fg$. Because of the dependence on the two fugacities $q=e^{\widehat{R}\E_1}$ and $t=e^{-\widehat{R}\E_2}$, the vortex character is a $qq$-character, in the denomination of \cite{Nekrasov:2015wsu}.\\

Two remarks are in order. First, similar $qq$-characters have been constructed in the related context of counting instantons in the presence of a 1/2-BPS Wilson loop on the manifold $\mathbb{C}^2\times S^1(\widehat{R})$.  There, the functional form of the character, meaning its dependence on the $Y$-operators $\left[{Y}_{1d}(\{z^{(a)}_\rho\})\right]_{\omega}$ and on the weights $\omega$, is identical to what we found here in the context of vortex counting. This is because the $Y$-operator dependence is entirely fixed by the choice of the algebra $\fg$ and the representation $(L^{(1)},\ldots, L^{(n)})$ in which the Wilson loop transforms. In particular, the functional form of the character does not depend on whether we study instanton or vortex counting, nor does it depend on the dimension of the manifold\footnote{Note this didn't have to be the case a priori: in the 5d context, the refinement parameters $\E_1$ and $\E_2$ are on a completely equal footing, and are both geometric by construction. In our context, these parameters have very different origins, the former being geometric and the latter being a twisted mass. Yet, the resulting $Y$-operators depend on $\E_1$ and $\E_2$ in an identical way in both 5d and 3d.}. Of course, there are still notable differences according to which gauge theory setup we study: this is encoded in the  expressions for the vevs $\langle \ldots \rangle$, and the functions $\mathcal{Q}^{(b)}$ in \eqref{character1d}. For instance, in the context of an instanton quantum mechanics, these functions $\mathcal{Q}^{(b)}$ are contributions of $\cN=4$ Fermi multiplets exclusively, while in the vortex context, they are contributions of $\cN=4$  chiral multiplets instead. Both result in characters, but with different twists.

Second, one can consider the limit where we shrink the circle size to zero. There are a priori many ways to take this limit, so we should be specific: here, we require that all flavor fugacities of the quantum mechanics remain fixed as we take $\widehat{R}\rightarrow 0$. In practice, all the trigonometric functions present in the 1-loop determinants of the quantum mechanics index will become rational functions of their arguments instead. The 3d gauge theory $G^{3d}$ turns into a 2d $\cN=(4,4)$ gauged sigma model on $\mathbb{C}^2$, and the loop defect wrapping $S^1(\widehat{R})$ becomes a 1/2-BPS point defect at the origin. Correspondingly, the index we computed should become a vortex $qq$-character of the 2d theory, whose existence was conjectured in \cite{Nekrasov:2017rqy}\footnote{In section 7 of that work, an expression was written in the case $q=1$, or equivalently $\E_1=0$, directly from two dimensions.}. Our results in this section therefore provide a refined microscopic derivation of the conjecture presented there.\\

It remains to prove that the index of $T^{1d}$ is indeed equal to the character expression \eqref{character1d}. Since a fully explicit proof would require the knowledge of a specific Higgs vacuum for the 3d theory, we find it more worthwhile to outline the universal features of the proof here in the general case, and showcase it in detail later when discussing an example in section \ref{sec:example}. The proof consists of two parts. First, recall that the contours of the index enclose the poles $\mathcal{M}_k$, for a given vortex number $k$. In contrast, the contours of the $Y$-operator \eqref{Yoperator1d} only enclose the poles $\mathcal{M}^{pure}_k$ of the index in the absence of  loop defect.
Because the set $\mathcal{M}_k\setminus\mathcal{M}^{pure}_k$ is non-empty for every $k$, it follows that the index has following expansion:
\beq\label{firstterm1d}
\left[\chi\right]^{(L^{(1)},\ldots, L^{(n)})}_{1d}(\{z^{(a)}_\rho\})=  \left\langle \prod_{a=1}^n \prod_{\rho=1}^{L^{(a)}}{Y_{1d}^{(a)}(z^{(a)}_\rho)} \right\rangle +\ldots\; ,
\eeq
where each term in the dots ``$...$" stands for a residue of $\left[\chi\right]^{(L^{(1)},\ldots, L^{(n)})}_{1d}$ at one of the poles \eqref{newpole1}, \eqref{newpole2}, or \eqref{newpole3}, in $\mathcal{M}_k\setminus\mathcal{M}^{pure}_k$. These extra poles making up the dotted terms need to be included, as dictated by the JK-prescription, and our first observation is that there is only a \emph{finite} number of them. This last point is highly nontrivial, and is derived from the explicit form of the integrand  \eqref{character1d}. As an example, suppose that there exists a pole of the first kind \eqref{newpole1}, at the locus $\phi^{(a)}_I - M^{(a)}_{\rho, *} - \epsilon_+ =0$, for some $I\in\{1,\ldots,k^{(a)}\}$. Then, there exist no pole at the locus $\phi^{(a)}_J - M^{(a)}_{\rho, *} - \epsilon_+ =0$ for any $J\neq I$. This is because there is a zero at the locus $\phi^{(a)}_I-\phi^{(a)}_J=0$, due to the numerator in $Z^{(a)}_{pure, vec}$. Similarly, the JK-residue prescription predicts a pole at the locus $\phi^{(a)}_J-\phi^{(a)}_I+\E_1=0$, due to the denominator of $Z^{(a)}_{pure, adj}$, and a pole at the locus $\phi^{(a)}_J-\phi^{(a)}_I+\E_2=0$, due to the denominator of $Z^{(a)}_{pure, vec}$. However, there is a zero at both loci, due to the zeros at the numerators of $Z^{(a)}_{defect}$. All in all, the locus \eqref{newpole1} contributes a \emph{single} new $M^{(a)}_{\rho, *}$-pole to the index. One can similarly show that  the only other $M^{(a)}_{\rho, *}$-poles are exclusively due to the loci \eqref{newpole2}, \eqref{newpole3}, and that this list of poles is bounded above for all $k$. Namely, for all vortex number $k$, the size of the set $\mathcal{M}_k\setminus\mathcal{M}^{pure}_k$ is always smaller or equal to some fixed integer $k'$. By carrying out the JK-residue procedure explicitly, we can determine $k'$ exactly: one finds that $k'+1$ is equal to the dimension of the finite-dimensional irreducible representation of the  quantum affine algebra $U_\hbar(\widehat{\fg})$ with highest weight $(L^{(1)},\ldots, L^{(n)})$. From this perspective, the first term in \eqref{firstterm1d} before the dots is nothing but the highest weight of the representation. This ends the first part of the proof.

It remains to show that each of the $k'$ terms in the dotted expansion  \eqref{firstterm1d} is precisely of the form \eqref{character1d}. This follows from a remarkable fact, which again can be proved by direct computation: for a given vortex number $k$, each contour enclosing $j'$ of the $k'$ poles in  $\mathcal{M}_k\setminus\mathcal{M}^{pure}_k$ can be traded for an integration contour which encloses $k-j'$ poles \textit{only}, where $j'=1,\ldots,k'$. The price to pay for such a trade of contours is the introduction in the integrand of extra $Y$-operator insertions, along with the residue at the $j'$ poles of the chiral matter factors  $Z^{(a)}_{pure, teeth}$. Performing this change of contours for all $k'$ dotted terms, we arrive at the advertised expression for the vortex $qq$-character.\\

\begin{figure}[h!]
	\emph{}
	\centering
	\includegraphics[trim={0 0 0 0cm},clip,width=0.99\textwidth]{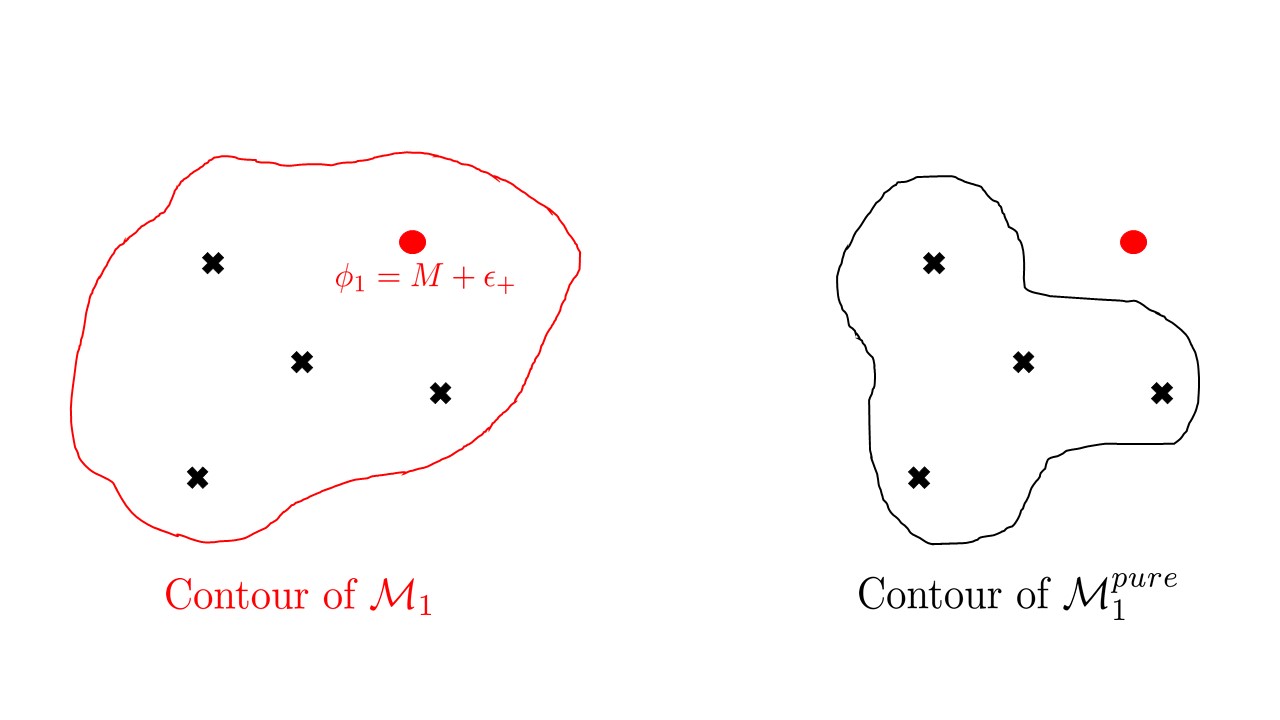}
	\vspace{-12pt}
	\caption{The black crosses denote poles in the set $\mathcal{M}^{pure}_k$, from the pure index, while the red dot denotes a pole in the set $\mathcal{M}_k\setminus \mathcal{M}^{pure}_k$. Such a pole is due to the factor $Z^{(a)}_{defect}$ in the integrand. On the left, we show a possible contour for the computation of the index at $k=1$. Note that by the JK prescription, we must in particular enclose the new pole in red. Remarkably, it is equivalent to trade this contour for the one on the right, which now only encloses the poles in the set $\mathcal{M}^{pure}_1$, but with a modified integrand; in the latter contour, the integrand will now contain insertions of additional $Y$-operators, with a vortex charge shift of one unit to account for the missing pole.} 
	\label{fig:contours}
\end{figure}

We emphasize  that at no point in the discussion did we need to know the content of the set $\mathcal{M}^{pure}_k$, that is to say the poles of the quantum mechanics index in the absence of defect. What is instead relevant here to derive the $qq$-character is the set of poles $\mathcal{M}_k\setminus \mathcal{M}^{pure}_k$, due entirely to the insertion of the defect. We now explain why this vortex character can be understood as a non-perturbative Schwinger-Dyson equation for the 3d theory $G^{3d}$.

\subsection{Physics of the Schwinger-Dyson Equations}
\label{ssec:schwingerphysics}

Let us focus on the case of a ``fundamental" defect on the $a$-th node of $G^{3d}$, meaning $L^{(a)}=1$ for some $a\in \{1,\ldots,n\}$ and $L^{(b)}=0$ for $b\neq a$. Correspondingly, the defect group is  $\widehat{G}_{defect}=U(1)$ in that case, parameterized by the mass  $z^{(a)}_1\equiv z$. 

The $Y$-operator we constructed mediates the change in vortex charge $k$ of the theory $G^{3d}$. More precisely, the vev $\left\langle Y^{(a)}_{1d} (z)\right\rangle$ represents the insertion of a line defect with associated mass $z$ on node $a$, enabling vortex particles to appear and disappear out of the bulk. This changes the topological sector of $G^{3d}$, and the $qq$-character \eqref{character1d} encodes a corresponding quantum affine symmetry of the theory. Put differently, the Schwinger-Dyson equation for the $Y$-operator vev is the statement that even though $\left\langle Y^{(a)}_{1d} (z)\right\rangle$ is singular in $z$, as is obvious from the explicit expression \eqref{Yoperator1d}, a particular Laurent series of $Y$-operator vevs, the $qq$-character, has regularity properties in $z$.  The precise statement of the Schwinger-Dyson equations is as follows:
\beq
\left[\widetilde{\chi}\right]^{(0,\ldots,0,1,0,\ldots,0)}_{1d}(z) \;\text{is regular in $z$ except for the poles of the function $\mathcal{Q}_{d_b^\omega}^{(b)}(z)$ in \eqref{character1d}.}\nonumber
\eeq
To prove this statement, one has to show that the residues at the various poles of the index do not develop singularities in the variable $z=e^{-M}$, other than at the denominators of  $\mathcal{Q}_{d_b^\omega}^{(b)}(z)$. A singularity in $z$ will arise if two poles of the integrand pinch one of the contours. It turns out that most potential singularities are canceled in a subtle manner by zeroes in the integrand, resulting in an \emph{almost} regular structure in $z$. Indeed, a finite set of $z$-singularities is produced after integration, and makes up the various poles of the function $\mathcal{Q}_{d_b^\omega}^{(b)}(z)$.

For instance, let $a\in\{1,\ldots,n\}$ and $I\in\{1,\ldots,k^{(a)}\}$, and consider the following two loci of poles in the integrand:
\beq
\phi^{(a)}_I - M - \E_+ = 0 \qquad \text{and}\qquad \phi^{(a)}_{I}- \widetilde{\mu}^{(a)}_j-\E_+ = 0 \; \text{for some mass}\; \{\widetilde{\mu}^{(a)}_j\}\; .
\eeq 
The first pole locus is due to the denominator of $Z^{(a)}_{defect}$, while the second pole locus is due to the denominator of $Z^{(a)}_{pure, teeth}$, the antifundamental chiral matter contribution. By the JK-residue prescription, the first locus is inside the integration contour, as we saw in \eqref{newpole1}.  Furthermore, the JK-residue instructs us to only enclose the $\{\mu^{(a)}_j\}$-poles coming from fundamental chiral multiplets \eqref{purepole3}, and none of the $\{\widetilde{\mu}^{(a)}_j\}$-poles coming from antifundamental chiral multiplets. It follows that the second locus is outside the integration contour. Then, the poles can freely coalesce and pinch the contour, resulting in the singular locus:
\beq
M=\widetilde{\mu}^{(a)}_j \; .
\eeq 
This singularity manifests itself as a simple pole in the function $\mathcal{Q}_{d_b^\omega}^{(b)}(z)$.

Given a generic theory, providing the comprehensive list of $z$-singularities in the index is a tedious exercise, though it presents no technical difficulties; one simply proceeds as above, analyzing the various sets of poles which can potentially pinch the contours. We will carry out this procedure in detail when presenting an example in section \ref{sec:example}.

As a last remark, note that this discussion straightforwardly generalizes to a  line defect with an arbitrary group $\widehat{G}_{defect}=\prod_{a=1}^n U(L^{(a)})$. In that case, the Schwinger-Dyson equations are still regularity conditions on the associated $qq$-character, but involving correlation functions of a higher number of $Y$-operators. Typically, the index  $\left[\widetilde{\chi}\right]^{(L^{(1)},\ldots,L^{(n)})}_{1d}(\{z^{(a)}_\rho\})$will develop new singularities in the defect masses $\{z^{(a)}_\rho\}$.\\

We now give an alternate derivation of the non-perturbative Schwinger-Dyson equations obeyed by 3d $\cN=4$ gauge theories, directly from 3 dimensions and without resorting to its vortex quantum mechanics.

\vspace{16mm}

\section{Schwinger-Dyson Equations: the 3-dimensional Perspective}
\label{sec:3dsection}

It is known that the half-index of the 3d $\cN=4$ quiver gauge theory $G^{3d}$ supported on  $\mathbb{C}\times S^1(\widehat{R})$ factorizes into perturbative and vortex contributions, where the latter precisely coincides with the Witten index of the quantum mechanics $T^{1d}_{pure}$ \cite{Hwang:2017kmk}. 
In this section we propose to couple $G^{3d}$ to the 1/2-BPS loop defect wrapping $S^1(\widehat{R})$, and show that a sensible 3d/1d half-index can be defined derive the vortex $qq$-character as we obtained it from the quantum mechanics $T^{1d}$.

\subsection{A Half-Index Presentation}
\label{ssec:3dindex}

Let us first review how to define a half-index for $G^{3d}$ on  $\mathbb{C}\times S^1(\widehat{R})$ in the absence of defect. This index is also referred to as a holomorphic block \cite{Beem:2012mb}\footnote{Formally, a holomorphic block is defined in the IR of the 3d $\cN=4$ theory: on the manifold $\mathbb{C}\times S^1(\widehat{R})$, one has to specify boundary conditions at infinity on $\mathbb{C}$ by choosing an IR vacuum. Alternatively, one can consider a ``half-index" on $D\times S^1(\widehat{R})$, where $D$ is a finite disk, with boundary conditions defined in the UV on the edge of the disk. The boundary conditions will flow to some boundary condition in the IR which may or may not agree with the one defined in the holomorphic block formalism. It would be important to explore these subtleties in our context.}.  We first consider the 3-manifold in the $\Omega$-background, to regularize the non-compactness of $\mathbb{C}$; namely, if we let $z$ be a complex coordinate on the complex line, we can view the 3-manifold as a $\mathbb{C}$-bundle over $S^1(\widehat{R})$, where as we go around the circle, we make the identification
\beq\label{omega}
z \sim  z\, e^{\widehat{R}\, \epsilon_1}\; ,\qquad  \epsilon_1\in\mathbb{R} \; .
\eeq
From now on, we denote the $\mathbb{C}$-line in this background as $\mathbb{C}_q$, with $q=e^{\widehat{R}\, \epsilon_1}$.
Then, the partition function of $G^{3d}$ is defined via the following half-index:
\beq
\label{index3d}
\left[\widetilde{\chi}\right]^{(0,\ldots, 0)}_{3d} =\text{Tr}\left[(-1)^F\,e^{-\widehat{R}\{Q,\overline{Q}\}}\, q^{S_1 - S_H}\, t^{-S_2 + S_H}\prod_{a=1}^n  \prod_{d=1}^{N^{(a)}_F} (x^{(a)}_d)^{\Pi^{(a)}_d}\,\right] \; .
\eeq
The trace is taken over the Hilbert space of states on  $\mathbb{C}_q$.
The index counts states in $Q$-cohomology, where $Q= Q^1_{1,1}$ and $\overline{Q}=Q^2_{2,2}$ were defined in  section \ref{ssec:review}. $F$ is the fermion number. $S_1$ is a rotation generator for $\mathbb{C}_q$, while $S_2$ is a R-symmetry generator; indeed, $S_2$ generates a $U(1)_C$ symmetry which is a subgroup of the $SU(2)_C$ R-symmetry acting on the vector multiplet scalars. Meanwhile, $S_H$ generates a $U(1)_H$ symmetry which is a subgroup of the $SU(2)_H$ R-symmetry acting on the hypermultiplet scalars. $\{\Pi^{(a)}_d\}$ are Cartan generators for the flavor group $G_F$, with conjugate fundamental masses $\{m^{(a)}_d\}$.  The field configurations which preserve the supersymmetries of the index are solutions to the vortex equations on $\mathbb{C}$.

So far we have not talked about the gauge symmetry group $G=\prod_{a=1}^{n} U(N^{(a)})$.
We start by treating it as a global symmetry, which we make abelian by breaking it to its maximal torus. The associated equivariant parameters are denoted collectively as ``$y$". We then gauge the symmetry by projecting to $G$-invariant states, which amounts to integrating over those parameters. Namely, 
\beq\label{Haar3d}
\oint d_{Haar}y=\oint \prod_{a=1}^{n}\prod_{i=1}^{N^{(a)}}\frac{dy^{(a)}_i}{y^{(a)}_i}\; .
\eeq
Above, the contour is chosen to project to states neutral under the $G$-symmetry. Because the parameters $\{y^{(a)}_i\}$ parameterize part of the Coulomb branch of $G^{3d}$, this presentation of the index is referred to as Coulomb branch localization:
\begin{align}\label{indexintegral3d}
\left[\widetilde{\chi}\right]^{(0,\ldots, 0)}_{3d} =  \oint_{\mathcal{M}^{bulk}} dy\,\left[I^{3d}_{bulk}(y)\, \right]  \; .
\end{align}
The choice of contours $\mathcal{M}^{bulk}$ determines a vacuum for $G^{3d}$. In this 3-dimensional setup, the contours are once again fixed by the JK-residue prescription, where we choose to work with the auxiliary vector $\eta=(1,\ldots,1)$, as we did before.
The integrand $I^{3d}_{bulk}(y)$ stands for the contribution of all the various multiplets to the index.  These can be read off directly from the 3d $\cN=4$ quiver description of the theory. This bulk contribution has the form \cite{Beem:2012mb,Aganagic:2017smx}: 
\begin{align}\label{bulk3d}
I^{3d}_{bulk}(y)=\prod_{a=1}^{n}\prod_{i=1}^{N^{(a)}}{y^{(a)}_i}^{\left(\zeta_{3d}^{(a)}-1\right)}\;I^{(a)}_{vec}(y)\cdot\prod_{b>a}I^{(a,b)}_{bif}(y) \cdot  I^{(a)}_{flavor}(y,\{x^{(a)}_d\})\; .
\end{align}
The factor 
\begin{align}\label{FI3d}
\prod_{a=1}^{n}\prod_{i=1}^{N^{(a)}}{y^{(a)}_i}^{\left(\zeta_{3d}^{(a)}\right)}
\end{align}
is the contribution of the 3d FI parameters. Namely, $\zeta_{3d}^{(a)}$ is the real FI parameter on node $a$, complexified by the holonomy of the corresponding background gauge field around $S^1(\widehat{R})$. Throughout this section, we will also make use of the K-theoretic notation  $\widetilde{\fq}^{(a)}=e^{\widehat{R}\,\zeta_{3d}^{(a)}}$, with the holonomy implicitly understood.

The factor
\begin{align}\label{vec3d}
I^{(a)}_{vec}(y)=\prod_{1\leq i\neq j\leq N^{(a)}}\frac{\left(y^{(a)}_{i}/y^{(a)}_{j};q\right)_{\infty}}{\left(t\, y^{(a)}_{i}/y^{(a)}_{j};q\right)_{\infty}}\;\prod_{1\leq i<j\leq N^{(a)}} \frac{\Theta\left(t\,y^{(a)}_{i}/y^{(a)}_{j};q\right)}{\Theta\left(y^{(a)}_{i}/y^{(a)}_{j};q\right)}
\end{align}
stands for the contribution of a ${\cN}=4$ vector multiplet for the gauge group $U(N^{(a)})$. Above, we use the following definitions of the $q$-Pochhammer symbol,
\beq
\left( x\, ; q\right)_{\infty}\equiv\prod_{l=0}^{\infty}\left(1-q^l\, x\right)\; ,
\eeq
and of the theta function,
\beq
\Theta\left(x\,;q\right)\equiv \left(x \,;\, q\right)_\infty\,\left(q/x \,;\, q\right)_\infty \; .
\eeq  
In particular, decomposing the $\cN=4$ vector multiplet as a $\cN=2$ vector multiplet and $\cN=2$ adjoint chiral multiplet, the numerator factor $\left(y^{(a)}_{i}/y^{(a)}_{j};q\right)_{\infty}$ is the contribution of the W-bosons in the $\cN=2$ vector multiplet, while the denominator factor $\left(t\, y^{(a)}_{i}/y^{(a)}_{j};q\right)_{\infty}$ is the contribution of the $\cN=2$ adjoint chiral multiplet\footnote{The ratio of theta functions has a natural interpretation when the manifold is thought of as $D\times S^1(\widehat{R})$, with $D$ the disk. The boundary of that manifold is $S^1\times S^1=T^2$, and one needs to specify the 2d theory on this torus $T^2$. A priori, any choice of 2d $\cN=(0,2)$ boundary conditions will do, as long as the theory is anomaly-free. In our context, one should specify the 3d chiral multiplet boundary conditions, which are either Dirichlet or Neumann. The gauge fields have Neumann boundary conditions, and the appearance of theta functions in the 3d vector multiplet is understood as the contribution of the 2d elliptic genus on the boundary torus. For details, see \cite{Gadde:2013wq,Yoshida:2014ssa,Dimofte:2017tpi}, and the related discussion in \cite{Aganagic:2017smx}.}.

The factor
\begin{align}
\label{bif3d}
I^{(a,b)}_{bif}(y)=\prod_{1\leq i \leq D^{(a)}}\prod_{1\leq j \leq D^{(b)}}\left [ \frac{(t\, v\, y^{(a)}_{i}/y^{(b)}_{j};q)_{\infty}}{(v \, y^{(a)}_{i}/y^{(b)}_{j};q)_{\infty}}\right]^{\Delta^{a b}}
\end{align}
is the contribution of $\cN=4$ bifundamental hypermultiplets. We use the same notations as introduced previously: $\Delta^{ab}$ is the incidence matrix of the Lie algebra $\fg$,  and $v=\sqrt{q/t}$.

The factor 
\begin{align}\label{matter3d}
I^{(a)}_{flavor}(y, \{x^{(a)}_d\}) =\prod_{d=1}^{N_F^{(a)}}\prod_{i=1}^{N^{(a)}}  \frac{\left(t\, v\, x^{(a)}_{d}/y^{(a)}_{i};q\right)_{\infty}}{\left(v\, x^{(a)}_{d}/y^{(a)}_{i};q\right)_{\infty}}
\end{align}
stands for the contribution of $\cN=4$ hypermultiplets in the fundamental representation of the $a$-th gauge group. The $\cN=4$ supersymmetry fixes the R-charge assignments of the various fugacities in the arguments of the $q$-Pochhammer symbols. In particular, note the presence of a cubic superpotential term due to the bifundamental/adjoint chiral multiplets in the $\cN=2$ language.

Let us briefly discuss the contours: there are three distinct sets of poles in the set $\mathcal{M}^{bulk}$, following the JK-residue prescription: the first is due to the denominators $\left(v\, x^{(a)}_{d}/y^{(a)}_{i};q\right)_{\infty}$ from the fundamental matter contribution \eqref{matter3d}, resulting in:
	\beq
	y^{(a)}_i =v\, x^{(a)}_d \, q^{s}\qquad\;\; ,\;\; s=0,1,2,\ldots\; , \;\;\; d\in\{1,\ldots,N^{(a)}_F\} \; .
	\eeq
	Second, there are the denominators $\left[(v \, y^{(a)}_{i}/y^{(b)}_{j};q)_{\infty}\right]^{\Delta^{a b}}$ from the bifundamentals \eqref{bif3d}, resulting in:
\begin{align}
&y^{(b)}_i =v\, y^{(a)}_j \, q^{s}\qquad ,\;\; s=0,1,2,\ldots,\; \text{if there is a link between nodes $a$ and $b>a$}\; .
\end{align}
Third, there are the denominators $\left(t\, y^{(a)}_{i}/y^{(a)}_{j};q\right)_{\infty}$ from the vector multiplets \eqref{vec3d}. However, such $t$-dependent poles turn out to have vanishing residue, as can be easily checked\footnote{The fact that poles of the third kind give a vanishing residue is characteristic of indices for 3d theories with $\cN=4$ supersymmetry. In particular, our argument does not apply to 3d theories with less supersymmetry. A similar phenomenon was noted in the 1d vortex quantum mechanics index.}. The above pole structure makes explicit the grading over the vortex charge, so we naturally denote the set of poles as $\mathcal{M}^{bulk}_k$, summed over all $k=\sum_{a=1}^n k^{(a)}$.\\

We now introduce a 1/2-BPS loop wrapping $S^1(\widehat{R})$, which is a codimension-2 defect from the point of view of $G^{3d}$. In particular, such a loop preserves the supersymmetries we wrote as $Q$ and $\overline{Q}$ in the half-index. We couple the 1-dimensional $\cN=4$ theory on the loop to the bulk 3-dimensional theory by considering the flavor symmetries of the 1d theory and gauging them with 3d $\cN=4$ vector multiplets. From the point of view of the index, this translates into gauging the 1d masses, turning them into the scalars of the corresponding 3d $\cN=4$ vector multiplets. When the vector multiplet is dynamical, the scalar becomes an eigenvalue $y$ to be integrated over, while in the case of a background vector multiplet, the scalar becomes a mass from the 3d point of view.
To achieve this, we start by defining a defect $Y$-operator vacuum expectation value, written as an integral over the Coulomb moduli of the 3d theory:
\begin{align}\label{3ddefectexpression}
\left\langle\left[{\widetilde{Y}^{(a)}_{3d\;  gauge/1d}}(z)\right]^{\pm 1} \right\rangle \equiv  \oint_{\mathcal{M}^{bulk}} d{y}\,\left[I^{3d}_{bulk}(y)\cdot\left[{\widetilde{Y}^{(a)}_{3d\; gauge/1d}}(y, z)\right]^{\pm 1}\right]  \; .
\end{align}
The defect factor in the integrand then corresponds to the chiral multiplet contribution 
\beq\label{3dWilsonfactor}
{\widetilde{Y}^{(a)}_{3d\; gauge/1d}}(y,z)=\prod_{i=1}^{N^{(a)}}\frac{1-t\, y^{(a)}_{i}/z}{1- y^{(a)}_{i}/z}\; .
\eeq
There is another piece of the defect $Y$-operator which is not integrated over, as it couples the loop to the flavor symmetry of $G^{3d}$; this  chiral multiplet contribution has the generic form
\beq\label{3dWilsonfactor2}
{\widetilde{Y}^{(a)}_{3d\; flavor/1d}}(\{x^{(b)}_{d}\}, z)=\prod_{b=1}^{n}\prod_{d=1}^{N_F^{(b)}}\frac{1- v^{\#^{(ab)}+1}\, x^{(b)}_{d}/z}{1- t\, v^{\#^{(ab)}+1}\, x^{(b)}_{d}/z}\; ,
\eeq
where $\#^{(ab)}$ is a non-negative integer equal to the number of links between nodes $a$ and $b$ in the Dynkin diagram of $\fg$.

Note that the contour definition for the above $Y$-operator vev \eqref{3ddefectexpression} is the same as the contour definition for the 3d index \eqref{indexintegral3d}  in the absence of defect. In particular, the contours are defined not to enclose the potential $z$-poles from the factor $\prod_{a=1}^n {\widetilde{Y}^{(a)}_{3d\; gauge/1d}}$.\\ 

Then, we \emph{define} the vortex character of $G^{3d}$ as a sum of 3d/1d half-indices:
\begin{empheq}[box=\fbox]{align}
\label{NEWcharacter3d}
&\left[\widetilde{\chi}\right]^{(L^{(1)},\ldots, L^{(n)})}_{3d}(\{z^{(a)}_\rho\}) \\
	&=\frac{1}{\left[{\chi}\right]^{(0,\ldots,0)}_{3d}}\sum_{\omega\in V(\lambda)}\prod_{b=1}^n \left({\widetilde{\fq}^{(b)}}\right)^{d_b^\omega}\; c_{d_b^\omega}(q, t)\; \left(\widetilde{\mathcal{Q}}_{d_b^\omega}^{(b)}(\{z_{\rho}^{(a)}\})\right)\,  \left[\widetilde{Y}_{3d\;  gauge/1d}(\{z^{(a)}_\rho\})\right]_{\omega} \, .\nonumber
\end{empheq}
In the above, the factor $\left[\widetilde{Y}_{3d\;  gauge/1d}(\{z^{(a)}_\rho\})\right]_{\omega}$ is defined as the vev of a rational function of the $Y$-operators $\left\langle \prod_a \left[\widetilde{Y}^{(a)}_{3d\; gauge/1d}\right]^{\pm 1} \right\rangle $ \eqref{3ddefectexpression}, and possible  derivatives thereof. All the other functions and notations appearing above are the same as were introduced in section \ref{ssec:1dqqcharacter}\footnote{There is one subtle difference: in the quantum mechanics $T^{1d}$, the function $\mathcal{Q}_{d_b^\omega}^{(b)}(z_{\rho}^{(a)})$ was the residue of $Z^{(b)}_{pure, teeth}$ at the various poles of  $\mathcal{M}_k\setminus \mathcal{M}^{pure}_k$. In the 3d setup used here, we have defined a function $\widetilde{\mathcal{Q}}_{d_b^\omega}^{(b)}(z_{\rho}^{(a)})$, which is still written as contributions of 1d $\cN=4$ chirals, but not quite the same expressions as in the quantum mechanics. Giving exact formulas would require specifying the theory $G^{3d}$, see section \ref{sec:example} for a detailed example.}. This implies that the index is once again a twisted $qq$-character of a finite dimensional irreducible representation $V(\lambda)$ of the quantum affine algebra $U_\hbar(\widehat{\fg})$, with highest weight $\lambda=\sum_{a=1}^m L^{(a)}\, \lambda_a$. 

The above definition of the 3d/1d half-index seems ad-hoc from the 3-dimensional perspective, but can be justified as follows: the vev $\left\langle{\widetilde{Y}^{(a)}_{3d\;  gauge/1d}}(z) \right\rangle$ has $z$-singularities, and the vortex character observable is constructed in such a way to precisely cancel those singularities. The representation theory of quantum affine algebras guarantees that this construction exists and is unique, resulting in a Laurent polynomial  in $Y$-operator vevs and possibly derivatives thereof. Alternatively, we will see next that the vortex character construction is in fact very natural in the light of the BPS/CFT correspondence.  We end this section by exhibiting the relation between the 3d/1d index \eqref{NEWcharacter3d} and the Witten index \eqref{character1d} of the 1d quantum mechanics: up to normalization, they turn out to be one and the same!

\subsection{Relation between the 3d and 1d Expressions for the $qq$-character}
\label{ssec:3dindexis1dindex}

As we reviewed, the choice of contour for the 3d half-index fixes a vacuum for $G^{3d}$. Let $T^{1d}_{pure}$ be the vortex quantum mechanics defined on that vacuum, and let  $T^{1d}$ be the vortex quantum mechanics in the presence of a defect. We now prove that the index of $T^{1d}$ is, up to a constant factor, the 3d/1d half-index introduced above. The proof rests on establishing a relation between the defect $Y$-operator vev $\left\langle{\widetilde{Y}^{(a)}_{3d\;  gauge/1d}}(z)\right\rangle$ and its quantum mechanical counterpart $\left\langle{{Y}^{(a)}_{1d}}(z)\right\rangle$. 
First recall that in defining the 1d $Y$-operator vev \eqref{Yoperator1d}, one sums over the poles  \eqref{polepure} in the set $\mathcal{M}^{pure}_k$ for each vortex charge $k=\sum_{a=1}^n k^{(a)}$,
\beq\label{polepureagain}
\phi^{(a)}_I = \mu^{(b)}_i - \E_- - (s_i-1) \E_1 + 2\,\#^{(ab)} \E_+ \; , \;\;\;\;  \text{with}\; s_i\in\{1,\ldots,k^{(a)}_i\},\;\; i\in\{1,\ldots,N^{(a)}\}\; ,
\eeq
for some mass index $b\in\{1,\ldots,n\}$. Recall that in the notation above, $(k^{(a)}_1, \ldots, k^{(a)}_{N^{(a)}})$ is a partition of the vortex charge $k^{(a)}$ into $N^{(a)}$ non-negative integers, and $\#^{(ab)}$ is a non-negative integer equal to the number of links between nodes $a$ and $b$ in the Dynkin diagram of $\fg$. We write the collection of all such $\phi$-loci as $\vec\phi_*$. After performing the residue sum, the $Y$-operator vev can be schematically written as:
\beq\label{residuesum1d}
\left\langle{{Y}^{(a)}_{1d}}(M)\right\rangle =\sum^{\infty}_{k=0}\sum_{\{\vec\phi_*\}\in\mathcal{M}^{pure}_k} Z^{1d}_{pure}(\vec\phi_*) \cdot Z^{(a)}_{defect}(\vec\phi_*,M) \; ,
\eeq
where we collected all the contributions independent of the defect inside a factor $Z^{1d}_{pure}$, while the remaining factor is due to the interaction \eqref{defect} between the loop and the vortices, rewritten here for convenience: 
\beq
\label{defectagain}
Z^{(a)}_{defect}(\phi^{{(a)}}_I,M) =  \prod_{I=1}^{k^{(a)}} \frac{\sh\left(\phi^{{(a)}}_I-M^{(a)}_\rho- \E_- \right)\, \sh\left(-\phi^{{(a)}}_I + M^{(a)}_\rho- \E_-\right)}{\sh\left(\phi^{{(a)}}_I-M^{(a)}_\rho- \E_+\right)\, \sh\left(-\phi^{{(a)}}_I + M^{(a)}_\rho - \E_+ \right)} \, .
\eeq
Meanwhile, in the 3-dimensional setup, we sum over the poles $\{\vec y_*\}\in\mathcal{M}^{bulk}$, and the $Y$-operator vev \eqref{3ddefectexpression} becomes:
\begin{align}\label{3ddefectexpressionagain}
\left\langle{\widetilde{Y}^{(a)}_{3d\;  gauge/1d}}(z) \right\rangle \equiv {\widetilde{Y}^{(a)}_{3d\; flavor/1d}}(\{x^{(b)}_d\}, z)\sum^{\infty}_{k=0}\sum_{\{\vec y_*\}\in\mathcal{M}^{bulk}_k} I^{3d}_{bulk}(\vec y_*)\cdot{\widetilde{Y}^{(a)}_{3d\; gauge/1d}}(\vec y_*, z) \; .
\end{align}
It is well known that in the absence of loop defect, the 3d index  $\sum^{\infty}_{k=0}\sum_{\{\vec y_*\}\in\mathcal{M}^{bulk}_k} I^{3d}_{bulk}(\vec y_*)=\left[\widetilde{\chi}\right]^{(0,\ldots, 0)}_{3d}$ is in fact equal to the quantum mechanical index $\sum^{\infty}_{k=0}\sum_{\{\vec\phi_*\}\in\mathcal{M}^{pure}_k} Z^{1d}_{pure}(\vec\phi_*)=\left[\widetilde{\chi}\right]^{(0,\ldots, 0)}_{1d}$, up to overall normalization. We refer the reader to the references \cite{Beem:2012mb,Bullimore:2014awa,Hwang:2017kmk,Aganagic:2017smx} for details. Because the contents of the sets $\mathcal{M}^{bulk}_k$ and $\mathcal{M}^{pure}_k$ are in one-to-one correspondence for all $k$, this implies in particular that the summands $I^{3d}_{bulk}(\vec y_*)$  and  $Z^{1d}_{pure}(\vec\phi_*)$ are the same, up to normalization.

We therefore need to simply investigate the remaining factor $Z^{(a)}_{defect}(\vec\phi_*,M)$ in \eqref{residuesum1d}. We switch to K-theoretic variables $z=e^{-M}$, $f^{(a)}_{i}=e^{-\mu^{(a)}_i}$, and let
\beq
y^{(a)}_{i,*} = f^{(b)}_i\, q^{k^{(a)}_i+1} \, v^{-2\#^{(ab)}}\;,\;\; i\in\{1,\ldots,N^{(a)}\}\;, \;\; b\in\{1,\ldots,n\} \; .
\eeq
We further renormalize the masses to make contact with their 3d definitions: 
\beq
f^{(b)}_{i} \equiv v^{3\#^{(ab)}+1}\,x^{(b)}_{i}  \; .
\eeq
After a finite number of telescopic cancellations, the 1d $Y$-operator at the locus \eqref{polepureagain} becomes:
\begin{align}\label{fromY1dtoY3d}
Z^{(a)}_{defect}({y}^{(a)}_{i,*},z) &=\prod_{i=1}^{N^{(a)}}\frac{1-t\, y^{(a)}_{i,*}/z}{1-y^{(a)}_{i,*}/z}\cdot\prod_{i=1}^{N^{(a)}}\frac{1- v^{\#^{(ab)}+1}\, x^{(b)}_{i}/z}{1- t\, v^{\#^{(ab)}+1}\, x^{(b)}_{i}/z} \; .
\end{align}
We recognize the first product as the 3d/1d contribution
\beq
{\widetilde{Y}^{(a)}_{3d\; gauge/1d}}(\vec y_*, z)= \prod_{i=1}^{N^{(a)}}\frac{1-t\, y^{(a)}_{i,*}/z}{1- y^{(a)}_{i,*}/z} \; .
\eeq 
Meanwhile, the latter product is part of the 3d/1d contribution
\beq
\widetilde{Y}^{(a)}_{3d\; flavor/1d}(\{x^{(b)}_d\}, z) = \prod_{i=1}^{N^{(a)}}\frac{1- v^{\#^{(ab)}+1}\, x^{(b)}_{i}/z}{1- t\, v^{\#^{(ab)}+1}\, x^{(b)}_{i}/z}\cdot c^{(a)}_{3d/1d}(\{x^{(b)}_d\}, z)\; ,
\eeq 
where the leftover factor have been collected in the expression $c_{3d/1d}(\{x^{(b)}_d\}, z)$; this factor can be determined exactly by simply comparing the above result to the contribution \eqref{3dWilsonfactor2}, if one wishes. 
Putting it all together, we have shown that the $Y$-operator vevs are proportional to each other:
\beq\label{Yequality}
\left\langle{\widetilde{Y}^{(a)}_{3d\;  gauge/1d}(z)} \right\rangle = c^{(a)}_{3d/1d}(\{x^{(b)}_d\}, z) \, \left\langle {Y_{1d}^{(a)}(z)} \right\rangle
\eeq
Now, the 3d/1d half-index is a Laurent series in $Y$-operator vevs, with the same functional form and number of terms as the index of $T^{1d}$. This does not yet guarantee the indices are the same, since $c^{(a)}_{3d/1d}$  appears as a relative factor between the various terms of the character. Remarkably, one can show after computing each term in the character  that they all share the \emph{same} proportionality factor $c^{(a)}_{3d/1d}$, so it can be factored out entirely. If one considers a general defect group $\widehat{G}_{defect}=\prod_{a=1}^n U(L^{(a)})$, the normalized characters of $G^{3d}$ and its quantum mechanics $T^{1d}$ therefore satisfy:
\begin{align}\label{CHIequality}
\left[\widetilde{\chi}\right]^{(L^{(1)},\ldots, L^{(n)})}_{3d}(\{z^{(a)}_\rho\})= c_{3d/1d}\cdot  \left[\widetilde{\chi}\right]^{(L^{(1)},\ldots, L^{(n)})}_{1d}(\{z^{(a)}_\rho\})\; .
\end{align}
The proportionality constant $c_{3d/1d}$ is simply:
\beq
c_{3d/1d}=\prod_{a=1}^{n}\prod_{\rho=1}^{L^{(a)}} c^{(a)}_{3d/1d}(\{x^{(b)}_d\}, z^{(a)}_\rho)\; .
\eeq


\vspace{16mm}

\section{Schwinger-Dyson Equations: the ${\cW}_{q,t}({\fg})$-Algebra Perspective}
\label{sec:deformedWsection}

The BPS/CFT correspondence predicts that Schwinger-Dyson equations for the gauge theory $G^{3d}$ should have a counterpart as a set of Ward identities for a conformal field theory, or a deformation thereof, on a Riemann surface. In this paper, we show that the vortex $qq$-character is a certain (chiral) correlator a deformed ${\cW}({\fg})$-algebra on the cylinder.

\subsection{The Deformed ${\cW}_{q,t}({\fg})$-Algebra} 
\label{ssec:qToda}

Let $\fg$ be a simply-laced Lie algebra.
In the work \cite{Frenkel:1998}, a deformation of ${\cW}({\fg})$-algebras was proposed, in free field formalism, based on a certain canonical deformation of the screening currents; this is the symmetry algebra of the so-called $\fg$-type $q$-Toda theory on a cylinder. See also \cite{Shiraishi:1995rp} for the special case $\fg=A_1$, and \cite{Feigin:1995sf,Awata:1995zk} for the case $\fg=A_n$. 
The starting point is to define a $(q,t)$-deformed Cartan matrix\footnote{In what follows, we follow the conventions of \cite{Frenkel:1998}. Other definitions of the deformed Cartan matrix are possible, by introducing explicit ``bifundamental masses" in the off-diagonal entries \cite{Kimura:2015rgi}.}:
\begin{align}\label{CartanToda}
C_{ab}(q,t)= \left(q\, t^{-1} +q^{-1}\, t\right) \, \delta_{ab}- [\Delta_{ab}]_q\; .
\end{align} 
Let us explain the notation: a number in square brackets is called a quantum number, defined as
\begin{align}\label{quantumnumber}
[n]_q = \frac{q^{n}-q^{-n}}{q-q^{-1}} \; ,
\end{align}
and the incidence matrix is $\Delta_{ab}= 2 \, \delta_{ab} - C_{ab}$. 
Later, we  will also need  the inverse of the Cartan matrix:
\beq\label{Mmatrixdef}
M(q,t)= C(q,t)^{-1}\; .
\eeq
On then constructs a deformed Heisenberg algebra, generated by $n$ positive simple roots, satisfying:
\begin{align}\label{commutatorgenerators}
[\alpha_a[k], \alpha_b[m]] = {1\over k} (q^{k\over 2} - q^{-{k\over 2}})(t^{{k\over 2} }-t^{-{k\over 2} })C_{ab}(q^{k\over 2} , t^{k\over 2} ) \delta_{k, -m} \; .
\end{align}
In the above, it is understood that the zero-th generator commutes with all others: $[\alpha_a[k], \alpha_b[0]]=0$, for $k$ an arbitrary integer.
The Fock space representation of this algebra is given by acting on the vacuum state $|\psi\rangle$:
\begin{align}
\alpha_a[0] |\psi\rangle &= \langle\psi, \alpha_a\rangle |\psi\rangle\label{eigenvalue}\nonumber\\
\alpha_a[k] |\psi\rangle &= 0\, , \qquad\qquad\;\; \mbox{for} \; k>0\; .
\end{align}
Then, we define deformed screening operators as\footnote{The screening operators we write down are called ``magnetic" in \cite{Frenkel:1998}, and are defined with respect to the parameter $q$. Another  set of ``electric" screening and vertex operators can be constructed using the parameter $t$ instead, but these will not enter our discussion. From the point of view of the 3d gauge theory, this amounts to having $G^{3d}$ defined on the manifold $\mathbb{C}_t\times S^1(\widehat{R})$ instead of the manifold $\mathbb{C}_q\times S^1(\widehat{R})$. We made the choice of working on the latter manifold in this paper, hence why we only make use of magnetic screenings here.}. 
\beq\label{screeningdef}
S^{(a)}(y) = y^{-\alpha_a[0]}\,: \exp\left(\sum_{k\neq 0}{ \alpha_a[k] \over q^{k\over 2} - q^{-\,{k\over 2}}} \, y^k\right): \; .
\eeq
Note all operators in this section are written up to a center of mass zero mode, whose effect is simply to shift the momentum of the vacuum. Up to redefinition of the vacuum $|\psi\rangle$, we safely ignore such factors.

The ${\cW}_{q,t}({\fg})$-algebra is defined as the associative algebra whose generators are Fourier modes of the operators commuting with the screening charges,
\beq\label{screeningchargedef}
Q^{(a)} =\oint dy\, S^{(a)}(y) \; .
\eeq 
We denote the generating currents as $W^{(s)}(z)$, labeled by their ``spin" $s$. We therefore have\footnote{Note that the generating currents must also commute with the set of electric screening charges we mentioned in the previous footnote.}:
\beq\label{generatorsdef}     
[W^{(s)}(z),Q^{(a)}]=0\; , \qquad \text{for all}\;\; a=1, \ldots, n, \;\;\text{and}\;\; s=2, \ldots,n+1\; .
\eeq
In this way, one finds that every generating current can be written as a Laurent polynomial in certain vertex operators, which we call $\cY$-operators for reasons that will soon be clear:
\beq\label{YoperatorToda}
{\cY^{(a)}}(z)= q^{w_a[0]}\,: \exp\left(-\sum_{k\neq 0} w_a[k]\, t^{-k/2}  \, z^k\right): \; .
\eeq
Degenerate vertex operators are constructed out of $n$ fundamental weight generators,
\begin{align}\label{commutator2}
[\alpha_a[k], w_b[m]] ={1\over k} (q^{k \over 2}  - q^{-{k \over 2} })(t^{{k \over 2}}-t^{-{k \over 2} })\,\delta_{ab}\,\delta_{k, -m} \, .
\end{align}
These are dual to the operators $\alpha_a[k]$. Put differently,
\beq\label{etow}
\alpha_a[k] = \sum_{b=1}^n C_{ab}(q^{k\over 2},t^{k \over 2})\, w_b[k]\; .
\eeq
For completeness, we also write the commutator of two coweight generators,
\begin{align}\label{commutator3}
[w_a[k], w_b[m]] ={1\over k} (q^{k\over 2}  - q^{-{k \over 2} })(t^{{k \over 2}}-t^{-{k \over 2} })\,M_{ab}(q^{k\over 2} , t^{k\over 2})\,\delta_{k, -m} \, .
\end{align}
Among the vertex operators of the theory, a distinguished class which will enter our story is the set of so-called \emph{fundamental} vertex operators \cite{Frenkel:1998}:
\beq\label{fundvertex}
V^{(a)}(x) = x^{w_a[0]}\,: \exp\left(-\sum_{k\neq 0}{ w_a[k] \over q^{k\over 2} - q^{-\,{k \over 2}}} \, x^k\right): \; .
\eeq

Before we proceed, let us briefly comment on an important limit: if we rescale $q= \exp(\widehat{R} \epsilon_1),\, t=\exp(-\widehat{R} \epsilon_2)$ and take $\widehat{R}\rightarrow 0$, the deformed ${\cW}_{q,t}({\fg})$-algebra becomes the standard ${\cW}({\fg})$-algebra\footnote{Note this is \emph{not} the limit which produces 2d vortex characters in gauged sigma models by circle reduction of the 3d gauge theory on the circle. See the work \cite{Nieri:2019mdl} for that limit instead.}, which is the symmetry algebra of $\fg$-type Toda conformal field theory\footnote{We presented ${\cW}_{q,t}({\fg})$-algebras in the free field formalism since it is the only known way to deform ${\cW}({\fg})$-algebras. This is sometimes called the Coulomb gas formalism, or the Dotsenko-Fateev formalism \cite{Dotsenko:1984nm}. For a modern treatment of the topic, we refer the reader to  \cite{Dijkgraaf:2009pc,Itoyama:2009sc,Mironov:2010zs,Morozov:2010cq,Maruyoshi:2014eja}.}; for an extensive review, see \cite{Bouwknegt:1992wg}. In particular, if we set $b\equiv -\epsilon_2/\epsilon_1$, the central charge of the theory is $c=m+12 \left\langle Q,Q\right\rangle$, where   $Q=\rho\, ( b + 1/b)$ is the background charge, $\rho$ the Weyl vector of $\fg$, and the bracket is the Cartan-Killing form.  For the Heisenberg algebra \eqref{commutatorgenerators} to keep making sense, the root (and weight) generators should also be rescaled by $\widehat{R}$ and $\epsilon_1$ in this limit.
The deformed screenings currents \eqref{screeningdef} become
\beq\label{screeningcurrent}
S^{(a)}(y) = \;:e^{\langle\alpha_{a}, \varphi(y)\rangle/b}: \; ,
\eeq
with $\alpha_a$ the $a$-th simple root of $\fg$, and  $\varphi$ a $n$-dimensional boson.
The deformed fundamental vertex operators \eqref{fundvertex} become vertex operators of unit momentum,
\beq\label{primary}
V^{(a)}(x) = \; :e^{\langle w_a, \varphi(x)\rangle/b}: \; ,
\eeq
with $w_a$ the $a$-th fundamental weight of $\fg$. Furthermore, in the limit $\widehat{R}\rightarrow 0$, the deformed generators $W^{(s)}(z)$ become the stress tensor and higher spin currents of the ${\cW}({\fg})$-algebra. The special case $\fg= A_1$ is called the Liouville CFT, and ${\cW}({A_1})$ is more commonly called the Virasoro algebra, generated by the spin 2 stress energy tensor $W^{(2)}(z)$. When $\fg$ is a higher rank algebra, the stress tensor $W^{(2)}(z)$ is still present, but there are also additional currents $W^{(s)}(z)$ of higher spin $s>2$.

As a concrete example, consider the deformed stress tensor of ${\cW}_{q,t}({A_1})$. It is a Laurent polynomial in the $\cY$ operators \eqref{YoperatorToda}: 
\beq\label{examplestress}
	W^{(2)}(z)= \cY(z) + \left[\cY(v^{-2}z)\right]^{-1} \; .
	\eeq
This can be checked explicitly by computing the commutator $[W^{(s)}(z),Q^{(a)}]$, and finding out that it indeed vanishes. In the limit $q= \exp(R \epsilon_1),\, t=\exp(-R \epsilon_2)$, and the further rescaling of the Heisenberg algebra generators, one finds
\beq\label{examplestress2}
	W^{(2)}(z) \qquad \longrightarrow \qquad-\frac{1}{2}:\left(\partial_z\phi(z)\right)^2:+ Q\, :\partial_z^2\phi(z):\; .
	\eeq
	The reader will recognize the Liouville stress energy tensor of the ${\cW}({A_1})$-algebra.

\subsection{The Vortex $qq$-character is a Deformed ${\cW}_{q,t}({\fg})$-Algebra Correlator}

We are interested in evaluating the following correlator:
\beq\label{correlatordef}
\left\langle \psi'\Bigg|\prod_{a=1}^{n}\prod_{d=1}^{N^{(a)}_f}V^{(a)}(x^{(a)}_d)\;  (Q^{(a)})^{N^{(a)}}\; \prod_{s=2}^{n+1}\prod_{\rho=1}^{L^{(s-1)}}W^{(s)}(z^{(s-1)}_\rho) \Bigg| \psi \right\rangle \, .
\eeq
In what follows, we use the shorthand notation  $\langle \mathellipsis\rangle$ for a vacuum expectation value. The incoming and outgoing states are written as $|\psi\rangle$ and  $|\psi'\rangle$ respectively, instead of the trivial vacuum $|0\rangle$.  Because the theory is defined in the free-field formalism, the above correlator can be evaluated using straightforward Wick contractions, as an integral over the positions $y$ of the $N^{(a)}$ screening currents.
Namely, after taking into account the normal ordering of the various operators, the correlator \eqref{correlatordef} becomes the integral
\beq\label{conf1def}
\oint d_{Haar}y \;I_{Toda}(y) \; ,
\eeq
where the Haar measure is given by
\beq\label{Haardef}
d_{Haar}y=\prod_{a=1}^{n}\prod_{i=1}^{N^{(a)}}\frac{dy^{(a)}_i}{y^{(a)}_i} \; .
\eeq
The integrand $I_{Toda}(y)$ is made up of various factors. First, we have 
\beq\label{TodaFI}
\prod_{a=1}^{n}\prod_{i=1}^{N^{(a)}} \left(y^{(a)}_i\right)^{\langle\psi, \alpha_a\rangle}\; ,
\eeq
where $\langle\psi, \alpha_a\rangle$ is the eigenvalue of the state $|\psi\rangle$, as we have defined it previously \eqref{eigenvalue}. In 3d gauge theory language, this is nothing but the FI term \eqref{FI3d} contribution to the index $\left[\widetilde{\chi}^{\fg}\right]^{3d}$.

There are also various two-point functions: for a given $a\in \{1,\ldots, n\}$, we find by direct computation:
\beq\label{screena}
\prod_{1\leq i< j\leq N^{(a)}}\left\langle S^{(a)}(y^{(a)}_i)\, S^{(a)}(y^{(a)}_j) \right\rangle = \prod_{1\leq i\neq j\leq N^{(a)}}\frac{\left(y^{(a)}_{i}/y^{(a)}_{j};q\right)_{\infty}}{\left(t\, y^{(a)}_{i}/y^{(a)}_{j};q\right)_{\infty}}\;\prod_{1\leq i<j\leq N^{(a)}} \frac{\Theta\left(t\,y^{(a)}_{j}/y^{(a)}_{i};q\right)}{\Theta\left(y^{(a)}_{j}/y^{(a)}_{i};q\right)}\; .
\eeq
We recognize the vector multiplet contribution \eqref{vec3d} to the index $\left[\widetilde{\chi}^{\fg}\right]^{3d}$.

For $a$ and $b$ two distinct nodes in the Dynkin diagram of $\fg$, we compute:
\beq\label{screenab}
\prod_{1\leq i \leq N^{(a)}}\prod_{1\leq j \leq N^{(b)}}\left\langle S^{(a)}(y^{(a)}_i)\; S^{(b)}(y^{(b)}_j) \right\rangle = \prod_{1\leq i \leq N^{(a)}}\prod_{1\leq j \leq N^{(b)}}\left [ \frac{(t\,v\, y^{(a)}_{i}/y^{(b)}_{j};q)_{\infty}}{(v\, y^{(a)}_{i}/y^{(b)}_{j};q)_{\infty}}\right]^{\Delta^{ab}}\; .
\eeq
We recognize the bifundamental contribution \eqref{bif3d} to the index $\left[\widetilde{\chi}^{\fg}\right]^{3d}$.

The two-point of a fundamental vertex operator with a screening current equals:
\beq\label{vertexscreening}
\prod_{i=1}^{N^{(a)}}\left\langle V^{(a)}(x^{(a)}_d)\; S^{(b)}(y^{(b)}_i)\right\rangle =\prod_{i=1}^{N^{(a)}} \left[\frac{\left(t\, v\, x^{(a)}_{d}/y^{(b)}_{i};q\right)_{\infty}}{\left(v\, x^{(a)}_{d}/y^{(b)}_{i};q\right)_{\infty}}\right]^{\delta^{ab}}\; .
\eeq
We recognize the flavor contributions \eqref{matter3d} to the index $\left[\widetilde{\chi}^{\fg}\right]^{3d}$.

We come to the two-point function of a screening current with a $\cY$-operator, which evaluates to:
\beq\label{Yopscreening}
\prod_{i=1}^{N^{(b)}}\left\langle S^{(b)}(y^{(b)}_i)\, \cY^{(a)}(z) \right\rangle = \prod_{i=1}^{N^{(b)}} \left[\frac{1-t\, y^{(b)}_{i}/z}{1- y^{(b)}_{i}/z}\right]^{\delta_{ab}}\; .
\eeq
We recognize part of the defect contribution \eqref{3dWilsonfactor} to the index $\left[\widetilde{\chi}\right]_{3d}$. Note the zero mode of the  $\cY$-operator \eqref{YoperatorToda} acts nontrivially on the vacuum $|\psi\rangle$. As a result, the two-point of a screening with a $\cY$-operator generates a relative shift of one unit of 3d FI parameter between the various terms of the generating current $W^{(s)}(z)$. 

The last missing ingredient is the two-point of a fundamental vertex operator with a $\cY$-operator, which at first sight takes a far less elegant form:
\beq\label{Yopvertex}
\left\langle V^{(b)}(x^{(b)}_d)\; \cY^{(a)}(z) \right\rangle = \exp\left(\sum_{k>0}\,M_{ba}(q^{k\over 2} , t^{k\over 2})\, \frac{t^k-1}{k} \, \left(\frac{x^{(a)}_d}{z}\right)^k\right)\; .
\eeq
Recall $M_{ba}$ is the inverse of the deformed Cartan matrix. Fortunately, this two-point can be rewritten as:
\beq\label{Vopvertex}
\left\langle V^{(b)}(x^{(b)}_d)\; \cY^{(a)}(z) \right\rangle = B(x^{(b)}_d, z)\; \frac{1- v^{\#^{(ab)}+1}\, x^{(b)}_{d}/z}{1- t\, v^{\#^{(ab)}+1}\, x^{(b)}_{d}/z}.
\eeq
where $\#^{(ab)}$ is a non-negative integer equal to the number of links between nodes $a$ and $b$ in the Dynkin diagram of $\fg$, and $B(x^{(b)}_d, z)$ is \emph{defined} by the above two equations: it is literally the exponential in \eqref{Yopvertex} divided by the ratio on the right-hand side of \eqref{Vopvertex}.

It may seem like we have not gained much by trading the exponential \eqref{Yopvertex} for a new seemingly artifical prefactor $B(x^{(b)}_d,z)$, but this is not so for two reasons: first, the flavor part of the defect $Y$-operator contribution \eqref{3dWilsonfactor2} in the 3d/1d half-index now appears explicitly. Second, a remarkable factorization comes into play when we consider not just the $\cY$-operator inside the correlator, but the full generating current $W^{(s)}(z)$, which is a Laurent polynomial in such $\cY$-operators. Namely, each term in this polynomial is of the form \eqref{Vopvertex}, with the \textit{same} prefactor $B(x^{(b)}_d,z)$ for each term. This implies that the prefactor $B(x^{(b)}_d,z)$ can be factorized out of the correlator integral altogether. Note that a related factorization had also been noticed in the gauge theory picture, see the discussion under \eqref{Yequality}.\\

To fully specify the correlator integral, we also need to make a choice of contour. Here, the contours are simply chosen to be the ones we used in defining the 3d half-index, enclosing the poles in $\mathcal{M}^{bulk}$. In particular, the contours will avoid all poles depending on the generating current fugacity $z$.\\  

For a general correlator, we can now claim:
\begin{empheq}[box=\fbox]{align}
&\frac{\left\langle \psi'\left|\prod_{a=1}^{n}\prod_{d=1}^{N^{(a)}_f}V^{(a)}(x^{(a)}_d)\;  (Q^{(a)})^{N^{(a)}}\;\prod_{s=2}^{n+1}\prod_{\rho=1}^{L^{(s-1)}}W^{(s)}(z^{(s-1)}_\rho) \right| \psi \right\rangle}{\left\langle \psi'\left|\prod_{a=1}^{n}\prod_{d=1}^{N^{(a)}_f}V^{(a)}(x^{(a)}_d)\;  (Q^{(a)})^{N^{(a)}} \right| \psi \right\rangle} \label{correlatoris3dindex} \\
&\qquad\qquad\qquad\qquad\qquad\qquad= B\left(\{x^{(b)}_d\},\{z^{(s-1)}_\rho\}\right)\, \left[\widetilde{\chi}\right]^{(L^{(1)},\ldots, L^{(n)})}_{3d}(\{z^{(s-1)}_\rho\})\; ,\nonumber
\end{empheq}
where the overall prefactor is 
\beq
B\left(\{x^{(b)}_d\},\{z^{(s-1)}_\rho\}\right) =\prod_{s=2}^{n+1}\prod_{\rho=1}^{L^{(s-1)}}\prod_{b=1}^{n}\prod_{d=1}^{N^{(b)}_f}B\left(x^{(b)}_d,z^{(s-1)}_\rho\right)\; .
\eeq
Naturally, $B(\{x_d\},\{z^{(s)}_\rho\})$ stands outside the correlator integrals, since it does not depend on the $y$-integration variables.

In the end, we find that the 3d/1d index of the gauge theory $G^{3d}$ with line defect is a deformed ${\cW}_{q,t}({\fg})$-algebra correlator, up to the constant  $B(\{x_d\},\{z^{(s)}_\rho\})$. Recall that this prefactor contains an exponential; we mention in passing that this ``phase" has a natural interpretation when $\fg=A_n$, in the general framework of Ding-Iohara-Miki (DIM) algebras \cite{Ding:1996,Miki:2007}; for a detailed study, see \cite{Mironov:2016yue}. Roughly speaking, in the DIM formalism, a vertex operator ${\mathbb{V}}^{(a)}(x^{(a)}_d)$ is built using intertwiners, as the product of a $V^{(a)}(x^{(a)}_d)$  vertex operator  from the ${\cW}_{q,t}({A_n})$-algebra, and another vertex operator coming from an additional Heisenberg algebra. This extra Heisenberg algebra comes with its own Fock space, and contributes to the correlator in a way to precisely cancel the  prefactor $B(\{x_d\},\{z^{(s)}_\rho\})$.\\

\vspace{16mm}

\section{Schwinger-Dyson Equations: the Little String Perspective}
\label{sec:littlestringsecion}

We have showcased 3 different physical frameworks where we can make sense of a vortex character for the 3d $\cN=4$ gauge theory $G^{3d}$. Moreover, we saw explicitly how the regularity properties of this character implies a non-perturbative type of Schwinger-Dyson identities for the theory. Ultimately, all the various perspectives are unified in a string theory picture. In the process of describing it, we will learn about the dynamics of new defects in $(2,0)$ little string theory. The literature on BPS-defects of the little string has been steadily growing in the last few years, with rich physical and mathematical implications: among them, we find codimension-2 defects \cite{Aganagic:2015cta,Haouzi:2016ohr,Haouzi:2016yyg,Haouzi:2017vec}, codimension-4 defects \cite{Aganagic:2016jmx,Aganagic:2017smx}, point and codimension-2 defects \cite{Haouzi:2019jzk}, and in this present work, point and codimension-4 defects.

\subsection{Little String Basics}
\label{ssec:basics}

We consider ten-dimensional type IIB string theory compactified on an $ADE$ surface $X$ times a circle, meaning type IIB on $X\times M_6$. The 6-manifold $M_6$ is the product of an infinite cylinder $\cC= \mathbb{R} \times S^1(R)$ of radius $R$ and two complex lines, which we distinguish using the subscript notation $\mathbb{C}_q$ and  $\mathbb{C}_t$, so that $M_6=\cC\times\mathbb{C}_q\times\mathbb{C}_t$. $X$ is a resolution of a ${\mathbb C}^2/\Gamma$ singularity, where $\Gamma$ is one of the discrete subgroups of $SU(2)$. By the McKay correspondence, such a discrete subgroup is labeled by one of the simply-laced Lie algebras ${\fg=A, D, E}$; we call $n$ the rank of $\fg$.  Explicitly, the singularity is resolved by blow-up: the exceptional divisor is a collection of 2-spheres $S_a$, $a=1,\ldots,n$, which organize themselves in the shape of the Dynkin diagram of $\fg$.

We focus our attention on a sector of the theory which has far less degrees of freedom than are present in the full IIB string. That is, we decouple gravity and focus only on the degrees of freedom supported near the origin of $X$ by sending the string coupling to $g_s\rightarrow 0$. In this limit, the type IIB string on $X$ becomes a 6-dimensional string theory on $M_6$, known as the $(2,0)$ little string of type $\fg=A, D, E$ \cite{Berkooz:1997cq,Seiberg:1997zk,Losev:1997hx}. It is not a local QFT \cite{Kapustin:1999ci}. It is instead a theory of strings proper (inherited from the ten-dimensional IIB strings), with finite tension $m_s^2$, the square of the string mass. There are a few good reviews in the literature, most notably \cite{Aharony:1999ks,Kutasov:2001uf}.

The moduli space of the $(2,0)$ little string is
\beq
\left(\mathbb{R}^4\times S^1\right)^{n}/W(\fg)\; ,
\eeq
where $W(\fg)$ is the Weyl group of $\fg$. The moduli come from periods of various 2-forms along the 2-cycles $S_a$ of the surface $X$: the $S^1$ modulus is the R-R 2-form $C^{(2)}$ of the ten-dimensional type IIB string theory integrated over $S_a$. Meanwhile, the $\mathbb{R}^4$ moduli come from the NS-NS B-field $B^{(2)}$, and a triplet of self-dual 2-forms $\omega_{I,J,K}$, which exist because $X$ is a hyperk\"ahler manifold. To get the correct R-R and NS-NS normalizations, one needs to recall the low energy action of the type IIB superstring. In particular, the R-R field is not accompanied by any power of $g_s$. Moreover, the mass dimension of a scalar in a theory of 2-forms should be 2. Then, in canonical normalization, we obtain:
\beq\label{periods}
\frac{m_s^4}{g_s}\int_{S_a}\omega_{I,J,K}\, ,\qquad \frac{m_s^2}{g_s}\int_{S_a}B^{(2)}\, , \qquad m_s^2\int_{S_a}C^{(2)}\, .
\eeq
The above periods remain fixed in the limit $g_s\rightarrow 0$.\\

As is, this background preserves 16 supercharges. Ultimately, we want to make contact with 3-dimensional physics and produce nontrivial dynamics. We can achieve both goals at once by introducing various supersymmetric branes. Since our construction originates in type IIB, we naturally consider adding certain D-branes, whose tension should remain finite in the $g_s\rightarrow 0$ limit. As we will argue, the relevant branes to consider here are D3 and D1 branes wrapping 2-cycles of the surface $X$, which we now turn to.

\subsection{The Effective Theory on the D-Branes}
\label{ssec:simplylaced}

To be more quantitative, we introduce some notations: According to the McKay correspondence, the second homology group $H_2(X, \mathbb{Z})$ of $X$ is identified with the root lattice $\Lambda$ of $\fg$. Then, $H_2(X, \mathbb{Z})$ is spanned by $n$ vanishing 2-cycles $S_a$, which we identify as the positive simple roots $\alpha_a$. The intersection pairing in homology is further identified with the Cartan Killing metric of $\fg$; explicitly,
\beq
\# (S_a \cap S_b)=-C_{ab}\; ,
\eeq
where $C_{ab}$ is the Cartan matrix of $\fg$.

We also consider the second relative homology group $H_2(X, \partial X, \mathbb{Z})$. This group is spanned by non-compact 2-cycles $S_a^*$, $a=1,\ldots, n$, where each $S_a^*$ is constructed as the fiber of the cotangent bundle $T^*S_a$ over a generic point on $S_a$. 
The group $H_2(X, \partial X, \mathbb{Z})$ is identified with the weight lattice $ \Lambda_*$ of $\fg$; correspondingly, the 2-cycle $S^*_a$ is identified with the $a$-th fundamental weight $\lambda_a$ of $\fg$. In particular, the following orthonormality relation holds in homology:
\beq
\# (S_a \cap S^*_b)=\delta_{ab}\; .
\eeq
Note that $H_2(X, \mathbb{Z})\subset H_2(X, \partial X, \mathbb{Z})$, since compact 2-cycles can be understood as elements of $H_2(X, \partial X, \mathbb{Z})$ with trivial boundary at infinity. This is just the homological version of the familiar statement  that the root lattice of $\fg$ is a sublattice of the weight lattice, $\Lambda\subset \Lambda_*$.\\

\begin{figure}[h!]
	\emph{}
	\centering
	\includegraphics[trim={0 0 0 3cm},clip,width=0.99\textwidth]{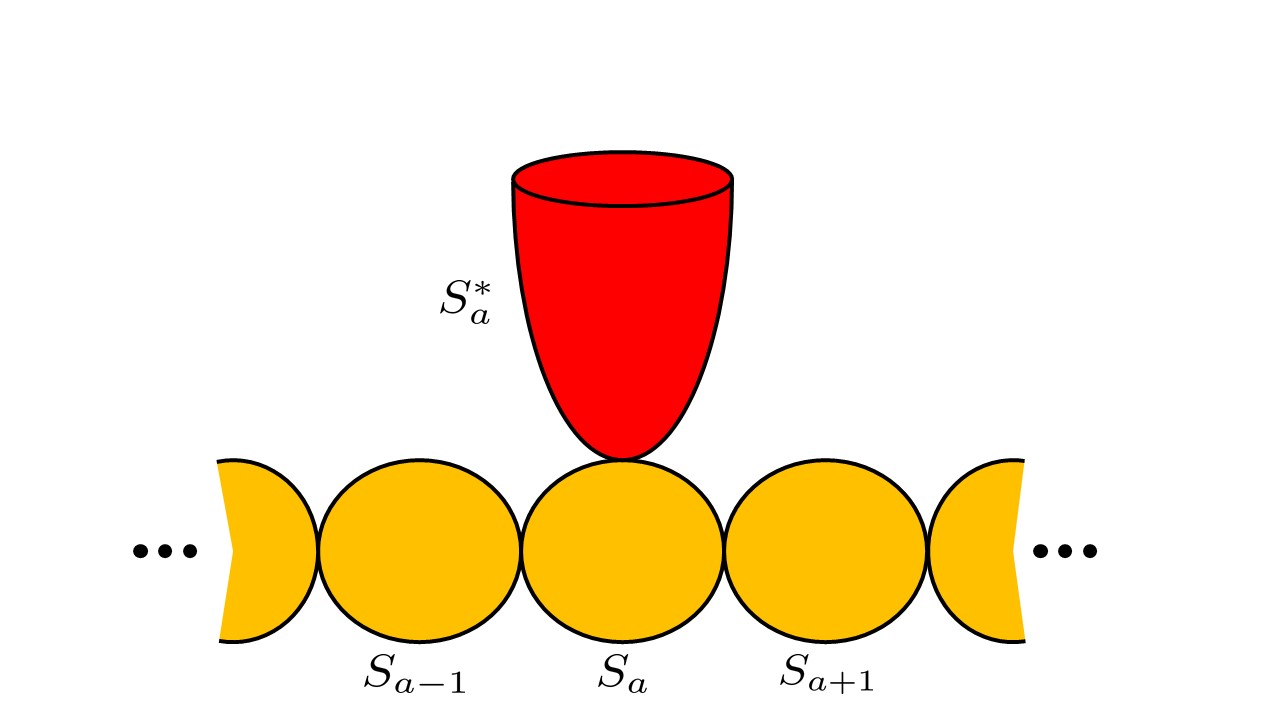}
	\vspace{-1pt}
	\caption{A vanishing 2-cycle of an $A_n$ singularity, labeled by $S_a$ (the black 2-sphere), and the dual non-compact 2-cycle $S_a^*$ (the black cigar).} 
	\label{fig:geometry}
\end{figure}

Consider a total of $N$ D3$_{gauge}$ branes wrapping the compact 2-cycles of $X$ and one of the complex lines $\mathbb{C}_q$ in $M_6$, while sitting at the origin of the transverse complex line $\mathbb{C}_t$. This results in a net non-zero D3$_{gauge}$ brane charge, measured by a class $[S]\in H_2(X, \mathbb{Z})$. We expand $[S]$ in terms of positive simple roots as
\beq\label{compact}
[S] = \sum_{a=1}^n  \,N^{(a)}\,\alpha_a\;\;  \in  \,\Lambda \; ,
\eeq
with $N^{(a)}$  non-negative integers. The $N$ D3$_{gauge}$ branes are points on the cylinder $\cC$, with coordinates $\{y^{(a)}_i\}$.\\

Next, we consider a total of $N_F$ D3$_{flavor}$ branes wrapping non-compact 2-cycles in $X$, along with the same complex line $\mathbb{C}_q$ in $M_6$, while also sitting at the origin of the transverse complex line $\mathbb{C}_t$. The charge for these branes is measured by a class $[S^*]\in H_2(X, \partial X, \mathbb{Z})$. We expand $[S^*]$ in terms of fundamental weights as
\beq\label{noncompact}
[S^*] = - \sum_{a=1}^n \, N^{(a)}_f \, \lambda_a \;\;  \in\, \Lambda_*\; ,
\eeq  
where $N^{(a)}_f$ are non-negative integers commonly called Dynkin labels. The $N_F$ D3$_{flavor}$ branes are points on the cylinder $\cC$, with coordinates $\{x^{(a)}_d\}$.\\


Lastly, we introduce $L$ D1$_{defect}$ branes wrapping the non-compact 2-cycles in $X$ and the transverse complex line  $\mathbb{C}_t$, while sitting at the origin of $\mathbb{C}_q$. The charge for these D1$_{defect}$ branes is  measured by a class $[S^*_{defect}]\in H_2(X, \partial X, \mathbb{Z})$, expanded in terms of fundamental weights as:
\begin{equation}
[S^*_{defect}] = - \sum_{a=1}^n \, L^{(a)} \, \lambda_a \;\;  \in\, \Lambda_*\; ,
\end{equation} 
where $L^{(a)}$ are non-negative integers, once again Dynkin labels. The $L$ D1$_{defect}$ branes are points on the cylinder $\cC$, with coordinates $\{z^{(a)}_\rho\}$.\\

\begin{figure}[h!]
	\emph{}
	\centering
	\includegraphics[width=1.0\textwidth]{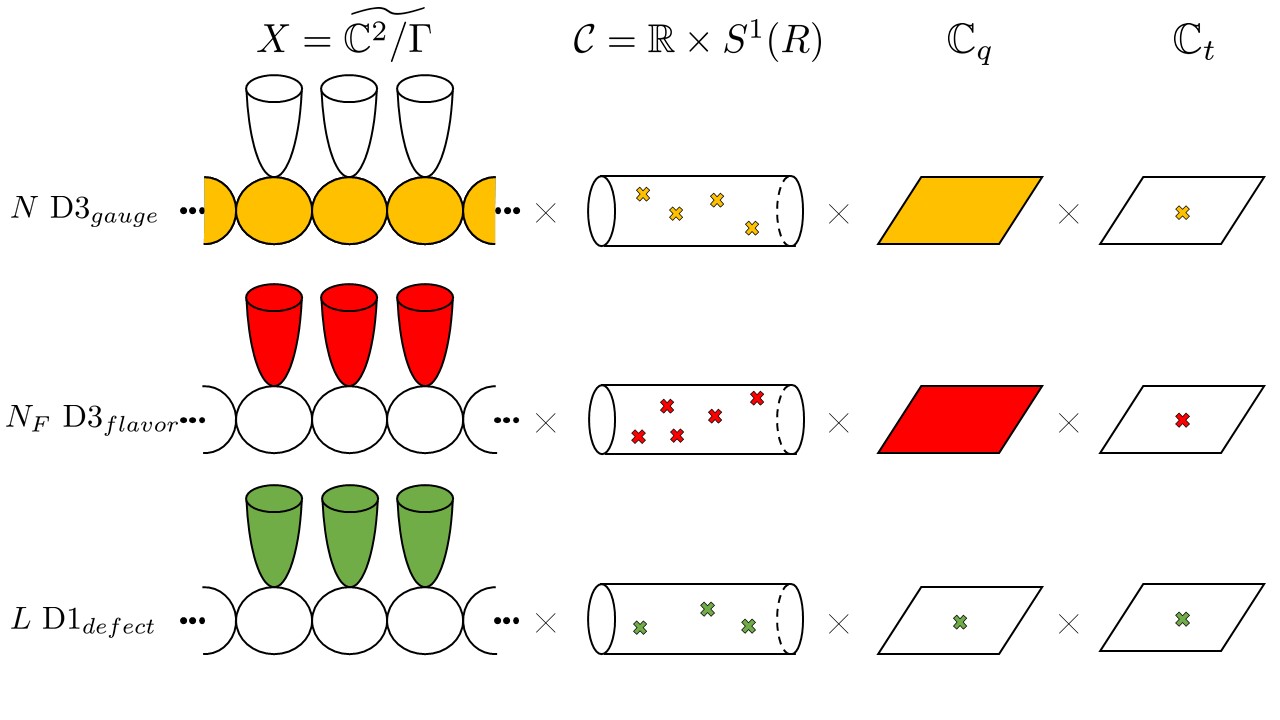}
	\vspace{-12pt}
	\caption{The brane configuration in type IIB: there are $N$ D3$_{gauge}$ branes wrapping compact 2-cycles $S_{a}$ and $\mathbb{C}_q$ (yellow),  $N_{f}$ D3$_{flavor}$ branes wrapping non-compact 2-cycles $S_{a}^{*}$'s  and $\mathbb{C}_q$ (red). There are also $L$ D1$_{defect}$ branes wrapping the non-compact 2-cycles $S_{a}^{*}$'s and $\mathbb{C}_q$ (green). All branes are points on the cylinder $\cC$. Later, we will also consider the quantum mechanics of $k$ D1$_{vortex}$ branes (not pictured) wrapping the compact 2-cycles $S_{a}$'s.} 
	\label{fig:branes}
\end{figure}

Let us first ignore the D1$_{defect}$ branes.
At energies $E$ well below the string scale, $E/m_s\ll 1$, the effective theory on the D3$_{gauge}$ branes is a 3-dimensional gauge theory with $\cN=4$ supersymmetry\footnote{Recall we have defined the periods of a triplet $\vec{\omega}=(\omega_I, \omega_J, \omega_K)$ of self-dual 2-forms \eqref{periods}. The D3$_{flavor}$ branes wrapping the  non-compact 2-cycles preserve the same supersymmetry only if the vectors $\int_{S^*_a} \vec{\omega}$ all point in the same direction, for all $a=1, \ldots, n$. Having made such a choice, we then have to worry about the supersymmetry preserved by the D3$_{gauge}$ branes wrapping the compact 2-cycles. This is determined by the periods of the 2-forms through the 2-cycles $S_a$, which is the choice of a metric on $X$. Then, it is always possible to choose D3 branes wrapping the compact and non-compact 2-cycles and which break the same supersymmetry.},  on the manifold $\mathbb{C}_q\times S^1(R)$. At first sight, our brane setup may suggest that the gauge theory we obtain should only be 2-dimensional, with  $\cN=(4,4)$ supersymmetry. However, it is not the case: the D3$_{gauge}$ branes are points on the cylinder $\cC=\mathbb{R}\times S^1(R)$, and strings wrap around the circle, resulting in a tower of Kaluza-Klein states on the T-dual circle of radius $\widehat{R}=1/(m^2_s\, R)$, which modifies the low-energy physics. Put differently, the $(2,0)$ little string compactified on $S^1(R)$ enjoys T-duality (inherited from type IIB), under which it becomes the $(1,1)$ little string theory compactified on $S^1(\widehat{R})$. Then, the D3$_{gauge}$ branes at points on the cylinder $\cC=\mathbb{R}\times S^1(R)$ in the $(2,0)$ little string are exactly the same as D4$_{gauge}$ branes  wrapping the circle of the T-dual cylinder $\cC'=\mathbb{R}\times S^1(\widehat{R})$, in the $(1,1)$ little string. It is clear in the second description that the low energy theory really is 3-dimensional, on $\mathbb{C}_q\times S^1(\widehat{R})$. We call this gauge theory $G^{3d}$. The choice of this denomination is not innocent, since we will now argue that the low energy theory on the branes is precisely the 3d theory we have studied in the rest of this paper.\\

The precise characterization of $G^{3d}$ was determined by Douglas and Moore \cite{Douglas:1996sw}: it is a quiver gauge theory of shape the Dynkin diagram of $\fg=ADE$\footnote{The original analysis of \cite{Douglas:1996sw} was carried out in the full type IIB background, and the quiver gauge theory there was labeled by an affine Dynkin diagram. Here, we are working in the little string limit $g_s\rightarrow 0$, and the affine node is decoupled.}. The gauge group is
\beq\label{gaugegroup}
G=\prod_{a=1}^n U(N^{(a)})\; ,
\eeq
where the ranks $N^{(a)}$ were defined in \eqref{compact} as the number of D3$_{gauge}$ branes wrapping the compact 2-cycle $S_a$,
\[
[S] = \sum_{a=1}^n  \,N^{(a)}\,\alpha_a\;\;  \in  \,\Lambda \; .
\]
The flavor symmetry is the gauge group
\beq\label{flavorgroup}
G_F=\prod_{a=1}^n U(N^{(a)}_f)\; ,
\eeq
where the ranks $N^{(a)}_f$ were defined in  \eqref{noncompact} as the number of D3$_{flavor}$ branes wrapping the non-compact 2-cycle $S^*_a$,
\[
[S^*] = - \sum_{a=1}^n \, N^{(a)}_f \, \lambda_a \;\;  \in\, \Lambda_*\; .
\] 
Note that because the 2-cycle $S^*_a$ are non-compact, the associated gauge fields of $U(N^{(a)}_f)$ are frozen. This produces $N^{(a)}_f$ hypermultiplets on node $a$, in the bifundamental representation $(N^{(a)}, \overline{N^{(a)}_f})$ of the group $U(N^{(a)})\times U(N^{(a)}_f)$. Such multiplets come about from quantizing the strings between the D3 branes wrapping the compact 2-cycle $S_a$ and the non-compact 2-cycle $S^*_a$.

Finally, for $a\neq b$, we have hypermultiplets coming from the intersection of 2-cycles $S_a$ and $S_b$ at a point. The intersection pairing $\#(S_a \cap S_b)$ in homology is identified with the incidence matrix of $\fg$.  Open strings with one end on the $a$-th D3$_{gauge}$ brane and the other end on the $b$-th D3$_{gauge}$ brane results in a hypermultiplet in the bifundamental representation $(N^{(a)}, \overline{N^{(b)}})$ of $U(N^{(a)})\times U(N^{(b)})$.

So far, the above stringy construction applies a priori to any configuration of  D3$_{gauge}$ and D3$_{flavor}$ branes. The resulting effective 3d $\cN = 4$ gauge theory then inherits an arbitrary gauge and flavor content. The aim of this work is to exhibit certain symmetries associated to BPS vortices, which sit at Higgs vacua. Therefore, from now on, we require the number $N_F$ of  D3$_{flavor}$ branes to be large enough, so that $G^{3d}$ possesses a Higgs branch and that its vacua be Higgs vacua.\\

Consider now adding the D1$_{defect}$ branes in the background. Recall that those branes wrap the non-compact 2-cycles of the geometry. As such, they are not dynamical. They are point defects on the $\mathbb{C}_q$-line, and break supersymmetry further by half. The number of such D1$_{defect}$ branes is $L=\sum_{a=1}^n L^{(a)}$. Correspondingly, the defects carry a background gauge group of their own,
\beq\label{defectflavorgroup}
\widehat{G}_{defect}=\prod_{a=1}^n U(L^{(a)})\;.
\eeq
\begin{figure}[h!]
	\emph{}
	\centering
	\includegraphics[trim={0 0 0 0cm},clip,width=0.99\textwidth]{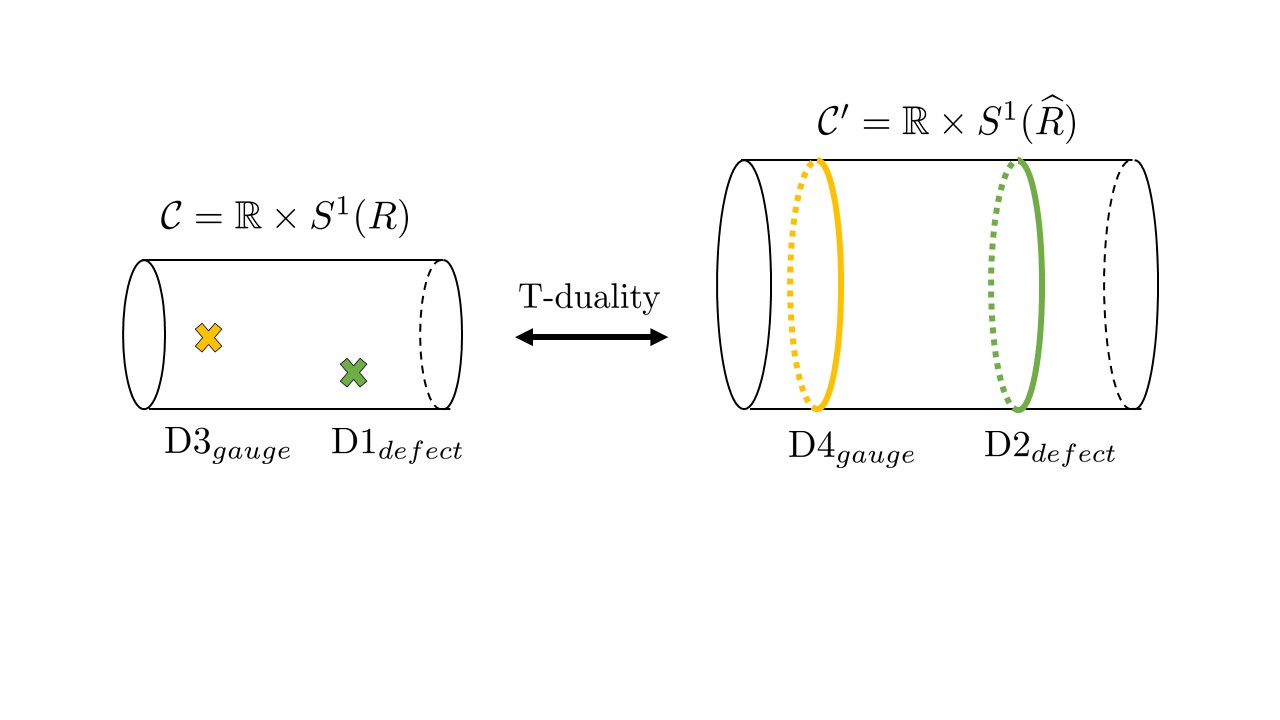}
	\vspace{-40pt}
	\caption{T-duality tells us that the D3 and D1 branes at points on the cylinder $\cC$ in the $(2,0)$ little string are the same as D4 and D2 branes wrapping the T-dual cylinder $\cC'$ in the $(1,1)$ little string.} 
	\label{fig:Tduality}
\end{figure}

The D1$_{defect}$ branes are points on $\cC$, or equivalently, following the same line of reasoning as before, they are D2$_{defect}$ branes wrapping the circle $S^1(\widehat{R})$ of the T-dual cylinder $\cC'$. In a nontrivial NS-NS B-field, these branes are 1/2-BPS. Thus, from the point of view of the gauge theory $G^{3d}$, the D1$_{defect}$ branes really are supersymmetric defects wrapping the $S^1(\widehat{R})$, and at the origin of $\mathbb{C}$; in other words, they make up a loop defect of $G^{3d}$.\\

Translating the geometry to gauge theory data, the periods  \eqref{periods}  of the  $(2,0)$ little string become parameters of $G^{3d}$. Namely, the modulus coming from the NS-NS $B^{(2)}$-field through the 2-cycle $S^2_a$ is identified with the gauge coupling on node $a$ of the quiver gauge theory. The triplet of self-dual two-forms $\omega_{I,J,K}$ are the FI parameters. The positions of the $N$ D3$_{gauge}$ branes on the cylinder $\cC$ are (part of the) Coulomb moduli of $G^{3d}$. The positions of the $N_F$ D3$_{flavor}$ branes on $\cC$ are mass parameters for the fundamental hypermultiplets of $G^{3d}$. Finally, the positions of the $L$ D1$_{defect}$ branes  on $\cC$ are the defect masses. All of the above moduli and parameters are complexified, due to the presence of the circle $S^1(\widehat{R})$ the 3d theory and the defect live on. This is precisely the gauge theory setup we studied previously.\\

We end this section with comments on the limit $m_s\rightarrow \infty$\footnote{As usual, there are a priori many ways to take this limit. The limit described here is the same one that turned deformed $\cW_{q,t}(\fg)$-algebras into the usual $\cW(\fg)$-algebras.}. In that regime, we lose the one scale of the theory and flow to a $(2,0)$ SCFT on $M_6$, labeled by the same simply-laced Lie algebra $\fg$ as the $(2,0)$ little string\footnote{In the T-dual setup, the $(1,1)$ little string in the low energy limit becomes 6d maximally supersymmetric Yang-Mills theory, with gauge group of type $\fg$.}. The various moduli of the little string are kept fixed in the limit, and become moduli of the SCFT. We further insist on keeping the Riemann surface we compactified the theory on fixed, along with the position of the various D3 branes on it; that is, the cylinder $\cC=\mathbb{R}\times S^1(R)$ remains fixed. Recall however that $G^{3d}$ is naturally defined on the T-dual cylinder  $\cC'=\mathbb{R}\times S^1(\widehat{R})$, where $\widehat{R}=1/(m^2_s R)$. Therefore, if $R$ is kept fixed, the dual radius $\widehat{R}$ vanishes in the SCFT limit and the theory on the branes becomes effectively 2-dimensional, with $\cN=(4,4)$ supersymmetry. The gauge coupling on the D3 branes becomes infinite, meaning the theory cannot be described as a gauge theory anymore. Meanwhile, the line defect wrapping $\widehat{R}$ becomes a 1/2-BPS point defect. In the rest of this paper, we keep $m_s$ finite.

\subsection{The Index of the Little String as a Vortex Character}

Our goal is to compute the partition function of the $(2,0)$ little string theory on $M_6=\cC\times\mathbb{C}_q\times\mathbb{C}_t$. Note this is fully equivalent to computing the partition function of the $(1,1)$ little string on  $M'_6=\cC'\times\mathbb{C}_q\times\mathbb{C}_t$, where $\cC'$ is the T-dual cylinder of radius $\widehat{R}$. In the latter setup, and in the absence of D-branes, the partition function is naturally expressed as a supersymmetric index: 

\beq
\label{index6d}
\text{Tr}\left[(-1)^F\, q^{S_1 - S_R}\, t^{-S_2 + S_R}\,\right] \; .
\eeq

The trace is defined over the circle $\widehat{R}$, and $q^{S_1 - S_R}\, t^{-S_2 + S_R}$ turns the manifold $M'_6$ into a twisted product. As we go around $S^1(\widehat{R})$, the line $\mathbb{C}_q$ is rotated by $q$, and the line $\mathbb{C}_t$ is rotated by $t^{-1}$, describing an $\Omega$-background. $S_1$ is the generator of the $\mathbb{C}_q$-rotations, while $S_2$ is the generator of the $\mathbb{C}_t$-rotations. $S_R$ is the generator of a $U(1)\subset SU(2)$ subgroup of the R-symmetry of the 6d theory. Without any branes, the index is trivial by pairwise cancellations of bosons and fermions, since the 6d theory has too much supersymmetry.\\

Working in the T-dual setup, we first add the D4$_{gauge}$ and D4$_{flavor}$ branes in the background. By a supersymmetric localization argument, the partition function of the bulk little string becomes the partition function on the defects. Indeed, supersymmetry is only broken near the locus of the defect branes, while the supersymmetries of the full $(1,1)$ little string are preserved away from the defect. It follows that the partition function on the D4 branes is precisely the half-index \eqref{index3d} of $G^{3d}$. 
In particular, the 3d FI parameter contribution \eqref{FI3d} comes from turning on the periods of the self-dual two-form $\omega_I$.  
The ${\cN}=4$ vector multiplet contribution on node $a$  \eqref{vec3d} comes about from quantizing the D4$_{gauge}$/D4$_{gauge}$ strings, with the D4 branes wrapping the $a$-th compact 2-cycle. 
The ${\cN}=4$ bifundamental hypermultiplets contribution between nodes $a$ and $b$ \eqref{bif3d} comes about from quantizing the D4$_{gauge}$/D4$_{gauge}$ strings, with one set of D4 branes wrapping the $a$-th compact 2-cycle, and the other set of D4 branes wrapping the $b$-th compact 2-cycle. 
The ${\cN}=4$ fundamental hypermultiplet contribution on node $a$ \eqref{matter3d} comes about from quantizing the D4$_{gauge}$/D4$_{flavor}$ strings, with the D4$_{gauge}$ branes wrapping the $a$-th compact 2-cycle, and the D4$_{flavor}$ branes wrapping the dual non-compact 2-cycle.\\

We now introduce the D2$_{defect}$ branes. These branes are nondynamical as they do not wrap $\mathbb{C}_q$, but they nonetheless modify the index. We conjecture here that the new string sectors realize the $Y$-operator defect in the gauge theory. Namely, \eqref{3dWilsonfactor} is the contribution of  D4$_{gauge}$/D2$_{defect}$ strings at node $a$, while \eqref{3dWilsonfactor2} is the contribution of D4$_{flavor}$/D2$_{defect}$ strings at node $a$. 
All in all, this implies that the index of the little string in the presence of all three types of branes localizes to a defect $\widetilde{Y}$-operator vev in the 3d gauge theory. The vortex character observable is constructed as a sum of such vevs, uniquely determined from the requirement that the defect singularities due to D4$_{gauge}$/D2$_{defect}$ strings should be removed. Again, this construction exists and is unique, as guaranteed from the representation theory of quantum affine algebras. The resulting character is as we wrote it in the 3d gauge theory language; we rewrite it here for convenience, normalized by the index in the absence of D2$_{defect}$ branes:

\begin{empheq}[box=\fbox]{align}
\label{NEWcharacter3dstring}
&\left[\widetilde{\chi}\right]^{(L^{(1)},\ldots, L^{(n)})}_{\text{D4}_g/\text{D4}_f/\text{D2}_d}(\{z^{(a)}_\rho\})\\
&=\frac{1}{\left[{\chi}\right]^{(0,\ldots,0)}_{\text{D4}_g/\text{D4}_f}}\sum_{\omega\in V(\lambda)}\prod_{b=1}^n \left({\widetilde{\fq}^{(b)}}\right)^{d_b^\omega} c_{d_b^\omega}(q, t)\; \left(\widetilde{\mathcal{Q}}_{d_b^\omega}^{(b)}(\{z_{\rho}^{(a)}\})\right)\,  \left[\widetilde{Y}_{3d\;  gauge/1d}(\{z^{(a)}_\rho\})\right]_{\omega} .\nonumber
\end{empheq}
The superscripts in the index designate the D2$_{defect}$ charge, while the subscripts indicate which types of branes are present (D4$_g$ for D4$_{gauge}$ and so on).\\

This above identification makes the dictionary to deformed $\cW$-algebras explicit: the $N$ screening charges \eqref{screeningchargedef} are  the $N$  D4$_{gauge}$ branes, the $N_F$ fundamental vertex operators \eqref{fundvertex} are the $N_F$  D4$_{flavor}$ branes, and the $L$ generating currents \eqref{generatorsdef} are the $L$ D2$_{defect}$ branes. The little string index can therefore be recast as the $q$-Toda correlator \eqref{correlatoris3dindex}.\\

To make contact with the vortex quantum mechanics $T^{1d}$, we need to do a little more work. Namely, we freeze the moduli of the D4$_{gauge}$ branes to be equal to the moduli of the D4$_{flavor}$ branes. This describes the root of the Higgs branch for $G^{3d}$. Geometrically, this means we can recombine the D4$_{gauge}$ branes with the D4$_{flavor}$ branes so that they exclusively make up a collection of $N_F$ D4'$_{flavor}$ branes wrapping the non-compact 2-cycles of $X$, and the theory is effectively massive.

\begin{figure}[h!]
	\emph{}
	\centering
	\includegraphics[trim={0 0 0 0cm},clip,width=0.99\textwidth]{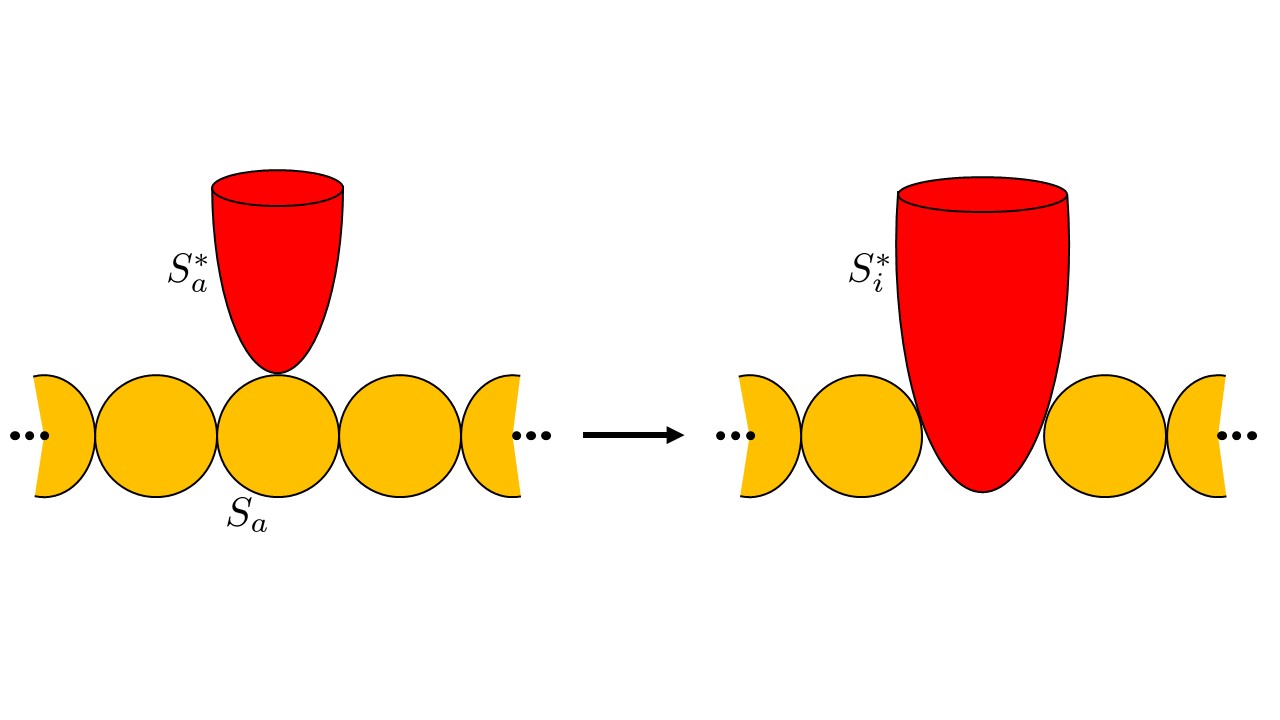}
	\vspace{-40pt}
	\caption{Illustration of the Higgsing procedure in the string theory picture. On the right side, the gauge and flavor branes have recombined exclusively into flavor branes.} 
	\label{fig:Higgsing}
\end{figure}

Now, we would like to introduce vortices for $G^{3d}$. First note that a generic collection of vortices is BPS if the 3d FI parameters are aligned in the same direction. For each $a=1,\ldots,n$, the triplet of FI terms $\int_{S_a}\omega_{I,J,K}$ transforms as a vector under the R-symmetry group $SU(2)_R$ rotating the hypermultiplet scalars. 
We identify $SU(2)_R$ as the $SU(2)$ R-symmetry of the little string. We then turn on the periods $\int_{S_a}\omega_{I}>0$, while setting the other periods to zero, $\int_{S_a}\omega_{J,K}=0$, for all $a$. Correspondingly, this turns on a real FI parameter on each node $a$, while the complex FI parameters are set to zero. This describes a generic point on the Higgs branch of $G^{3d}$, and $SU(2)_R$ is broken to $U(1)_R$, which acts by rotating the periods of $\omega_J$ and $\omega_K$. This background indeed allows for 1/2-BPS vortex solutions: they are D2$_{vortex}$ branes wrapping the compact 2-cycles of $X$ and the circle of the cylinder $\cC'$ in the $(1,1)$ little string. Alternatively, they are D1$_{vortex}$ branes wrapping the compact 2-cycles of $X$ at a point on  $\cC$ in the $(2,0)$ little string. With all branes present, only two supercharges are preserved in the background.

The effective theory on $k$ D2$_{vortex}$ branes is precisely the quantum mechanics $T^{1d}$. It follows that the index of the $(1,1)$ little string in the presence of the D4'$_{flavor}$ branes, the D2$_{defect}$ branes, and the new D2$_{vortex}$ branes, is the 1d Witten index \eqref{index}. In its integral representation \eqref{vortexintegral}, the index is comprised of 1-loop determinants, all of which can be attributed to the various strings stretching between the branes. In particular, the interaction of the vortices with the defect is attributed to D2$_{vortex}$/D2$_{defect}$ strings on node $a$. It provides the degrees of freedom for 1d $\cN=4$ chiral multiplets; these 1-loop determinants were also worked out in T-dual setups, see \cite{Haouzi:2019jzk,Kim:2016qqs,Nekrasov:2016gud}.  

All in all, the Witten index of $T^{1d}$ is a little string index, which can be naturally expressed once again as a vortex character:
\begin{empheq}[box=\fbox]{align}
\label{character1dstring}
&\left[\widetilde{\chi}\right]^{(L^{(1)},\ldots, L^{(n)})}_{\text{D2}_v/\text{D4'}_f/\text{D2}_d}(\{z^{(a)}_\rho\})\\
&=\frac{1}{\left[{\chi}\right]^{(0,\ldots,0)}_{\text{D2}_v/\text{D4'}_f}}\sum_{\omega\in V(\lambda)}\prod_{b=1}^n \left({\widetilde{\fq}^{(b)}}\right)^{d_b^\omega} c_{d_b^\omega}(q, t)\; \left(\mathcal{Q}_{d_b^\omega}^{(b)}(\{z_{\rho}^{(a)}\})\right)\,  \left[{Y}_{1d}(\{z^{(a)}_\rho\})\right]_{\omega}  .\nonumber
\end{empheq}
Again, the superscripts indicate the D2$_{defect}$ charge, while the subscripts indicate which types of branes are present  in the background. In this picture, each term stands for a D2$_{defect}$ brane recombining with one of the D2$_{vortex}$ branes when they coalesce on the cylinder; these are $z$-pole loci in the integrand of the quantum mechanics. The number of ways the D2$_{defect}$ brane can recombine with a vortex brane is precisely the size of the set $\left|\mathcal{M}_k\setminus \mathcal{M}^{pure}_k\right|$, for all vortex numbers $k$.

Finally, recall that this index depends on a choice of sign for the 1d FI parameter. In the  little string context, this is the sign of the period for the NS-NS $B$-field  $\int_{S_a}B^{(2)}$. In particular, as we will argue next, changing the sign of this period gives a little string realization of 3d Seiberg duality.

\vspace{16mm}

\section{Discussion}
\label{sec:discussionsection}

\subsection{Seiberg Duality of the Vortex Character}

It can happen that two distinct UV vortex theories as we have described them in the previous Sections flow to the same point in the IR, a phenomenon called Seiberg duality\footnote{The 3d Seiberg duality is only strictly true on the Higgs branch, where the vortex solutions are defined; at a generic point  on the moduli space of vacua, theories we identify as Seiberg-dual can disagree globally.  For physical considerations on this point, see  \cite{Assel:2017jgo,Dey:2017fqs,Bashkirov:2013dda,Gaiotto:2013bwa,Yaakov:2013fza}. For mathematical considerations, see \cite{Bullimore:2015lsa,Nakajima:2015txa,Braverman:2016wma}.}$^{,}$\footnote{A description of the various theories one encounters in the IR can be found in \cite{Gaiotto:2008ak,Kapustin:2010mh}.} \cite{Seiberg:1994pq}. We would like to understand how 3d Seiberg duality acts on our $qq$-character observable.\\ 

In the absence of defect, 3d $\cN=4$ gauge theories which are Seiberg-dual to each other are described by different quivers in the UV, but turn out to have the same partition function. The duality has an elegant interpretation in the vortex quantum mechanics $T^{1d}_{pure}$: two Seiberg-dual theories only differ by the choice of 1d FI parameter-chamber for the $\zeta^{(a)}_{1d}$ parameters\footnote{Mathematically, this is a choice of stability conditions for the quiver $T^{1d}_{pure}$.}. In particular, the 1d quivers of Seiberg-dual theories are identical\footnote{The 1d quivers are the same in dual theories, but the vortex numbers $k^{(a)}=\frac{1}{2\pi}\int \text{Tr} F^{(a)}$ are not invariant. In this paper, we have been identifying the rank of the $a$-th  gauge group in the 1d quiver with the vortex number $k^{(a)}$; such an identification is always possible, given a choice of FI parameter chamber. Throughout the paper, we fixed all FI parameters to be generic and positive; with such a choice of chamber, our 1d quivers are known as ``canonical" vortex theories, in the terminology of \cite{Hwang:2017kmk}.} \cite{Hwang:2017kmk}.

Choosing different signs for the parameters $\zeta^{(a)}_{1d}$ translates to different choices of contours for the Witten index integral \eqref{vortexintegral}. The new poles to be enclosed are perfectly tractable, for instance via the JK-residue prescription. At least for $A_n$ quivers, it was shown in \cite{Hwang:2017kmk} that even though the contours are different, the residue sums of Seiberg-dual theories agree (up to sign redefinition of the various masses). It is conjectured to also be true for $D_n$ and $E_n$ quiver theories.  There is one caveat, however: the FI parameters are continuous, and as we change chambers in FI parameter space, wall crossing phenomena can happen at the boundary between two chambers. This typically results in new states from the Coulomb branch contributing to the Witten index. Such contributions can be identified explicitly by computing the residues at $\phi^{(a)}\rightarrow\pm \infty$ in the vortex integral, and are known to factorize out of the vortex partition function.\\

After coupling the 3d theory to the line defect, the vortex quantum mechanics  $T^{1d}$ only differs from $T^{1d}_{pure}$ by the additional chiral multiplets contributing $\prod_{a=1}^n Z^{(a)}_{defect}$  to the index. Then, in a Seiberg-dual frame where the sign of $\zeta^{(a)}_{1d}$ is flipped, the contour prescription is simply to enclose the poles in the denominator of $Z^{(a)}_{defect}$ which are complementary to the ones we have been considering so far. Carrying out the JK residue prescription, we checked for $A_n$ quivers where $n\leq 3$ that Seiberg-duality maps a $qq$-character to another $qq$-character, in the \emph{same} representation of the $U_\hbar(\widehat{A_n})$ quantum affine algebra. We suspect this result to be true more generally for all simply-laced Lie algebras $\fg$. However, recall that all vortex characters are further twisted by the contributions of fundamental chiral matter to the index. Such twisted contributions are not invariant under Seiberg duality, and are uniquely determined by the details of the 3d quiver gauge theory one studies in the UV. A detailed analysis of the case $\fg=A_1$ is given in Section \ref{sec:example}. It would be important to reinterpret the different twists which appear in different dual frames directly from the point of view of the representation theory of the quantum affine algebra $U_\hbar(\widehat{\fg})$, or perhaps via its \emph{shifted} version  \cite{2017arXiv170801795F}.

\subsection{Future Directions}

There are many important questions to investigate. Let us list a few pressing open problems:\\ 

-- One important question would be to understand what the D1$_{defect}$ branes of the little string mean in geometry, most notably in the language of quantum K-theory for Nakajima quiver varieties \cite{Aganagic:2017smx,Aganagic:2017gsx}.\\

-- Defining the vortex characters should be possible for classical gauge groups  using orientifold arguments \cite{Haouzi:2020yxy}, and for more general quiver theories than the $ADE$ ones. For instance, so-called ``fractional" quiver theories, which include the non simply-laced $BCFG$ algebras, should be obtained by folding \cite{Aganagic:2017smx,Haouzi:2019jzk,Kimura:2017hez}. Characters of affine quivers theories can likewise be defined in a straightforward way using our construction.\\

-- The vortex characters we constructed are naturally defined for 3d theories on the manifold $\mathbb{C}\times S^1(\widehat{R})$. As we mentioned in the text, reducing the theory on the circle produces vortex characters for 2d $\cN=(4,4)$ gauged sigma models \cite{Nekrasov:2017rqy}. In the other direction, one can likewise define an elliptic version of the  vortex characters \cite{Kimura:2016dys}, where the underlying gauge theory has 4d $\cN=2$ supersymmetry, supported on the manifold $\mathbb{C}\times T^2$. The partition function should coincide with the elliptic genus on $T^2$, following the arguments of our work.\\

-- Recently,  a vortex $qq$-character was defined for specific  3d $\cN=2$ gauge theories of handsaw-type \cite{Haouzi:2019jzk} obtained from the Higgsing a 5-dimensional theory, using a similar construction to the one we presented  for $\cN=4$ theories. In the  $\cW_{q,t}(\fg)$-algebra formalism, the characters arise as correlators close to the ones  studied here, but with the insertion of deformed ``primary" vertex operators at points on the cylinder rather than the fundamental vertex operators \eqref{fundvertex} compatible with $\cN=4$ supersymmetry.
It would be important to construct the characters for more generic 3d $\cN=2$ theories, and understand their realization in the language of $\cW(\fg)$-algebras and string theory, as well as the action of Seiberg-like duality \cite{Aharony:1997gp,Benini:2011mf}.\\

-- Since our defect loops are preserved by the A-twist, they should coincide with the vortex line defects recently studied in \cite{Dimofte:2019zzj}; in particular, it would be important to understand the categorical aspect of our non-perturbative Schwinger-Dyson identities, where the vortex lines should be understood as objects in a braided tensor category. 
Another mathematical application concerns the geometric Langlands program: our Schwinger-Dyson identities are intimately related to the construction of $ADE$-type \emph{opers}, and more specifically their quantization  \cite{Frenkel:2020iqq,Elliott:2018yqm}; we leave the investigation of this important point to future work.\\

-- We note that our string theory realization of the loops in terms of branes a priori differs from the one studied in \cite{Assel:2015oxa}; we think that the discrepancy is due to the presence of nonzero B-fields in our setup, but it would be important to clarify this point in the future.

\vspace{16mm}

\section{A Case Study: 3d $\cN=4$ SQCD}
\label{sec:example}

In this section, we illustrate in detail all the statements made in the paper for the Lie algebra $\fg=A_1$. Namely, consider the 3d $\cN=4$ gauge theory $G^{3d}$ with gauge group $G=U(N)$ and flavor group $G_F=U(N_F)$, on the manifold $\mathbb{C}\times S^1(\widehat{R})$, with a 1/2-BPS line defect at the origin of $\mathbb{C}$ and wrapping $S^1(\widehat{R})$.

\vspace{8mm}

------- \emph{The 1d Quantum Mechanics} -------\\

Let us first describe the theory in the absence of defect. We freeze each equivariant parameter of $G$ to be one of the equivariant parameters of $G_F=U(N_F)$, describing the root of the Higgs branch of $G^{3d}$. We turn on the FI parameter $\zeta_{3d}>0$,  and consider the moduli space of $k$ vortices. Let $T^{1d}_{pure}$ be the quantum mechanics on the vortices. It is a theory with (the reduction from 2d $\cN=(2,2)$ to) 1d $\cN=4$ supersymmetry on $S^1(\widehat{R})$, with gauge group $\widehat{G}=U(k)$. The flavor symmetry is $\widehat{G_F}=U(N)\times U(N_F - N)$, where the first group is the symmetry of $N$ fundamental chiral multiplets, while the second group is the symmetry of $N_F-N$ antifundamental chiral multiplets.

\begin{figure}[h!]
	\emph{}
	\centering
	\includegraphics[trim={0 0 0 0cm},clip,width=0.99\textwidth]{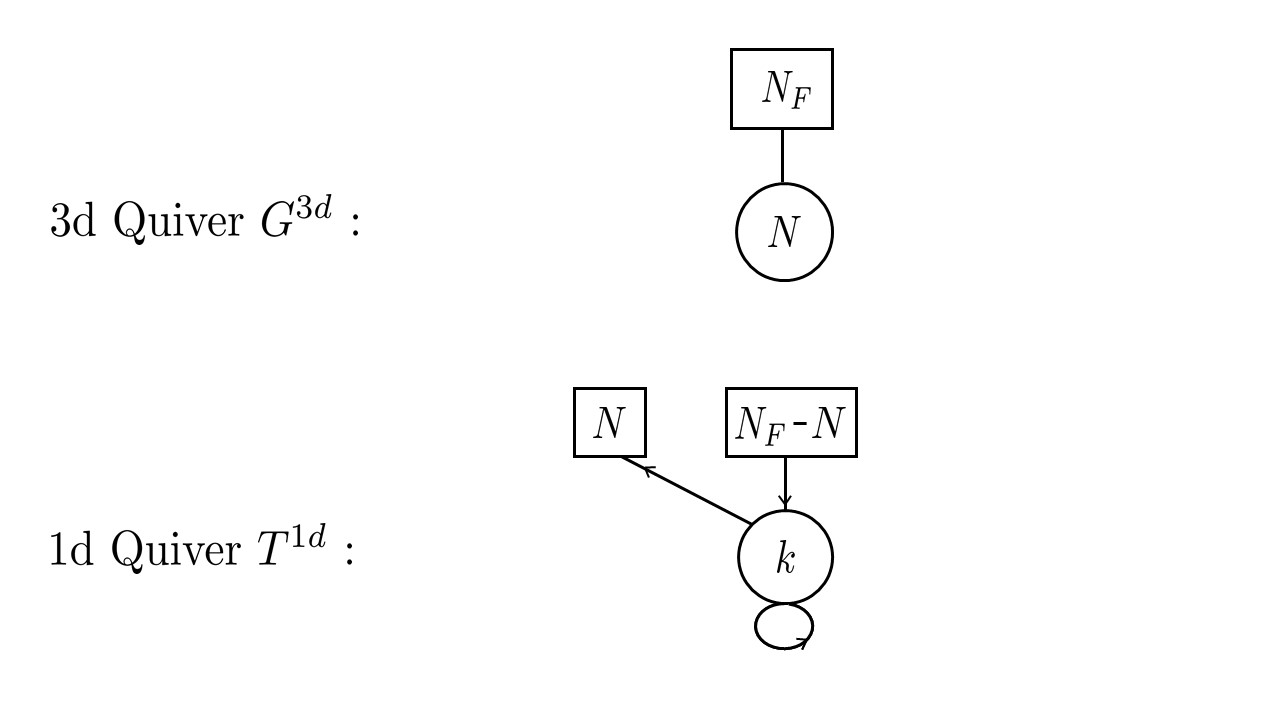}
	\vspace{-12pt}
	\caption{The 3d gauge theory $G^{3d}$ and the vortex quantum mechanics $T^{1d}_{pure}$.} 
	\label{fig:A1examplepure}
\end{figure}

The Witten index \eqref{index} of the quantum mechanics is expressed as the following integral:

{\allowdisplaybreaks
	\begin{align}
	\label{vortexintegralA1pure}
	&\left[\chi\right]^{(0)}_{1d}  =\sum_{k=0}^{\infty}\frac{e^{\zeta_{3d}\, k}}{k!} \oint_{\mathcal{M}^{pure}_k}   \left[\frac{d\phi_I}{2\pi i}\right]Z_{pure, vec}\cdot Z_{pure,  adj}\cdot Z_{pure, teeth}\; , \\
	&Z_{pure, vec} = \frac{\prod_{\substack{I\neq J\\ I,J=1}}^{k}\sh\left(\phi_I-\phi_J\right)}{\prod_{I, J=1}^{k} \sh\left(\phi_{I}-\phi_{J}+\E_2 \right)}\nonumber\\
	&Z_{pure, adj} = \prod_{I, J=1}^{k}\frac{\sh\left(\phi_I-\phi_J+\E_1+\E_2\right)}{\sh\left(\phi_{I}-\phi_{J}+\E_1 \right)}\nonumber\\
	&Z_{pure, teeth} = \prod_{I=1}^{k}\prod_{i=1}^{N}\frac{\sh\left(\phi_I-m_i+\E_-+\E_2\right)}{\sh\left(\phi_{I}-m_i+\E_-\right)}\prod_{j=N+1}^{N_F}\frac{\sh\left(-\phi_I+m_j+\E_++\E_2\right)}{\sh\left(-\phi_{I}+ m_j+\E_+ \right)}\nonumber\, .
	\end{align}}
We made use of the notations $\sh(x)=2\, \sinh(\widehat{R}\, x/2)$, $\E_+=(\E_1+\E_2)/2$ and $\E_-=(\E_1-\E_2)/2$.
Crucially, the index depends on the sign of the 1d FI parameter, which we take here to be $\zeta_{1d}>0$. After applying the JK-residue prescription in that FI-chamber, the poles that end up contributing at vortex charge $k$ to the $T^{1d}_{pure}$ index make up a set $\mathcal{M}^{pure}_k$. The elements of this set satisfy:
\begin{align}
&\phi_I = \phi_J - \E_1 \; , \label{purepole1A1}\\
&\phi_I = \phi_J - \E_2 \; , \label{purepole2A1}\\
&\phi_I = m_i - \E_- \; , \;\;\;\; i\in\{1,\ldots, N\} \; .\label{purepole3A1}
\end{align}
The poles \eqref{purepole1A1} arise from the adjoint chiral factor $Z_{pure, adj}$, the poles \eqref{purepole2A1}  arise from the vector multiplet $Z_{pure, vec}$, and the poles \eqref{purepole3A1} arise from flavor factor $Z_{pure, teeth}$. Most notably, the last set of contours only encloses poles originating from the fundamental chiral multiplets, and none of the antifundamental chiral multiplets. Furthermore, the residues at the locus \eqref{purepole2} are all zero, thanks to the numerator of $Z_{pure, teeth}$. 
Putting it all together, the various poles which end up contributing with nonzero residue are of the form
\beq\label{purepolesA1}
\phi_I = m_i - \E_- - (s_i-1) \E_1 \; , \qquad \text{with}\; s_i\in\{1,\ldots,k_i\}\; ,\qquad i\in\{1,\ldots,N\}\; .
\eeq
In this notation, $(k_1, \ldots, k_N)$ is a partition of $k$ into $N$ non-negative integers, and the pair of integers $(i, s_i)$ is assigned to one of the integers $I\in\{1,\ldots,k\}$ exactly once.

Performing the residue integral, one finds the following well-known expression \cite{Kim:2012uz}:
\begin{align}
\left[\chi\right]^{(0)}_{1d} =\sum_{k=0}^\infty e^{\zeta_{3d}\, k} &\sum_{\substack{\sum_i k_i=k \\ k_i\geq 0}} \;  \left[\prod_{i,j=1}^{N}\prod_{s=1}^{k_i}\frac{\sh\left(m_i-m_j+\E_2- (s-k_j-1)\, \E_1\right)}{\sh\left(m_i-m_j - (s-k_j-1)\, \E_1\right)}\right]\nonumber\\ &\qquad\qquad\times\left[\prod_{i=N+1}^{N_F}\prod_{j=1}^{N}\prod_{p=1}^{k_j}\frac{\sh\left(m_i-m_j+\E_2 + p\, \E_1\right)}{\sh\left(m_i-m_j + p\, \E_1\right)}\right]\label{examplepure1dA1}
\end{align}
Let us now consider the inclusion of the defect, with associated group  $\widehat{G}_{defect}=U(1)$ and corresponding mass fugacity $M$. The inclusion of the loop defect modifies the vortex quantum mechanics, which we now call $T^{1d}$.

\begin{figure}[h!]
	\emph{}
	\centering
	\includegraphics[trim={0 0 0 0cm},clip,width=0.99\textwidth]{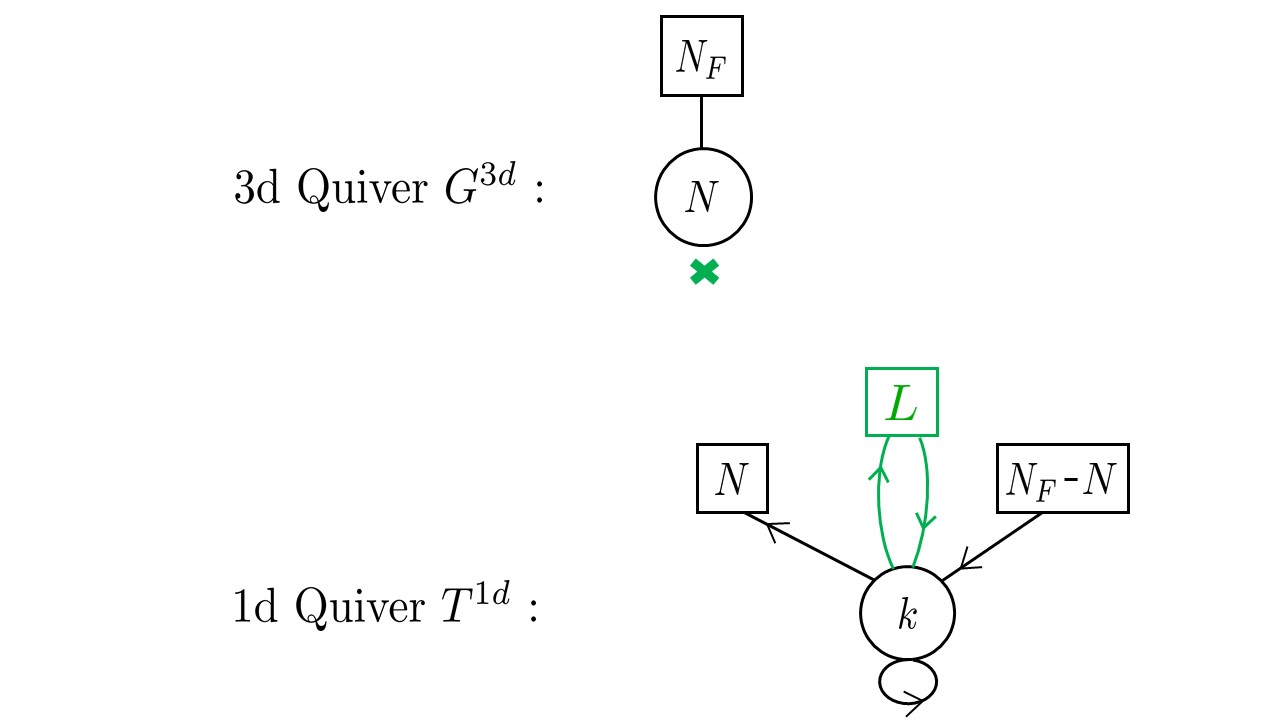}
	\vspace{-12pt}
	\caption{The 3d gauge theory $G^{3d}$ with a loop defect and the vortex quantum mechanics $T^{1d}$. The notations are as in Figure \ref{fig:flagdefect}. In the text, we study the defect corresponding to $L=1$.} 
	\label{fig:A1exampledefect}
\end{figure}

Its Witten index is given by:
{\allowdisplaybreaks
	\begin{align}
	\label{vortexintegralA1}
	&\left[\chi\right]^{(1)}_{1d}  =\sum_{k=0}^{\infty}\frac{e^{\zeta_{3d}\, k}}{k!} \oint_{\mathcal{M}_k}  \left[\frac{d\phi_I}{2\pi i}\right]Z_{pure, vec}\cdot Z_{pure,  adj}\cdot Z_{pure, teeth}\cdot   Z_{defect} \; , \\
	&Z_{defect} =  \prod_{I=1}^{k} \frac{\sh\left(\phi_I-M- \E_- \right)\, \sh\left(-\phi_I + M-\E_-\right)}{\sh\left(\phi_I-M- \E_+\right)\, \sh\left(-\phi_I + M - \E_+ \right)}\, .\nonumber
	\end{align}}
We once again work in the FI-chamber $\zeta_{1d}>0$. For a given vortex charge $k$, the set of poles to be enclosed by the contours is called $\mathcal{M}_k$. This set contains the set  $\mathcal{M}^{pure}_k$ we specified in \eqref{purepolesA1} for the theory $T^{1d}_{pure}$, and an extra locus depending on the defect  mass $M$:
\begin{align}\label{newpoleA1}
\phi_I=M + \epsilon_+ \; ,\;\;\;\text{for some}\; I\in\{1,\ldots,k\} \; .
\end{align}
No other pole depending on $M$ exists, by the following argument: consider the pole at $\phi_I=M+\epsilon_+$ for $I$ fixed. The JK-prescription indicates there is a pole at the locus $\phi_J-M-\epsilon_+=0$, with  $J \neq I$. But the residue there is zero, because of the factor $\sh(\phi_I - \phi_J)$ in the numerator of $Z_{pure, vec}$. Similarly, the JK prescription requires us to include the hyperplanes $\phi_{J}-\phi_{I}+\epsilon_1=0$ and $\phi_{J}-\phi_{I}+\epsilon_2=0$. But the numerators of $Z_{defect}$ guarantee a zero residue at this loci, so we have succeeded in showing that there is exactly one $M$-dependent pole in $\mathcal{M}_k$. Put differently,  $|\mathcal{M}_k|=|\mathcal{M}^{pure}_k|+1$, for all $k$. We conclude that for a given vortex charge $k$, we choose one of the contours to pick up the unique $M$-dependent pole in $\mathcal{M}_k$, namely \eqref{newpoleA1}, and the $k-1$ other poles are to be chosen in the set $\mathcal{M}^{pure}_{k-1}$, according to \eqref{purepolesA1}. 

We introduce a  1/2-BPS codimension-2 defect operator for the loop, with associated vev:
\begin{align}
\label{Yoperator1dA1}
&\left\langle \left[Y_{1d}(M)\right]^{\pm 1} \right\rangle \equiv
\sum_{k=0}^\infty\frac{e^{\zeta_{3d}\, k}}{k!}\oint_{\mathcal{M}^{pure}_k}  \left[\frac{d\phi_I}{2\pi i}\right]Z_{pure,  vec}\cdot Z_{pure,  adj}\cdot Z_{pure,  teeth}\cdot \left[Z_{defect}(M)\right]^{\pm 1}\, . 
\end{align}
Note the contour is defined to exclusively enclose poles in the set $\mathcal{M}^{pure}_k$, thereby avoiding the pole at  $\phi_I=M+\epsilon_+$.
In what follows, we will freely make use of K-theoretic notations,
\begin{align}\label{fugacitiesKtheoryA1}
\widetilde{\fq}&= e^{\zeta_{3d}},\\
q &= e^{\epsilon_1},\qquad t= e^{-\epsilon_2},\qquad v= e^{\epsilon_+}=\sqrt{q/t},\qquad u= e^{\epsilon_-}=\sqrt{q\, t},\nonumber\\
f_d& = e^{-m_d},\qquad z=e^{-M}\; .\nonumber
\end{align} 
Furthermore, for ease of presentation, we renormalize the index by the index of the vortex quantum mechanics $T^{1d}_{pure}$ in the absence of loop defect:
\beq
\label{normalizedA1}
\left[\widetilde{\chi}\right]^{(1)}_{1d}(z) \equiv \frac{\left[\chi\right]^{(1)}_{1d}(z)}{\left[{\chi}\right]^{(0)}_{1d}}
\eeq
Then, we find that the normalized index can be expressed in terms of the $Y$-operators, as a sum of exactly two terms:
\begin{align}
\label{A1pure}
\left[\widetilde{\chi}\right]^{(1)}_{1d}(z) = \frac{1}{\left[{\chi}\right]^{(0)}_{1d}}\left[ \left\langle Y_{1d}(z) \right\rangle + \widetilde{\fq} \; \prod_{i=1}^N\frac{1-q\,t^{-1}\,f_i/z}{1-q\,f_i/z} \prod_{j=N+1}^{N_F} \frac{1-t\,f_j/z}{1-f_j/z}\left\langle\frac{1}{Y_{1d}(z\,v^{-2})}\right\rangle\right]\; .
\end{align}
This is a twisted $qq$-character of the fundamental representation of the quantum affine algebra $U_\hbar(\widehat{A_1})$, with $\hbar\equiv v^2=q/t$.
The meaning of the above expression is the following: The first term on the right-hand side encloses almost all the ``correct" poles in the index integrand, but it is missing exactly one: the extra pole at $\phi_I-M-\epsilon_+=0$. The second term on the right-hand side makes up for this missing pole, and relies on a key observation: we can trade a contour enclosing this extra pole for a contour which does not enclose it, at the expense of inserting the operator $Y(z\, v^{-2})^{-1}$ inside the vev. This result is derived at once from the integral expression \eqref{vortexintegralA1}, and the $Y$-operator definition \eqref{Yoperator1dA1}. Finally, note the presence of the 3d FI parameter $\widetilde{\fq}$ in the second term; it counts exactly one vortex, to make up for the missing $M$-pole, consistent with the fact that $|\mathcal{M}_k|=|\mathcal{M}^{pure}_k|+1$.\\

It is instructive to recast the above result in terms of the general expression for the index presented in the main text:
\begin{align}\label{character1dA1}
\left[\widetilde{\chi}\right]^{(1)}_{1d}(z)=\frac{1}{\left[{\chi}\right]^{(0)}_{1d}}\sum_{\omega\in V(\lambda)} \left({\widetilde{\fq}}\right)^{d^\omega}\; c_{d^\omega}(q, t)\,\mathcal{Q}_{d^\omega}(z)\,  \left[{Y}_{1d}(z)\right]_{\omega} \, .
\end{align}
In that notation, the sum is over two weights exactly, the highest weight $\omega_1=[1]$ of the spin-1/2 representation of $A_1$, and the $\omega_2=[-1]$, obtained by lowering $\omega_1$ by the positive simple root $\alpha$ of $A_1$: $[1]-[2] = [-1]$. The coefficients $c_{d^\omega}(q, t)$ are all 1, while 
\beq
\label{QfunctionA1}
\mathcal{Q}_{d^{[1]}}(z)=1 \; ,\qquad\qquad \mathcal{Q}_{d^{[-1]}}(z)= \prod_{i=1}^N\frac{1-q\,t^{-1}\,f_i/z}{1-q\,f_i/z} \prod_{j=N+1}^{N_F} \frac{1-t\,f_j/z}{1-f_j/z}\; ,
\eeq
and
\beq
\label{YfunctionA1}
\left[{Y}_{1d}(z)\right]_{[1]}= \left\langle Y_{1d}(z) \right\rangle  \; ,\qquad\qquad\left[{Y}_{1d}(z)\right]_{[-1]}= \left\langle \frac{1}{Y_{1d}(z\, v^{-2})} \right\rangle \; .
\eeq
Non-perturbative Schwinger-Dyson identities for $G^{3d}$ follow from the regularity conditions of the vortex character in the defect fugacity $z=e^{-M}$. Namely, a $z$-singularity will arise if two poles of the integrand pinch one of the contours. 
Let us list such pairs of poles, where the poles on the left are inside the contours and the poles on the right are outside the contours, by the JK-prescription. We let $I\in\{1,\ldots,k\}$ and find:
\begin{align}
&\phi_I - M - \E_+ = 0 \qquad \text{and}\qquad \phi_{I}- m_j-\E_+ = 0 \;\;\;\;\; (j\in\{N+1, \ldots, N_F\}) \label{pinch1}\\
&\phi_I - M - \E_+ = 0 \qquad \text{and}\qquad \phi_{I}- \phi_{J}-\E_1 = 0  \;\;\;\;\; (J\neq I) \label{pinch2}\\
&\phi_I - M - \E_+ = 0 \qquad \text{and}\qquad \phi_{I}- \phi_{J}-\E_2 = 0 \;\;\;\;\; (J\neq I) \label{pinch3}\\
&\phi_{I}- m_i+\E_- = 0 \qquad \text{and}\qquad \phi_I - M + \E_+ = 0 \;\;\;\;\; (i\in\{1, \ldots, N\}) \label{pinch4}\\
&\phi_{I}- \phi_{J}+\E_1 = 0 \qquad \text{and}\qquad \phi_I - M + \E_+ = 0  \;\;\;\;\; (J\neq I) \label{pinch5}\\
&\phi_{I}- \phi_{J}+\E_2 = 0 \qquad \text{and}\qquad \phi_I - M + \E_+ = 0 \;\;\;\;\; (J\neq I) \label{pinch6}
\end{align}
The sets of poles \eqref{pinch2}, \eqref{pinch3}, \eqref{pinch5} and \eqref{pinch6} pinch the contour, but the singularity is canceled by a zero in the integrand. For instance, the set \eqref{pinch5} implies a singularity at the locus $\phi_J-M-\E_- =0$, but there is a zero there due to the numerator of $Z_{defect}$. The sets of poles \eqref{pinch1} and \eqref{pinch4} genuinely pinch the contours, and result in singularities
\begin{align}
&M=m_j \;\;\;\;\;\;\;\;\;\;\;\; (j\in\{N+1, \ldots, N_F\})\; ,\label{locus1}\\
&M=m_i+\E_2 \;\;\;\;\; (i\in\{1, \ldots, N\}). \label{locus2}
\end{align}
These two loci are features of 3d $\cN=4$ theories, meaning these singularities are generically always present. If one insists, these singularities can be removed by inserting additional flavor Wilson loops in the 3d theory, transforming in the fundamental representation of the flavor subgroups $U(N)\subset U(N_F)$ and $U(N_f-N)\subset U(N_F)$; this would result in extra Fermi multiplet contributions to the index, which ultimately would break supersymmetry to $\cN=2$, but create zeros at the loci \eqref{locus1} and \eqref{locus2}. We decided against adding these flavor Wilson loops in our discussion. Ultimately, the index as defined simply means that we have to live with these flavor singularities. The Schwinger-Dyson identities are the statement that these are the \emph{only} singularities present.

\vspace{8mm}

------- \emph{The 3d Gauge Theory Perspective} -------\\

In the absence of defect, the half-index of the 3d theory reads
\begin{align}\label{indexintegral3dA1}
\left[\widetilde{\chi}\right]^{(0)}_{3d} =  \oint_{\mathcal{M}^{bulk}} dy\,\left[I^{3d}_{bulk}(y)\, \right]  \; .
\end{align}
where the bulk contribution reads
\begin{align}\label{bulk3dA1}
I^{3d}_{bulk}(y)=\prod_{i=1}^{N}{y_i}^{\left(\zeta_{3d}-1\right)}\;I_{vec}(y) \cdot  I_{flavor}(y,\{x_d\})\; .
\end{align}
The factor 
\begin{align}\label{FI3dA1}
\prod_{i=1}^{N}{y_i}^{\left(\zeta_{3d}\right)}
\end{align}
is the contribution of the 3d FI parameter. 

The factor
\begin{align}\label{vec3dA1}
I_{vec}(y)=\prod_{1\leq i\neq j\leq N}\frac{\left(y_{i}/y_{j};q\right)_{\infty}}{\left(t\, y_{i}/y_{j};q\right)_{\infty}}\;\prod_{1\leq i<j\leq N} \frac{\Theta\left(t\,y_{i}/y_{j};q\right)}{\Theta\left(y_{i}/y_{j};q\right)}
\end{align}
stands for the contribution of the ${\cN}=4$ vector multiplet for the gauge group $G=U(N)$.

The factor 
\begin{align}\label{matter3dA1}
I_{flavor}(y, \{x_d\}) =\prod_{d=1}^{N_F}\prod_{i=1}^{N}  \frac{\left(t\, v\, x_{d}/y_{i};q\right)_{\infty}}{\left(v\, x_{d}/y_{i};q\right)_{\infty}}
\end{align}
stands for the contribution of the $\cN=4$ hypermultiplets in the fundamental representation of the $a$-th gauge group. 
The set of poles to be enclosed by the contours is denoted as $\mathcal{M}^{bulk}$. Following the JK-residue prescription, the poles which contribute with nonzero residue are located at
	\beq
	y_i =v\, x_d \, q^{s}\qquad\;\; ,\;\; s=0,1,2,\ldots\; , \;\;\; d\in\{1,\ldots,N_F\} \; ,
	\eeq
where each integer $\{i\}$ gets mapped uniquely to one of the integers $\{d\}$. Note that the $t$-dependent poles coming from the vector multiplet denominators $\left(t\, y^{(a)}_{i}/y^{(a)}_{j};q\right)_{\infty}$ are allowed by the JK-prescription, but they contribute with zero residue due to the fundamental hypermultiplet numerators $\left(t\, v\, x_{d}/y_{i};q\right)_{\infty}$. Having identified the poles, one can perform the residue integral to recover the Witten index \eqref{examplepure1dA1} of the quantum mechanics $T^{1d}_{pure}$ (up to normalization by irrelevant infinite quantum dilogarithm factors).

We now introduce the 1/2-BPS loop defect wrapping $S^1(\widehat{R})$ via gauging its 1d degrees of freedom, as explained in the main text. The corresponding defect $Y$-operator vev is written as an integral over the Coulomb moduli of the 3d theory:
\begin{align}\label{3ddefectexpressionA1}
\left\langle\left[{\widetilde{Y}_{3d\;  gauge/1d}}(z)\right]^{\pm 1} \right\rangle \equiv\oint_{\mathcal{M}^{bulk}} d{y}\,\left[I^{3d}_{bulk}(y)\cdot\left[{\widetilde{Y}_{3d\; gauge/1d}}(y, z)\right]^{\pm 1}\right]  \; ,
\end{align}
with
\beq\label{3dWilsonfactorA1}
{\widetilde{Y}_{3d\; gauge/1d}}(y,z)=\prod_{i=1}^{N}\frac{1-t\, y_{i}/z}{1- y_{i}/z}\; .
\eeq
There is also the flavor part of the defect which we have to include,
\beq\label{3dWilsonfactor2A1}
{\widetilde{Y}_{3d\; flavor/1d}}(\{x_{d}\}, z)=\prod_{d=1}^{N_F}\frac{1- v\, x_{d}/z}{1- t\, v\, x_{d}/z}\; .
\eeq
Then, the (normalized) index of $G^{3d}$ in the presence of a defect is \emph{defined} as the following vortex character:
\begin{align}\label{character1dA1letsgo}
\left[\widetilde{\chi}\right]^{(1)}_{3d}(z)=\frac{1}{\left[{\chi}\right]^{(0)}_{3d}}\left[\prod_{d=1}^{N_F}\frac{1- v\, x_{d}/z}{1- t\, v\, x_{d}/z}\left\langle \widetilde{Y}_{3d\;  gauge/1d}(z) \right\rangle+ \widetilde{\fq} \left\langle\frac{1}{\widetilde{Y}_{3d\;  gauge/1d}(z\, v^{-2})}\right\rangle\right] \, .
\end{align}
This is once more a twisted $qq$-character of the fundamental representation of $U_\hbar(\widehat{A_1})$. It is again instructive to recast the above result in terms of the general expression for the index presented in the main text:
\begin{align}\label{character1dA1again}
\left[\widetilde{\chi}\right]^{(1)}_{3d}(z)=\frac{1}{\left[{\chi}\right]^{(0)}_{3d}}\sum_{\omega\in V(\lambda)} \left({\widetilde{\fq}}\right)^{d^\omega}\; c_{d^\omega}(q, t)\,\widetilde{\mathcal{Q}}_{d^\omega}(z)\,  \left[{Y}_{3d}(z)\right]_{\omega} \, .
\end{align}
Just as in the case of the quantum mechanics presentation, the sum is over two weights exactly: the highest weight $\omega_1=[1]$ of the spin-1/2 representation of $A_1$, and the $\omega_2=[-1]$, obtained by lowering $\omega_1$ by the positive simple root $\alpha$ of $A_1$: $[1]-[2] = [-1]$. The coefficients $c_{d^\omega}(q, t)$ are all 1, while
\beq
\label{QfunctionA13d}
\widetilde{\mathcal{Q}}_{d^{[1]}}(z)={\widetilde{Y}_{3d\; flavor/1d}}(\{x_{d}\}, z) \; ,\qquad\qquad \widetilde{\mathcal{Q}}_{d^{[-1]}}(z)= 1\; ,
\eeq
and
\beq
\label{YfunctionA13d}
\left[\widetilde{Y}_{3d\;  gauge/1d}(z)\right]_{[1]}= \left\langle \widetilde{Y}_{3d\;  gauge/1d}(z) \right\rangle  \; ,\qquad\left[\widetilde{Y}_{3d\;  gauge/1d}(z)\right]_{[-1]}= \left\langle \frac{1}{\widetilde{Y}_{3d\;  gauge/1d}(z\, v^{-2})} \right\rangle \; .
\eeq
The correctness of this definition can be checked by comparing it to the Witten index of the vortex theory; recall that we previously derived:
\begin{align}
\label{A1pureagain}
\left[\widetilde{\chi}\right]^{(1)}_{1d}(z) = \frac{1}{\left[{\chi}\right]^{(0)}_{1d}}\left[ \left\langle Y_{1d}(z) \right\rangle + \widetilde{\fq} \; \prod_{i=1}^N\frac{1-q\,t^{-1}\,f_i/z}{1-q\,f_i/z} \prod_{j=N+1}^{N_F} \frac{1-t\,f_j/z}{1-f_j/z}\left\langle\frac{1}{Y_{1d}(z\,v^{-2})}\right\rangle\right]\; .
\end{align}
We can perform the residue integral over the poles \eqref{purepolesA1} explicitly, and define ``3d variables" as:
\beq
y_{i,*} = f_i\, q^{k_i+1} \;,\;\; i\in\{1,\ldots,N\} \; ,
\eeq
along with the rescaling of the 1d masses to define them in terms of the 3d masses,
\beq
f_i=v\, x_i \;,\;\;\;\; i=1,\ldots,N_F \; .
\eeq
The defect residue in the quantum mechanics becomes:
\begin{align}\label{fromY1dtoY3dA1}
Z_{defect}({y}_{i,*},z) &=\prod_{i=1}^{N}\frac{1-t\, y_{i,*}/z}{1-y_{i,*}/z}\cdot\prod_{i=1}^{N}\frac{1- f_i/z}{1- t\,  f_{i}/z}\nonumber\\
 &=\prod_{i=1}^{N}\frac{1-t\, y_{i,*}/z}{1-y_{i,*}/z}\cdot\prod_{i=1}^{N}\frac{1- v\, x_i/z}{1- t\, v\, x_{i}/z}\nonumber\\
  &=\left[{\widetilde{Y}_{3d\; gauge/1d}}(\vec y_*, z)\right]\cdot \left[\widetilde{Y}_{3d\; flavor/1d}(\{x_d\}, z)\right]\cdot \prod_{i=N+1}^{N_F}\frac{1-t\, v\, x_i/z}{1-  v\, x_{i}/z}
\end{align}
In other terms, we find that the $Y$-operator vev as defined in the quantum mechanics can be written in terms of the $Y$-operator vev as defined in the 3d theory.

Let us now look at the second term in the 1d vortex character:
\begin{align}\label{fromY1dtoY3dA12}
 &\prod_{i=1}^N\frac{1-q\,t^{-1}\,f_i/z}{1-q\,f_i/z} \prod_{j=N+1}^{N_F} \frac{1-t\,f_j/z}{1-f_j/z}\;\frac{1}{Z_{defect}({y}_{i,*},z\, v^{-2})}\nonumber\\
 &=\cancel{\prod_{i=1}^N\frac{1-q\,t^{-1}\,f_i/z}{1-q\,f_i/z}} \prod_{j=N+1}^{N_F} \frac{1-t\,f_j/z}{1-f_j/z}\prod_{i=1}^{N}\frac{1-v^2\, y_{i,*}/z}{1-t\, v^2\, y_{i,*}/z}\cdot\cancel{\prod_{i=1}^{N}\frac{1- q\,  f_{i}/z}{1- q\,t^{-1}\,f_i/z}}\nonumber\\
  &=\prod_{i=1}^{N}\frac{1-v^2\, y_{i,*}/z}{1-t\, v^2\, y_{i,*}/z} \prod_{j=N+1}^{N_F} \frac{1-t\,v\,x_j/z}{1-v\, x_j/z}\nonumber\\
&=\frac{1}{\left[{\widetilde{Y}_{3d\; gauge/1d}}(\vec y_*, z\, v^{-2})\right]}\cdot \prod_{i=N+1}^{N_F}\frac{1-t\, v\, x_i/z}{1-  v\, x_{i}/z}
\end{align}
After the above remarkable cancellations, we find that the characters are in fact proportional to each other! Denoting the proportionality factor as 
\beq
c_{3d/1d}=\prod_{i=N+1}^{N_F}\frac{1-t\, v\, x_i/z}{1-  v\, x_{i}/z} \; ,
\eeq
we proved that
\begin{align}\label{CHIequalityA1}
\left[\widetilde{\chi}\right]^{(1)}_{3d}(z)= c_{3d/1d}\cdot  \left[\widetilde{\chi}\right]^{(1)}_{1d}(z)\, .
\end{align}

\vspace{8mm}

------- \emph{The ${\cW}_{q,t}(A_1)$-Algebra} -------\\

$q$-Liouville theory on the cylinder $\cC$ enjoys a ${\cW}_{q,t}(A_1)$-algebra symmetry, which is generated by the deformed stress tensor $W^{(2)}(z)$. This generating current is constructed as the commutant of the screening charge. We find:
\beq\label{A1stress}
W^{(2)}(z) = :\cY(z): + :\left[\cY(v^{-2}z)\right]^{-1}: \; ,
\eeq
where $\cY$ is the operator defined in \eqref{YoperatorToda}.
We consider the correlator:
\beq\label{A1correlatordef}
\left\langle \psi'\left|\prod_{d=1}^{N_F} V(x_d)\; Q^{N}\; W^{(2)}(z) \right| \psi \right\rangle\, .
\eeq
The contours are specified as to not  enclose any pole in the $z$ variable.
The state $|\psi\rangle$ is defined such that:
\begin{align}
\alpha[0] |\psi\rangle &= \langle\psi, \alpha\rangle |\psi\rangle\label{eigenvalueA1}\\
\alpha[k] |\psi\rangle &= 0\, , \qquad\qquad\;\; \mbox{for} \; k>0,\nonumber
\end{align}
where the $\alpha[k]$ generate a $q$-deformed Heisenberg algebra:
\beq\label{A1commutator}
[\alpha[k], \alpha[m]] = {1\over k} (q^{k\over 2} - q^{-{k\over 2}})(t^{{k\over 2} }-t^{-{k\over 2} })(v^{k}+ v^{-k}) \delta_{k, -m} \; .
\eeq
We compute the various two-points making up the correlator; first, the bulk contributions
\begin{align}
&\prod_{1\leq i< j\leq N}\left\langle S(y_i)\, S(y_j) \right\rangle = \prod_{1\leq i\neq j\leq N} \frac{\left(y_{i}/y_{j};q\right)_{\infty}}{\left(t\, y_{i}/y_{j};q\right)_{\infty}}\;\prod_{1\leq i<j\leq N} \frac{\Theta\left(t\,y_{j}/y_{i};q\right)}{\Theta\left(y_{j}/y_{i};q\right)}\\
&\prod_{i=1}^{N}\left\langle V(x_d)\, S(y_i) \right\rangle = \prod_{i=1}^{N} \frac{\left(t\,v\, x_{d}/y_{i};q\right)_{\infty}}{\left(v\, x_{d}/y_{i};q\right)_{\infty}}\; .
\end{align}
The two-point of fundamental vertex operators with themselves will drop out after normalization, so we omit writing it.

We now come to the contributions involving the loop defect. First, the two-point of the stress tensor with the screening currents
\begin{align}
&\prod_{i=1}^{N}\left\langle S(y_i)\, W^{(2)}(z) \right\rangle = \prod_{i=1}^{N} \frac{1-t\, y_{i}/z}{1- y_{i}/z}+\prod_{i=1}^{N} \frac{1-v^2\, y_{i}/z}{1- t\, v^2\, y_{i}/z}
\end{align}
Note that in the actual correlator, the vacuum is labeled by $|\psi\rangle$ instead of $|0\rangle$, resulting in a relative shift of $\widetilde{\fq}$ between the two terms. A more involved computation is the two-point of  the fundamental vertex operator with the stress tensor:
\begin{align}
&\left\langle  V(x_d)\, W^{(2)}(z) \right\rangle = \exp\left(\sum_{k>0}\, \frac{1}{k} \,\frac{t^k-1}{v^k+v^{-k}}\left(\frac{x_d}{z}\right)^k\right)\nonumber\\
&\qquad\qquad+ \exp\left(-\sum_{k>0}\, \frac{1}{k} \,\frac{t^k-1}{v^k+v^{-k}}\left(\frac{v^2\,x_d}{z}\right)^k\right)\nonumber\\
&\;\;\;= \exp\left(-\sum_{k>0}\, \frac{1}{k} \,\frac{t^k-1}{v^k+v^{-k}}\left(\frac{v^2\,x_d}{z}\right)^k\right)\cdot\left(\frac{1-v\, x_d/z}{1-t\,v\,x_d/z}+1\right)\nonumber\\
&\;\;\;= B\left(x_d, z\right)\cdot\left(\frac{1-v\, x_d/z}{1-t\,v\,x_d/z}+1\right)
\end{align}
In the first line, we used the commutator
\begin{align}
[w[k], w[n]] = {1\over k} (q^{k\over 2} - q^{-{k\over 2}})(t^{{k\over 2} }-t^{-{k\over 2} })\frac{1}{v^{k}+ v^{-k}}\delta_{k, -n} \; ,
\end{align}
which is dual to the relation \eqref{A1commutator}. In the second line we used the identity $\exp(-\sum_{k>0}\frac{x^k}{k})=(1-x)$. In the third line, we gave a name to the overall exponential factor,
\beq\label{BdefinedA1}
B\left(x_d, z\right)\equiv \exp\left(-\sum_{k>0}\, \frac{1}{k} \,\frac{t^k-1}{v^k+v^{-k}}\left(\frac{v^2\,x_d}{z}\right)^k\right) \; .
\eeq
All in all, the normalized ${\cW}_{q,t}(A_1)$ correlator comes out to be proportional to the 3d vortex character,
\begin{align}\label{A1correlatoris3dindex}
\frac{\left\langle \psi'\left|\prod_{d=1}^{N_F} V(x_d)\; Q^{D}\; W^{(2)}(z) \right| \psi \right\rangle}{\left\langle \psi'\left|\prod_{d=1}^{N_F} V(x_d)\; Q^{D} \right| \psi \right\rangle}
= B(\{x_d\}, z)\, \left[\widetilde{\chi}\right]^{(1)}_{3d}(z)\; .
\end{align}
with
\beq\label{BdefinedA1again}
B\left(\{x_d\}, z\right)\equiv\prod_{d=1}^{N_F} \exp\left(-\sum_{k>0}\, \frac{1}{k} \,\frac{t^k-1}{v^k+v^{-k}}\left(\frac{v^2\,x_d}{z}\right)^k\right) \; ,
\eeq
As explained in the main text, this factor can be naturally canceled out in the Ding-Iohara-Miki formalism, where it arises as an extra $U(1)$ due to an auxiliary Heisenberg algebra \cite{Mironov:2016yue}.

The non-perturbative Schwinger-Dyson equation for $G^{3d}$ manifests itself here as a Ward identity. It is interpreted as a statement about the regularity in the fugacity $z$ of the correlator $\left\langle[\dots\ldots]\, W^{(2)}(z)\right\rangle$.\\

It is straightforward to generalize this discussion to the case of a defect in a higher spin representation, by considering a group $\widehat{G}_{defect}=U(L)$.  The JK-residue prescription dictates that for each $\rho\in\{1, 2, \ldots, L\}$, the contours of the quantum mechanics should enclose a pole at 
\begin{equation}
\phi_I-M_\rho-\epsilon_+=0 \; .
\end{equation}
Once again, the partition function can be expressed as a $qq$-character of $U_\hbar(\widehat{A_1})$, with highest weight $[L]$ (the spin $L/2$ representation). 
In the $q$-Liouville picture, one would simply consider a deformed $\cW_{q,t}(A_1)$-algebra correlator with $L$ insertions of the deformed stress tensor:
\beq\label{A1correlatoris3dindexMORE}
\frac{\left\langle \psi'\left|\prod_{d=1}^{N_F} V(x_d)\; Q^{D}\; \prod_{\rho=1}^{L} W^{(2)}(z_{\rho}) \right| \psi \right\rangle}{\left\langle \psi'\left|\prod_{d=1}^{N_F} V(x_d)\; Q^{D} \right| \psi \right\rangle} = B(\{x_d\}, \{z_\rho\})\, \left[\widetilde{\chi}\right]^{(L)}_{3d}(\{z_\rho\})\; .
\eeq

\vspace{15mm}

------- \emph{Defects of the $A_1$ $(2,0)$ Little String} -------\\

Let $X$ be a resolved $A_1$ singularity, and consider type IIB string theory on $X\times\cC\times\mathbb{C}_q\times\mathbb{C}_t$, with $\cC= \mathbb{R} \times S^1(R)$ an infinite cylinder of radius $R$, and $\mathbb{C}_q$ and  $\mathbb{C}_t$ two complex lines. We introduce $N$ D3$_{gauge}$ branes wrapping the compact 2-cycle $S$ of $X$ and $\mathbb{C}_q$. We further introduce $N_F$ D3$_{flavor}$ branes  wrapping the dual non-compact 2-cycle $S^*$ and $\mathbb{C}_q$. We also add to this background $L$ D1$_{defect}$ branes wrapping that same 2-cycle $S^*$ and $\mathbb{C}_t$. This background preserves 4 supercharges. We send the string coupling to $g_s\rightarrow 0$; the tensions of the various D3 branes survive in the limit. Then, this amounts to studying the $(2,0)$ $A_1$ little string on $\cC\times\mathbb{C}_q\times \mathbb{C}_t$ in the presence of various codimension-4 defects. 
At energies below the string scale, the dynamics  are fully captured by the theory on the D3$_{gauge}$ branes: the effective theory on the branes is the 3d gauge theory $G^{3d}$, with gauge group $G=U(N)$, defined on the manifold $\mathbb{C}_q\times S^1(\widehat{R})$. Note that this is the T-dual circle to the original circle $S^1(R)$ of the cylinder, meaning $\widehat{R}=1/(m^2_s\, R)$. The D3$_{flavor}$ branes realize the fundamental matter content $G_F=U(N_F)$.  From the 3d gauge theory point of view, the $L$ D1$_{defect}$ branes in the nontrivial NS-NS B-field make up a 1/2-BPS loop defect wrapping $S^1(\widehat{R})$ and sitting at the origin of $\mathbb{C}_q$. Let us focus on the case of a single defect brane, $L=1$.

\begin{figure}[h!]
	\emph{}
	\centering
	\includegraphics[trim={0 0 0 3cm},clip,width=0.99\textwidth]{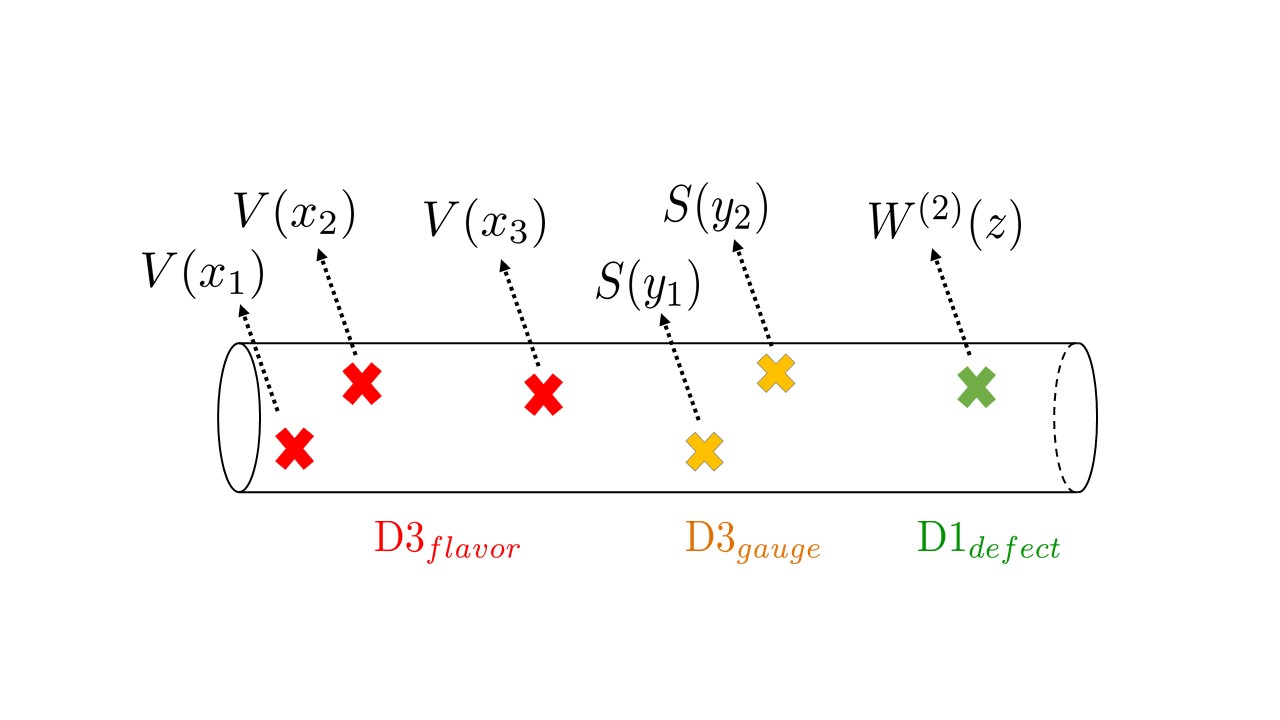}
	\vspace{-40pt}
	\caption{Example of a correlator in $q$-Liouville, along with the corresponding D-branes at points on the cylinder. The specific correlator pictured here is $\langle \psi' |  \prod_{d=1}^3 V(x_d)  \, (Q)^{2} \, W^{(2)}(z)   | \psi \rangle$.}
	\label{fig:cylinderbranes}
\end{figure}

The index of the $(2,0)$ little string in this background is a 3d/1d half-index corresponding to the $\widetilde{Y}$-operator vev definition in the 3d gauge theory. The vortex character observable is constructed as a sum of such vevs, uniquely determined from the requirement that the defect singularities due to D3$_{gauge}$/D1$_{defect}$ strings should be removed. All in all, we recover the 3d expression  \eqref{character1dA1letsgo}:
\begin{align}\label{character1dA1letsgoagain}
\left[\widetilde{\chi}\right]^{(1)}_{\text{D3}_g, \text{D3}_f, \text{D1}_d}(z)=\frac{1}{\left[{\chi}\right]^{(0)}_{\text{D3}_g, \text{D3}_f}}\left[\prod_{d=1}^{N_F}\frac{1- v\, x_{d}/z}{1- t\, v\, x_{d}/z}\left\langle \widetilde{Y}_{3d\;  gauge/1d}(z) \right\rangle+ \widetilde{\fq} \left\langle\frac{1}{\widetilde{Y}_{3d\;  gauge/1d}(z\, v^{-2})}\right\rangle\right] .
\end{align}
Up to an overall normalization, this also happens to be the $q$-Liouville correlator \eqref{A1correlatoris3dindex} on the cylinder, see Figure \ref{fig:cylinderbranes}.

We break $G$ by freezing the D3$_{gauge}$ moduli, and reorganize the branes to make up a set of D3'$_{flavor}$ branes exclusively. We turn on the period  $\int_{S}\omega_{I}>0$, which is the 3d FI parameter, and study the vortex solutions on the Higgs branch of $G^{3d}$. These are D1$_{vortex}$ branes wrapping the 2-cycle $S$ in the $(2,0)$ string. The partition function localizes to the Witten index \eqref{A1pure} of the quantum mechanics $T^{1d}$ on the D1$_{vortex}$ branes. Up to an overall normalization, this is again the same vortex character:
\begin{align}
\label{A1pureletsgoagain}
\left[\widetilde{\chi}\right]^{(1)}_{\text{D3}'_f, \text{D1}_d, \text{D1}_v}(z) = \frac{1}{\left[{\chi}\right]^{(0)}_{\text{D3}'_f, \text{D1}_v}}\left[ \left\langle Y_{1d}(z) \right\rangle + \widetilde{\fq} \; \prod_{i=1}^N\frac{1-q\,t^{-1}\,f_i/z}{1-q\,f_i/z} \prod_{j=N+1}^{N_F} \frac{1-t\,f_j/z}{1-f_j/z}\left\langle\frac{1}{Y_{1d}(z\,v^{-2})}\right\rangle\right].
\end{align}
The second term corresponds to the D1$_{defect}$ brane recombining with one of the D1$_{vortex}$ branes when they coalesce on the cylinder, which is a $z$-pole locus in the integrand of the quantum mechanics. The fact this happens only once is the string-theoretic version of the statement $\left|\mathcal{M}_k\setminus \mathcal{M}^{pure}_k\right|=1$, for all vortex numbers $k$.

\vspace{8mm}

------- \emph{Seiberg Duality of the Vortex Character} -------\\

Let us first go back to the index of the quantum mechanics $T^{1d}_{pure}$, in the absence of defect, which we rewrite here for convenience:

{\allowdisplaybreaks
	\begin{align}
	\label{vortexintegralA1pureSeiberg}
	&\left[\chi\right]^{(0)}_{1d}  =\sum_{k=0}^{\infty}\frac{e^{\zeta_{3d}\, k}}{k!} \oint_{\mathcal{M}^{pure}_k}   \left[\frac{d\phi_I}{2\pi i}\right]Z_{pure, vec}\cdot Z_{pure,  adj}\cdot Z_{pure, teeth}\; , \\
	&Z_{pure, vec} = \frac{\prod_{\substack{I\neq J\\ I,J=1}}^{k}\sh\left(\phi_I-\phi_J\right)}{\prod_{I, J=1}^{k} \sh\left(\phi_{I}-\phi_{J}+\E_2 \right)}\nonumber\\
	&Z_{pure, adj} = \prod_{I, J=1}^{k}\frac{\sh\left(\phi_I-\phi_J+\E_1+\E_2\right)}{\sh\left(\phi_{I}-\phi_{J}+\E_1 \right)}\nonumber\\
	&Z_{pure, teeth} = \prod_{I=1}^{k}\prod_{i=1}^{N}\frac{\sh\left(\phi_I-m_i+\E_-+\E_2\right)}{\sh\left(\phi_{I}-m_i+\E_-\right)}\prod_{j=N+1}^{N_F}\frac{\sh\left(-\phi_I+m_j+\E_++\E_2\right)}{\sh\left(-\phi_{I}+ m_j+\E_+ \right)}\nonumber\, .
	\end{align}}
Once again, we used the notations $\E_+=(\E_1+\E_2)/2$ and $\E_-=(\E_1-\E_2)/2$.
We now study the Witten index in a chamber with a negative  1d FI parameter,  $\zeta_{1d}<0$. We then apply the JK-residue prescription in that FI-chamber\footnote{Just as before, the JK-residue requires us to define an auxiliary vector $\eta$ of size $k$, and we once again choose $\eta=\zeta_{1d}$ to remove contributions from $\phi$-poles at $\pm \infty$. We find the choice $\eta=(-1,\ldots,-1)$ convenient here.}. For each vortex charge $k$, the poles that end up contributing  make up the set $\mathcal{M}^{pure}_k$. The elements of this set satisfy:
\begin{align}
&\phi_I = \phi_J + \E_1 \; , \label{purepole1A1Seiberg}\\
&\phi_I = \phi_J + \E_2 \; , \label{purepole2A1Seiberg}\\
&\phi_I = m_j + \E_+ \; , \;\;\;\; j\in\{N+1,\ldots, N_F\} \; .\label{purepole3A1Seiberg}
\end{align}
The poles \eqref{purepole1A1Seiberg} arise from the adjoint chiral factor $Z_{pure, adj}$, the poles \eqref{purepole2A1Seiberg}  arise from the vector multiplet factor $Z_{pure, vec}$, and the poles \eqref{purepole3A1Seiberg} arise from flavor factor $Z_{pure, teeth}$. The last set of contours now encloses poles originating from the \emph{antifundamental} chiral multiplets, and none of the fundamental chiral multiplets. Furthermore, the residue at the locus \eqref{purepole2A1Seiberg} is zero. Putting it all together, the various poles which end up contributing with nonzero residue are of the form:
\beq\label{purepolesA1Seiberg}
\phi_I = m_j + \E_+ + (s_i-1) \E_1 \; , \qquad \text{with}\;\; s_i\in\{1,\ldots,k_i\}\; ,\qquad i\in\{N+1,\ldots,N_F\}\; .
\eeq
In this notation, $(k_1, \ldots, k_N)$ is a partition of $k$ into $N_F-N$ non-negative integers, and the pair of integers $(i, s_i)$ is assigned to one of the integers $I\in\{1,\ldots,k\}$ exactly once. 
Performing the residue integral, we get the closed-form expression:
\begin{align}\label{examplepure1dA1Seiberg}
\left[\chi\right]^{(0)}_{1d,\; \zeta_{1d}<0} =\sum_{k=0}^\infty \left(-e^{\zeta_{3d}}\right)^k &\sum_{\substack{\sum_i k_i=k \\ k_i\geq 0}} \;  \left[\prod_{i,j=N+1}^{N_F}\prod_{s=1}^{k_i}\frac{\sh\left(m_i-m_j+\E_2- (s-k_j-1)\, \E_1\right)}{\sh\left(m_i-m_j - (s-k_j-1)\, \E_1\right)}\right]\nonumber\\ &\;\;\qquad\times\left[\prod_{i=N+1}^{N_F}\prod_{j=1}^{N}\prod_{p=1}^{k_j}\frac{\sh\left(m_i-m_j+\E_2 + p\, \E_1\right)}{\sh\left(m_i-m_j + p\, \E_1\right)}\right]\; .
\end{align}
After flipping the signs of the $N_F$ masses $\{m_d\}\rightarrow \{-m_d-\E_2\}$ (the shift by $-\E_2$ is inconsequential at this stage, but will matter later) and the sign of the 3d FI parameter $\widetilde{\fq}\rightarrow  -\widetilde{\fq}$, we recognize  the index of a 3d $U(N_F - N)$ gauge theory with $N_F$ fundamental flavors. As predicted, changing the sign of the 1d FI parameter in the quantum mechanics realizes 3d Seiberg duality \cite{Hwang:2017kmk}. For comparison, we rewrite the index of the $U(N)$ gauge theory with $N_F$ fundamental flavors we previously derived in the chamber $\zeta_{1d}>0$:

\begin{align}\label{examplepure1dA1again}
\left[\chi\right]^{(0)}_{1d, \; \zeta_{1d}>0} =\sum_{k=0}^\infty e^{\zeta_{3d}\, k}&\sum_{\substack{\sum_i k_i=k \\ k_i\geq 0}} \;  \left[\prod_{i,j=1}^{N}\prod_{s=1}^{k_i}\frac{\sh\left(m_i-m_j+\E_2- (s-k_j-1)\, \E_1\right)}{\sh\left(m_i-m_j - (s-k_j-1)\, \E_1\right)}\right]\nonumber\\ &\;\;\qquad\times\left[\prod_{i=N+1}^{N_F}\prod_{j=1}^{N}\prod_{p=1}^{k_j}\frac{\sh\left(m_i-m_j+\E_2 + p\, \E_1\right)}{\sh\left(m_i-m_j + p\, \E_1\right)}\right]\; .
\end{align}
Having reviewed the pure case, let us now introduce the loop defect. Recall that the Witten index of the quantum mechanics $T^{1d}$ now reads:

{\allowdisplaybreaks
	\begin{align}
	\label{vortexintegralA1Seiberg}
	&\left[\chi\right]^{(1)}_{1d}  =\sum_{k=0}^{\infty}\frac{e^{\zeta_{3d}\, k}}{k!} \oint_{\mathcal{M}_k}  \left[\frac{d\phi_I}{2\pi i}\right]Z_{pure, vec}\cdot Z_{pure,  adj}\cdot Z_{pure, teeth}\cdot   Z_{defect} \; , \\
	&Z_{defect} =  \prod_{I=1}^{k} \frac{\sh\left(\phi_I-M- \E_- \right)\, \sh\left(-\phi_I + M- \E_-\right)}{\sh\left(\phi_I-M- \E_+\right)\, \sh\left(-\phi_I + M - \E_+ \right)}\, .\nonumber
	\end{align}}
In the FI-chamber $\zeta_{1d}<0$, we have a new pole at the locus
\begin{align}\label{newpoleA1Seiberg}
\phi_I=M - \epsilon_+ \; ,\;\;\;\text{for some}\; I\in\{1,\ldots,k\} \; .
\end{align}
No other pole depending on $M$ exists, by the same arguments invoked in the case $\zeta_{1d}>0$. For each vortex charge $k$, the set of poles $\mathcal{M}_k$ is therefore the set $\mathcal{M}^{pure}_k$, augmented by the pole \eqref{newpoleA1Seiberg}. The  codimension-2 $Y$-operator is defined as before, avoiding this new pole: 
\begin{align}
\label{Yoperator1dA1Seiberg}
&\left\langle \left[Y_{1d}(M)\right]^{\pm 1}\right\rangle \equiv
\sum_{k=0}^\infty\frac{e^{\zeta_{3d}\, k}}{k!}\oint_{\mathcal{M}^{pure}_k}  \left[\frac{d\phi_I}{2\pi i}\right]Z_{pure,  vec}\cdot Z_{pure,  adj}\cdot Z_{pure,  teeth}\cdot \left[Z_{defect}(M)\right]^{\pm 1}\, . \nonumber
\end{align}
We renormalize the index by the defect-free index of the vortex quantum mechanics $T^{1d}_{pure}$:
\beq
\label{normalizedA1Seiberg}
\left[\widetilde{\chi}\right]^{(1)}_{1d, \;\zeta_{1d}<0}(z) \equiv \frac{\left[\chi\right]^{(1)}_{1d,\; \zeta_{1d}<0}(z)}{\left[{\chi}\right]^{(0)}_{1d,\; \zeta_{1d}<0}}\; ,
\eeq
and derive at once the vortex character in the  FI-chamber $\zeta_{1d}<0$, written in K-theoretic notation as:
\begin{align}
\label{A1pureSeiberg}
\left[\widetilde{\chi}\right]^{(1)}_{1d, \;\zeta_{1d}<0}(z) = \frac{1}{\left[{\chi}\right]^{(0)}_{1d, \;\zeta_{1d}<0}}\left[ \left\langle Y_{1d}(z) \right\rangle - \widetilde{\fq} \; \prod_{i=1}^N\frac{1-f_i/z}{1-t\,f_i/z} \prod_{j=N+1}^{N_F} \frac{1-t^2\,q^{-1}f_j/z}{1-t\,q^{-1}f_j/z}\left\langle\frac{1}{Y_{1d}(z\,v^{2})}\right\rangle\right]\; .
\end{align}
If we flip the sign of the $N_F$ masses $\{m_d\}\rightarrow \{-m_d-\E_2\}$ (or $\{f_d\}\rightarrow \{t^{-1}\,f^{-1}_d\}$ in the new variables), the defect  mass as $M\rightarrow  -M$ (or $z\rightarrow z^{-1}$ in the new variables),  and the 3d FI parameter as $\widetilde{\fq}\rightarrow  -\widetilde{\fq}$, we recognize the vortex character of a 3d $U(N_F - N)$ gauge theory with $N_F$ fundamental flavors. Note the nontrivial rescaling of the $N_F$ flavor masses by $t^{-1}$.

For comparison, we also rewrite the vortex character of the 3d $U(N)$ gauge theory with $N_F$ fundamental flavors \eqref{A1pure}:
\begin{align}
\label{A1pureagainagain}
\left[\widetilde{\chi}\right]^{(1)}_{1d,  \;\zeta_{1d}>0}(z) = \frac{1}{\left[{\chi}\right]^{(0)}_{1d, \;\zeta_{1d}>0}}\left[ \left\langle Y_{1d}(z) \right\rangle + \widetilde{\fq} \; \prod_{i=1}^N\frac{1-q\,t^{-1}\,f_i/z}{1-q\,f_i/z} \prod_{j=N+1}^{N_F} \frac{1-t\,f_j/z}{1-f_j/z}\left\langle\frac{1}{Y_{1d}(z\,v^{-2})}\right\rangle\right]\; .
\end{align}

As a last remark, note that the index $\left[\chi\right]^{(0)}_{1d, \; \zeta_{1d}>0}$ \eqref{examplepure1dA1again} in the positive FI chamber is not equal to the index  $\left[\chi\right]^{(0)}_{1d, \; \zeta_{1d}<0}$  \eqref{examplepure1dA1Seiberg} in the negative FI chamber. This is because new states appear and contribute to the index at $\zeta_{1d}=0$, due to the opening of the Coulomb branch there. The vortex mechanics $T^{1d}_{pure}$ experiences wall-crossing, and the BPS index of the extra states can be computed explicitly by identifying the residues at asymptotic infinity, enclosing the $\phi$-poles of the integrand  \eqref{vortexintegralA1pureSeiberg} at $\pm \infty$. A quick computation shows that such residues can be summed up exactly to give the contribution of $2N-N_F$ decoupled twisted hypermultiplets, which do exist on the Coulomb branch of $G^{3d}$ \cite{Gaiotto:2008ak,Kim:2012uz,Yaakov:2013fza,Gaiotto:2013bwa,Hwang:2017kmk}. Explicitly, the wall-crossing contribution can be written as a plethistic exponential:
\beq\label{extrahyper}
\text{PE}\left[\frac{\sh(2\E_+)\; \sh((2N-N_F)\E_2)}{\sh(\E_1)\;\sh(\E_2)}\; \widetilde{\fq}\right]
\eeq
Note that such a contribution vanishes when $N_F = 2 N$, in which case the indices agree: $\left[\chi\right]^{(0)}_{1d, \; \zeta_{1d}<0}= \left[\chi\right]^{(0)}_{1d, \; \zeta_{1d}>0}$. 

We can carry out the same computation for the index $\left[\chi\right]^{(1)}_{1d}$ in the presence of the defect \eqref{vortexintegralA1Seiberg}, to find that the extra contributions due to $\phi$-poles at $\pm \infty$ are the same as above: there are $2N-N_F$ extra decoupled twisted hypermultiplets, resulting in a decoupled factor \eqref{extrahyper}. Because the vortex character observable is the index $\left[\chi\right]^{(1)}_{1d}$ normalized by the pure index $\left[\chi\right]^{(0)}_{1d}$, the twisted hypermultiplets contributions cancel out at any rate in our context.

\section*{Acknowledgments}
We thank Mina Aganagic for remarks made at a UC Berkeley String-Math Seminar. and Nikita Nekrasov for elucidating some aspects of his instanton $qq$-character construction. The research of N. H. is supported  by the Simons Center for Geometry and Physics.
\vspace{16mm}

	\begin{appendix}
	
	\section{Some Examples of Vortex Characters}

We write some explicit expressions for the vortex $qq$-character observables $\left[{\chi}\right]^{(L^{(1)},\ldots,L^{(n)})}_{3d}$  for some 3d gauge theories, in the 3d/1d half-index formalism. It is straightforward to write the observables in the quantum mechanics or $q$-Toda variables instead, if one wishes. The normalized characters $\left[\widetilde{\chi}\right]^{(L^{(1)},\ldots,L^{(n)})}_{3d}$ in the main text are obtained by  dividing the index by the pure index $\left[{\chi}\right]^{(0,\ldots,0)}_{3d}$, which we omitted doing here not to overburden the formulas. Furthermore, we define the shorthand notation ${\widetilde{Y}^{(a)}_{3d}}(z)\equiv {\widetilde{Y}^{(a)}_{3d\; gauge/1d}}(y,z)$, where the expression ${\widetilde{Y}^{(a)}_{3d\; gauge/1d}}(y,z)=\prod_{i=1}^{N^{(a)}}\frac{1-t\, y^{(a)}_{i}/z}{1- y^{(a)}_{i}/z}$ was defined in the main text.

\begin{figure}[h!]
	\emph{}
	\centering
	\includegraphics[trim={0 0 0 0cm},clip,width=0.85\textwidth]{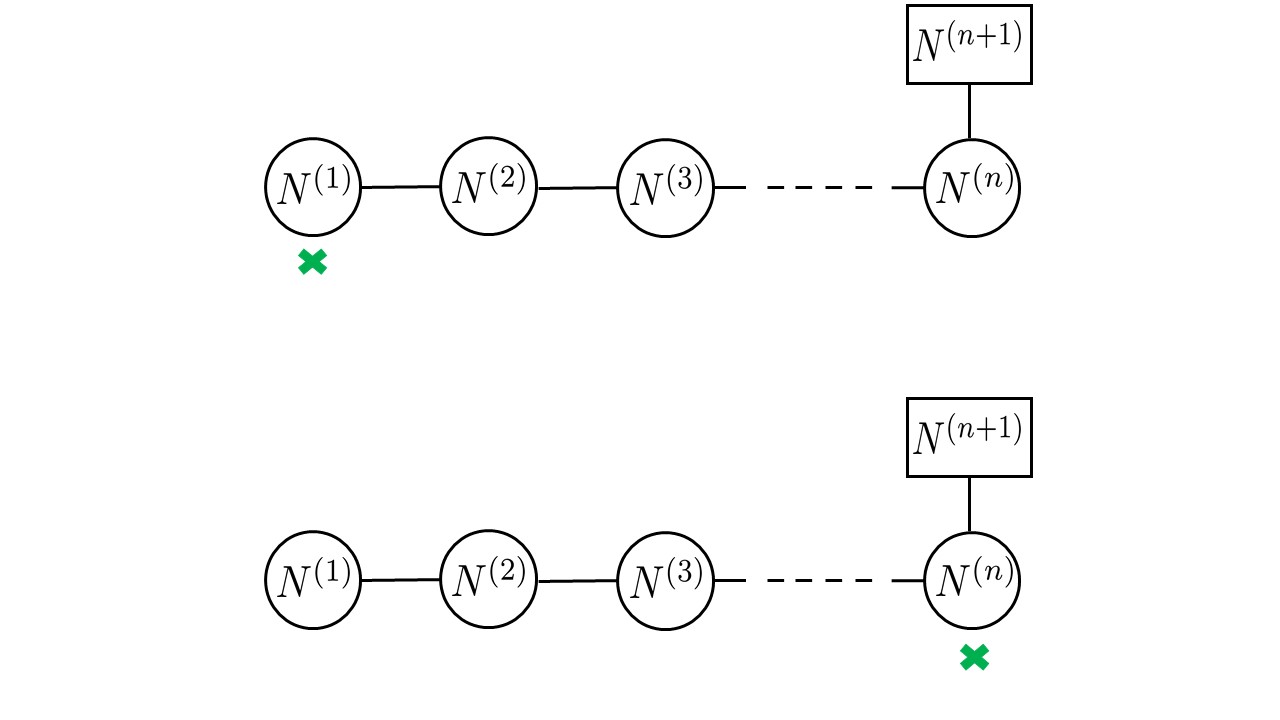}
	\vspace{-15pt}
	\caption{The $T_\rho[SU(N^{(n+1)})]$ theory, with a loop defect producing the first fundamental vortex character (top), and a loop defect producing the $n$-th fundamental vortex character (bottom).}
	\label{fig:Anexamples}
\end{figure}

For the $T_\rho[SU(N^{(n+1)})]$ theory on top of Figure \ref{fig:Anexamples}, we compute:	
\begin{align}
&\left[{\chi}\right]^{(1,0,\ldots,0)}_{3d}(z) =\prod_{d=1}^{N^{(n+1)}}\frac{1- v^n\, x_{d}/z}{1- t\, v^n\, x_{d}/z}\left\langle\widetilde{Y}^{(1)}_{3d}(z)  \right\rangle\label{3dpartitionfunctionAmSIMPLE1}\\
&\qquad\qquad + \widetilde{\fq}^{(1)} \, \prod_{d=1}^{N^{(n+1)}}\frac{1- v^n\, x_{d}/z}{1- t\, v^n\, x_{d}/z} \left\langle\frac{\widetilde{Y}^{(2)}_{3d}(z\, v^{-1})}{\widetilde{Y}^{(1)}_{3d}(z\, v^{-2})}\right\rangle\nonumber\\
&\qquad\qquad + \widetilde{\fq}^{(1)}\widetilde{\fq}^{(2)} \,\prod_{d=1}^{N^{(n+1)}}\frac{1- v^n\, x_{d}/z}{1- t\, v^n\, x_{d}/z}  \left\langle\frac{\widetilde{Y}^{(3)}_{3d}(z\, v^{-2})}{\widetilde{Y}^{(2)}_{3d}(z\, v^{-3})}\right\rangle\nonumber\\
&\qquad\qquad+\ldots\nonumber\\
&\qquad\qquad + \prod_{a=1}^{n-1} \widetilde{\fq}^{(a)} \,\prod_{d=1}^{N^{(n+1)}}\frac{1- v^n\, x_{d}/z}{1- t\, v^n\, x_{d}/z}  \left\langle\frac{\widetilde{Y}^{(n)}_{3d}(z\, v^{-n+1})}{\widetilde{Y}^{(n-1)}_{3d}(z\, v^{-n})}\right\rangle\nonumber\\
&\qquad\qquad+\prod_{a=1}^n \widetilde{\fq}^{(a)}\, \left\langle\frac{1}{\widetilde{Y}^{(n)}_{3d}(z\, v^{-n-1})}\right\rangle\; .\nonumber
\end{align}
For the $T_\rho[SU(N^{(n+1)})]$ theory on the bottom of Figure \ref{fig:Anexamples}, we compute:	
\begin{align}
&\left[{\chi}\right]^{(0,\ldots,0,1)}_{3d}(z) =\prod_{d=1}^{N^{(n+1)}}\frac{1- v\, x_{d}/z}{1- t\, v\, x_{d}/z}\left\langle\widetilde{Y}^{(n)}_{3d}(z)  \right\rangle\label{3dpartitionfunctionAmSIMPLE2}\\
&\qquad\qquad + \widetilde{\fq}^{(n)} \,\left\langle\frac{\widetilde{Y}^{(n-1)}_{3d}(z\, v^{-1})}{\widetilde{Y}^{(n)}_{3d}(z\, v^{-2})}\right\rangle\nonumber\\
&\qquad\qquad + \widetilde{\fq}^{(n)}\widetilde{\fq}^{(n-1)} \,  \left\langle\frac{\widetilde{Y}^{(n-2)}_{3d}(z\, v^{-2})}{\widetilde{Y}^{(n-1)}_{3d}(z\, v^{-3})}\right\rangle\nonumber\\
&\qquad\qquad+\ldots\nonumber\\
&\qquad\qquad+\prod_{a=1}^n \widetilde{\fq}^{(a)}\, \left\langle\frac{1}{\widetilde{Y}^{(1)}_{3d}(z\, v^{-n-1})}\right\rangle\; .\nonumber
\end{align}

\begin{figure}[h!]
	\emph{}
	\centering
	\includegraphics[trim={0 0 0 3cm},clip,width=0.85\textwidth]{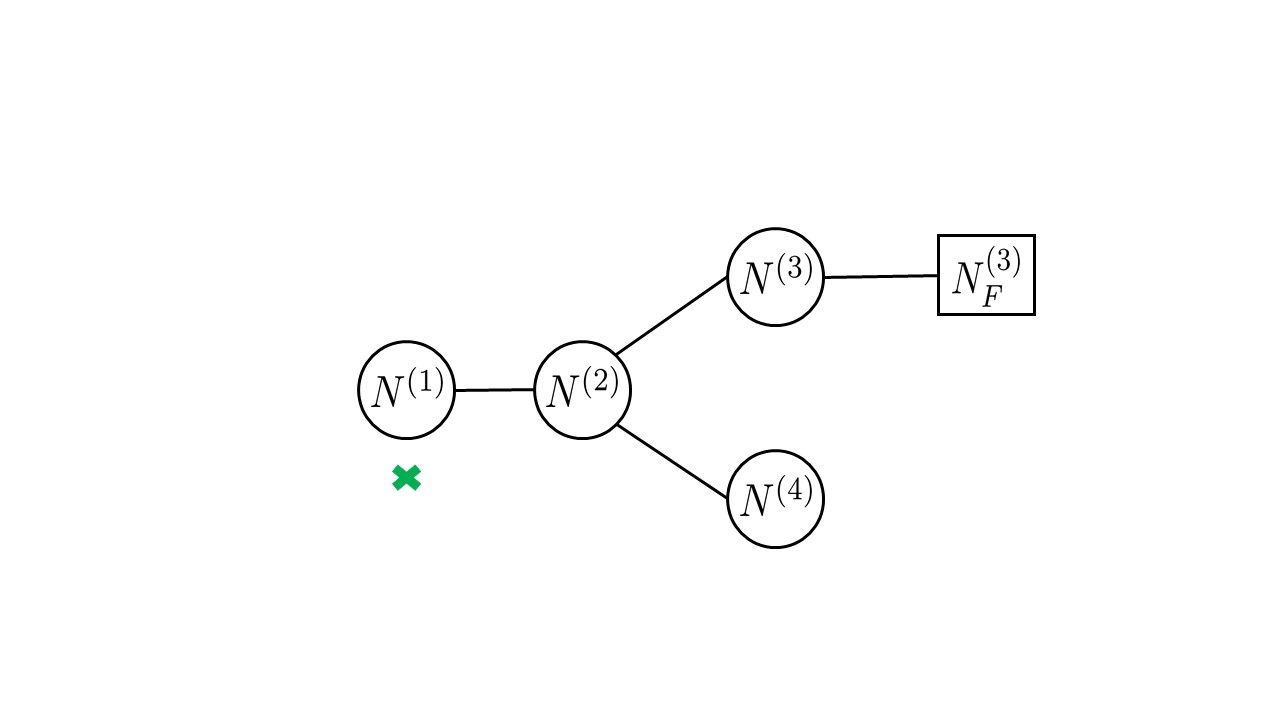}
	\vspace{-15pt}
	\caption{A $D_4$ theory with fundamental matter on node 3, with a loop defect producing the first fundamental vortex character.}
	\label{fig:D4example}
\end{figure}

For the $D_4$ theory in Figure \ref{fig:D4example}, we compute:
\begin{align}
&\left[{\chi}\right]^{(1,0,0,0)}_{3d}(z) =\prod_{d=1}^{N^{(3)}_F}\frac{1- v^3\, x_{d}/z}{1- t\, v^3\, x_{d}/z}\left\langle\widetilde{Y}^{(1)}_{3d}(z)  \right\rangle\label{3dpartitionfunctionD4}\\
&\qquad\qquad + \widetilde{\fq}^{(1)} \, \prod_{d=1}^{N^{(3)}_F}\frac{1- v^3\, x_{d}/z}{1- t\, v^3\, x_{d}/z} \left\langle\frac{\widetilde{Y}^{(2)}_{3d}(z\, v^{-1})}{\widetilde{Y}^{(1)}_{3d}(z\, v^{-2})}\right\rangle\nonumber\\
&\qquad\qquad + \widetilde{\fq}^{(1)}\widetilde{\fq}^{(2)} \,\prod_{d=1}^{N^{(3)}_F}\frac{1- v^3\, x_{d}/z}{1- t\, v^3\, x_{d}/z}  \left\langle\frac{\widetilde{Y}^{(3)}_{3d}(z\, v^{-2})\, \widetilde{Y}^{(4)}_{3d}(z\, v^{-2})}{\widetilde{Y}^{(2)}_{3d}(z\, v^{-3})}\right\rangle\nonumber\\
&\qquad\qquad + \widetilde{\fq}^{(1)}\widetilde{\fq}^{(2)}\widetilde{\fq}^{(3)} \,  \left\langle\frac{\widetilde{Y}^{(4)}_{3d}(z\, v^{-2})}{\widetilde{Y}^{(3)}_{3d}(z\, v^{-4})}\right\rangle\nonumber\\
&\qquad\qquad + \widetilde{\fq}^{(1)}\widetilde{\fq}^{(2)}\widetilde{\fq}^{(4)} \,\prod_{d=1}^{N^{(3)}_F}\frac{1- v^3\, x_{d}/z}{1- t\, v^3\, x_{d}/z}  \left\langle\frac{\widetilde{Y}^{(3)}_{3d}(z\, v^{-2})}{\widetilde{Y}^{(4)}_{3d}(z\, v^{-4})}\right\rangle\nonumber\\
&\qquad\qquad + \widetilde{\fq}^{(1)}\widetilde{\fq}^{(2)}\widetilde{\fq}^{(3)}\widetilde{\fq}^{(4)} \, \left\langle\frac{\widetilde{Y}^{(2)}_{3d}(z\, v^{-3})}{\widetilde{Y}^{(3)}_{3d}(z\, v^{-4})\, \widetilde{Y}^{(4)}_{3d}(z\, v^{-4})}\right\rangle\nonumber\\
&\qquad\qquad + \widetilde{\fq}^{(1)}\left[\widetilde{\fq}^{(2)}\right]^2\widetilde{\fq}^{(3)}\widetilde{\fq}^{(4)} \, \left\langle\frac{\widetilde{Y}^{(1)}_{3d}(z\, v^{-4})}{\widetilde{Y}^{(2)}_{3d}(z\, v^{-5})}\right\rangle\nonumber\\
&\qquad\qquad + \left[\widetilde{\fq}^{(1)}\right]^2\left[\widetilde{\fq}^{(2)}\right]^2\widetilde{\fq}^{(3)}\widetilde{\fq}^{(4)} \,  \left\langle\frac{1}{\widetilde{Y}^{(1)}_{3d}(z\, v^{-6})}\right\rangle\; . \nonumber
\end{align}

\end{appendix}

\newpage
\bibliography{summaryqqVortexJHEP}
\bibliographystyle{JHEP}

\end{document}